\documentclass[preprint]{aastex}

\usepackage{emulateapj5,times,mathptm}

\def\asca{{\it ASCA\/}}

\def\chandra{{\it Chandra\/}}

\def\genx{{\it Generation-X\/}}

\def\rosat{{\it ROSAT\/}}

\def\sax{{\it BeppoSAX\/}}

\def\xeus{{\it XEUS\/}}
\def\heao{{\it HEAO 1\/}}
\def\xmm{{\it XMM-Newton\/}}

\def\etal{{et\,al.\,}}

\def\ltsima{$\; \buildrel < \over \sim \;$}
\def\simlt{\lower.5ex\hbox{\ltsima}}
\def\gtsima{$\; \buildrel > \over \sim \;$}
\def\simgt{\lower.5ex\hbox{\gtsima}}

\journalinfo{The Astronomical Journal, 2003 August, astro-ph/0304392}

\begin{document}

%
\title{The Chandra Deep Field-North Survey. XIII. 2~Ms Point-Source Catalogs}
%

\author{D.M.~Alexander,$^1$ 
F.E.~Bauer,$^1$ 
W.N.~Brandt,$^1$
D.P.~Schneider,$^1$
A.E.~Hornschemeier,$^2$ 
C.~Vignali,$^1$
A.J.~Barger,$^{3,4,5}$ 
P.S.~Broos,$^1$ 
L.L.~Cowie,$^5$ 
G.P.~Garmire,$^1$ 
L.K.~Townsley,$^1$
M.W.~Bautz,$^6$
G.~Chartas,$^1$
and W.L.W.~Sargent$^7$ 
}

\footnotetext[1]{Department of Astronomy \& Astrophysics, 525 Davey Laboratory, 
The Pennsylvania State University, University Park, PA 16802}

\footnotetext[2]{Chandra Fellow, Department of Physics and Astronomy, 
Johns Hopkins University, 3400 North Charles Street, Baltimore, MD 21218}

\footnotetext[3]{Department of Astronomy, University of Wisconsin-Madison,
475 N. Charter Street, Madison, WI 53706}

\footnotetext[4]{University of Hawaii, 2505 Correa Road, Honolulu, HI 96822} 

\footnotetext[5]{Institute for Astronomy, University of Hawaii, 
2680 Woodlawn Drive, Honolulu, HI 96822} 

\footnotetext[6]{Massachusetts Institute of Technology, Center for Space Research, 70 Vassar Street, Building 37, Cambridge, MA 02139}

\footnotetext[7]{Palomar Observatory, California Institute of Technology, Pasadena, CA 91125}

\shorttitle{2~MS CHANDRA POINT-SOURCE CATALOGS}

\shortauthors{ALEXANDER ET AL.}

\slugcomment{Received 2003 January 3; accepted 2003 April 21}

%
\begin{abstract}
%

We present point-source catalogs for the $\approx$~2~Ms exposure of
the \chandra\ Deep Field-North, currently the deepest \hbox{X-ray}
observation of the Universe in the 0.5--8.0~keV band. Five hundred and
three (503) \hbox{X-ray} sources are detected over an
$\approx$~448~arcmin$^2$ area in up to seven \hbox{X-ray}
bands. Twenty (20) of these X-ray sources lie in the central
$\approx$~5.3~arcmin$^2$ Hubble Deep Field-North
($13600^{+3800}_{-3000}$ sources deg$^{-2}$). The on-axis sensitivity
limits are $\approx 2.5\times 10^{-17}$~erg~cm$^{-2}$~s$^{-1}$
(0.5--2.0~keV) and $\approx 1.4\times 10^{-16}$~erg~cm$^{-2}$~s$^{-1}$
(2--8~keV). Source positions are determined using matched-filter and
centroiding techniques; the median positional uncertainty is $\approx
0\farcs 3$. The \hbox{X-ray} colors of the detected sources indicate a
broad variety of source types, although absorbed AGNs (including a
small number of possible Compton-thick sources) are clearly the
dominant type. We also match lower significance \hbox{X-ray} sources
to optical counterparts and provide a list of 79 optically bright
($R\simlt$~23) lower significance \chandra\ sources. The majority of
these sources appear to be starburst and normal galaxies.

The average backgrounds in the 0.5--2.0~keV and 2--8~keV bands are
0.056 and 0.135 counts Ms$^{-1}$ pixel$^{-1}$, respectively. The
background count distributions are very similar to Poisson
distributions. We show that this $\approx$~2~Ms
exposure is approximately photon limited in all seven \hbox{X-ray}
bands for regions close to the aim point, and we predict that
exposures up to $\approx$~25~Ms (0.5--2.0~keV) and $\approx$~4~Ms
(2--8~keV) should remain nearly photon limited. We demonstrate that
this observation does not suffer from source confusion within
$\approx$~6$\arcmin$ of the aim point, and future observations are
unlikely to be source-confusion limited within $\approx$~3$\arcmin$ of
the aim point even for source densities exceeding 100,000
deg$^{-2}$. These analyses directly show that
\chandra\ can achieve significantly higher sensitivities in an
efficient nearly photon-limited manner and be largely free of source
confusion.

To allow consistent comparisons, we have also produced point-source
catalogs for the $\approx$~1~Ms \chandra\ Deep Field-South (CDF-S). Three
hundred and twenty-six (326) X-ray sources are included in the main
\chandra\ catalog, and an additional 42 optically bright X-ray sources
are included in a lower significance \chandra\ catalog. We find good
agreement with the photometry of the previously published CDF-S
catalogs; however, we provide significantly improved positional
accuracy.

\end{abstract}



%
\section{Introduction}
%

One of the primary scientific goals behind the construction of the
{\it Chandra X-ray Observatory\/} (hereafter {\it Chandra}; Weisskopf
et~al. 2000) was to perform the deepest possible X-ray studies of the
Universe. Great advances in this direction were made with the
completion of two $\approx$~1~Ms surveys: the Chandra Deep Field-North
(CDF-N; Brandt et~al. 2001a, hereafter B01), and the Chandra Deep
Field-South (CDF-S; Giacconi et~al. 2002, hereafter G02). These
ultra-deep {\it Chandra} surveys resolve the bulk of the
\hbox{0.5--8.0}~keV background, providing the deepest views of the
Universe in this band (e.g.,\ Campana et~al. 2001; Cowie et~al. 2002;
Rosati et~al. 2002; Moretti \etal 2003). A broad variety of source
types is detected, including (in approximately decreasing
source-density order) absorbed and unabsorbed Active Galactic Nuclei
(AGN), starburst galaxies, normal galaxies, stars, galaxy groups, and
galaxy clusters (see Brandt
\etal 2002 and Hasinger
\etal 2002 for reviews of the \chandra\ Deep Field results, and
Barger \etal 2002 and Szokoly \etal 2003 for the \chandra\ Deep Field
optical catalogs and spectroscopy).

These 1~Ms \chandra\ surveys are $\approx$~2 orders of magnitude
more sensitive in the \hbox{0.5--2.0}~keV and \hbox{2--8}~keV bands than any
X-ray survey performed before the launch of \chandra\ and
\xmm\ (Jansen \etal 2000). They are also at least three times more sensitive
than any other current \chandra\ or \xmm\ survey (see Figure~1). With
a small on-axis point-spread function (PSF) and low background, an
ultra-deep \chandra\ observation is unlikely to be background or
source-confusion limited, and longer \chandra\ exposures should
achieve significantly higher sensitivities in an efficient nearly
photon limited manner. Since several analyses have suggested a steep
rise in the number counts of X-ray sources around 0.5--2.0~keV fluxes
of $\approx$~(3--7)~$\times 10^{-18}$~erg~cm$^{-2}$~s$^{-1}$ (e.g.,\
Ptak et~al. 2001; Hornschemeier et~al. 2002; Miyaji \& Griffiths 2002;
Ranalli \etal 2003), longer \chandra\ exposures should also detect
significantly more sources. Achieving such sensitivities is one of the
main goals of future X-ray observatories (e.g.,\ \xeus\ and
\genx). However, X-ray stacking analysis results have directly shown
that \chandra\ can reach these flux levels with exposures of several
Ms (e.g.,\ Brandt \etal 2001c; Hornschemeier et~al. 2002).

The CDF-N was recently awarded a second $\approx$~1~Ms of \chandra\
exposure, bringing the total exposure in this field to
$\approx$~2~Ms. In this paper we present two point-source catalogs
derived from the full $\approx$~2~Ms exposure: a main catalog that
includes high-significance \chandra\ sources and a supplementary
catalog that includes lower significance
\chandra\ sources that are matched to optically bright ($R\simlt$~23) 
counterparts. We provide basic analyses of the detected sources;
however, we defer detailed analyses to other papers. For instance, Barger
\etal (2003a) and Hornschemeier \etal (2003) investigate the overall properties
of the X-ray detected sources, Vignali \etal (2002a,b), Alexander
\etal (2003), and Bauer \etal (2003) perform X-ray spectral analyses
of various X-ray source populations, and Barger \etal (2003b) place
constraints on the number of $z>5$ AGNs. We also include analyses of
the background and X-ray sensitivity across the field, determine the
fraction of the field that is close to photon limited, and calculate
the source-confusion limit. We use this information to discuss the
prospects for longer \chandra\ exposures. In order to allow consistent
comparisons between the CDF-N and CDF-S observations, we have also
produced point-source catalogs for the CDF-S (see the Appendix).

The CDF-N catalogs presented here supersede those presented in \hbox{B01}. The
Galactic column density along this line of sight is
\hbox{(1.6$\pm$0.4)}$\times 10^{20}$~cm$^{-2}$ (Stark et~al. 1992).
The coordinates throughout this paper are J2000. The Hubble Deep
Field-North (HDF-N; Williams \etal 1996) lies close to the aim point
of the CDF-N and is centered at $\alpha_{2000}=$~12$^{\rm h}$ 36$^{\rm
m}$ 49\fs 4,
$\delta_{2000}=$~$+62^\circ$12$^{\prime}$58$^{\prime\prime}$
($l=125\fdg 888$, $b=54\fdg 828$). $H_0=65$~km~s$^{-1}$ Mpc$^{-1}$,
$\Omega_{\rm M}=1/3$, and $\Omega_{\Lambda}=2/3$ are adopted.

%
\section{Observations and Data Reduction}
%

The observational procedure and data processing were similar to those
described in B01. For completeness, we provide all of the details
here.

\subsection{Observations and Observing Conditions}

The Advanced CCD Imaging Spectrometer (ACIS; Garmire et~al. 2003) was
used for all of the \chandra\ observations.\footnote{For additional
information on the ACIS and \chandra\ see the \chandra\ Proposers'
Observatory Guide at http://cxc.harvard.edu/udocs/docs.} ACIS consists
of ten CCDs (each $1024\times1024$ pixels) designed for efficient
X-ray detection and spectroscopy. Four of the CCDs (ACIS-I; CCDs
I0--I3) are arranged in a $2\times 2$ array with each CCD tipped
slightly to approximate the curved focal surface of the \chandra\ High
Resolution Mirror Assembly (HRMA). The remaining six CCDs (ACIS-S;
CCDs S0--S5) are set in a linear array and are tipped to approximate
the Rowland circle of the objective gratings that can be inserted
behind the HRMA. The CCD which lies on-axis in ACIS-I is I3.

The full ACIS-I field of view is $16\farcm 9\times 16\farcm 9$
($\approx$~285 arcmin$^2$), and the ACIS pixel size is $\approx
0\farcs 492$. The PSF is smallest at the softest energies and for
sources at small off-axis angles. For instance, the 95\%
encircled-energy radius at 1.5~keV for off-axis angles of
$0^{\prime}$--$8^{\prime}$ is $\approx 1\farcs 8$--$7\farcs 5$
(Feigelson, Broos, \& Gaffney 2000; Jerius et~al. 2000; M.~Karovska
and P.~Zhao 2001, private communication).\footnote{Feigelson
et~al. (2000) is available at
http://www.astro.psu.edu/xray/acis/memos/memoindex.html.} The PSF is
approximately circular at small off-axis angles, broadens and
elongates at intermediate off-axis angles, and becomes complex at
large off-axis angles. The four \hbox{ACIS-I} CCDs were operated in
all of the 20 observations that comprise the $\approx$~2~Ms exposure,
while the ACIS-S CCD S2 was used for the first 12 observations. Due to
its large off-axis angle, and consequently its low sensitivity, CCD S2
is not used in this analysis.

The second $\approx$~1~Ms of \chandra\ exposure consisted of eight
separate observations taken between 2001 Nov 16 and 2002 Feb 22. These
eight new observations and the 12 observations that comprised the
initial $\approx$~1~Ms exposure are described in Table~1. The HDF-N
was placed near the aim point of the ACIS-I array for the majority of
the observations. The eight new observations were taken in Very Faint
mode to improve the screening of background events and thus increase
the sensitivity of ACIS in detecting faint X-ray sources.\footnote{For
more information on the Very Faint mode see
http://cxc.harvard.edu/udocs/vf.html and Vikhlinin (2001).} The
focal-plane temperature was $-120^\circ$\,C for all of the eight new
observations.

The background light curves for all 20 observations were inspected
using {\sc event browser} in the Tools for ACIS Real-time Analysis
({\sc tara}; Broos et~al. 2000) software package.\footnote{{\sc tara}
is available at http://www.astro.psu.edu/xray/docs.} All but two are
free from strong flaring due to ``space weather'' and are stable to
within $\approx 20$\%. The two observations with substantial flaring
are 2344 and 3389. Due to high solar activity, the background was
$\simgt 2$ times higher than nominal for $\approx 2.5$~ks of
observation 2344, and these data have been removed. The background
rose dramatically toward the end of observation 3389, and the
observation was cut short; $\approx 17.0$~ks of the data at the end of
this observation have been removed. The total exposure time for the 20
observations is 1.945~Ms.

Due to the different pointings required to keep the HDF-N near the aim
point and the roll constraints of \chandra, the total region covered
by these 20 observations is 447.8~arcmin$^2$, considerably larger than
the \hbox{ACIS-I} field of view. Combining the 20 observations, the
average aim point weighted by exposure time is
$\alpha_{2000}=$~12$^{\rm h}$ 36$^{\rm m}$ $45\fs 7$,
$\delta_{2000}=$~$+62^\circ$13$^{\prime}$58$^{\prime\prime}$ (see
Table~1). The aim points of the individual observations are separated
from the average aim point by $1\farcm 1$--$3\farcm 8$; most are
within $1\farcm 5$ of the average aim point. Due to the two main roll
angles employed througout these observations, and our requirement to
keep the HDF-N close to the aim point, none of these observations lies
closer to the average aim point.

\subsection{Data Reduction}

The versions of the \chandra\ X-ray Center (hereafter CXC) pipeline
software used for basic processing of the data are listed in
Table~1. The reduction and analysis of the data used
\chandra\ Interactive Analysis of Observations ({\sc ciao}) Version~2.2.1
tools whenever possible;\footnote{See http://cxc.harvard.edu/ciao/ for
details on {\sc ciao}.} however, the {\sc tara} software package and
custom software were also used extensively.

All data were corrected for the radiation damage sustained by the CCDs
during the first few months of \chandra\ operations using the Charge
Transfer Inefficiency (CTI) correction procedure of Townsley
et~al. (2000, 2002).\footnote{The software associated with the
correction method of Townsley et~al. (2000, 2002) is available at
http://www.astro.psu.edu/users/townsley/cti/.} In addition to
correcting partially for the positionally dependent grade distribution
due to CTI effects, this procedure also partially corrects for quantum
efficiency losses (see Townsley et~al. 2000, 2002 for further details).

All bad columns, bad pixels, and cosmic ray afterglows were removed
using the ``status'' information in the event files, and we only used
data taken during times within the CXC-generated good-time
intervals. The {\sc ciao} tool {\sc acis\_process\_events} was used to
remove the standard pixel randomization and to identify potential
background events for the observations taken in Very Faint mode.

%
\section{Production of the Point-source Catalogs}
%

The production of the two point-source catalogs closely followed the
procedure described in \S3 of B01. The main differences in the
catalog production procedure used here are the use of a matched-filter
technique to improve the accuracy of the X-ray source positions, the
addition of a further three standard source-detection bands, and the
correlation of optically bright sources with lower significance
\chandra\ sources; the latter modification is only relevant for the
supplementary optically bright \chandra\ source catalog. For
completeness, we provide all of the catalog production details here.

\subsection{Image and Exposure Map Creation}

Each observation was registered to observation 3293 using 7--19 X-ray
sources (typically with $>$50 counts) detected within $\approx
6\arcmin$ of the aim points in the individual observations;
observation 3293 was chosen because it is one of the longest
observations and the raw coordinates are closely matched to the radio
astrometric frame (see \S3.3). Image registration was performed
using the {\it align\_evt} tool written by
T.~Aldcroft\footnote{Further information about the {\it align\_evt}
tool can be obtained from
http://cxc.harvard.edu/cal/ASPECT/align\_evt/.}; registration is
accurate to within $\approx 0\farcs 4$. The \chandra\ sources used in
the registration were identified with the sliding-cell source
detection program {\sc celldetect} (Calderwood \etal 2001).

We constructed images using both the standard \asca\ grade set and the
``restricted ACIS grade set'' for seven standard bands: 0.5--8.0~keV
(full band; FB), 0.5--2.0~keV (soft band; SB), 2--8~keV (hard band;
HB), 0.5--1~keV (SB1), 1--2~keV (SB2), 2--4~keV (HB1), and 4--8~keV
(HB2); see Table~2. The restricted ACIS grade set improves our
ability to detect faint sources in some cases; see \S2.1 in Brandt
et~al. (2001b). In Figures 2 and 3, we show the full-band raw and
adaptively smoothed \asca-grade images.\footnote{Raw and adaptively
smoothed \asca-grade images for all of the seven standard bands (see
Table~2) are available from
http://www.astro.psu.edu/users/niel/hdf/hdf-chandra.html.} The
adaptively smoothed images were not used for source detection, but
they do show many of the detected X-ray sources more clearly than the
raw images. In Figure 4 we show a color composite of the adaptively
smoothed 0.5--2.0~keV, 2--4~keV, and 4--8~keV images. Soft sources
appear red, moderately hard sources appear green, and the hardest
sources appear blue.

We constructed exposure maps in the seven standard bands. In Figure 5,
we show the full-band exposure map.\footnote{Exposure maps for all of
the seven standard bands (see Table~2) are available from the World
Wide Web site listed in Footnote 8.} These were created following the
basic procedure outlined in \S3.2 of Hornschemeier et~al. (2001) and
are normalized to the effective exposure of a source located at the
aim point. Briefly, this procedure takes into account the effects of
vignetting, gaps between the CCDs, bad column filtering, and bad pixel
filtering. However, it does not take into account the $\approx$~2--5\%
CCD ``dead time'' due to cosmic ray blooming since the magnitude of
this effect is not well quantified. A photon index of $\Gamma=1.4$,
the slope of the X-ray background in the $\approx$~0.5--8.0~keV band
(e.g.,\ Marshall \etal 1980; Gendreau \etal 1995; Kushino \etal 2002),
was assumed in creating the exposure maps. These exposure maps need to
be convolved with the background maps and off-axis angle dependent PSF
to generate sensitivity maps; see \S4.2.

In Figure 6, we show a cumulative plot of the survey solid angle as a
function of full-band effective exposure. Approximately 51\%
($\approx$~230 arcmin$^2$) of the CDF-N field has $>1$~Ms of effective
exposure.

\subsection{Point-source Detection}

We extended the number of standard source-detection bands over that
presented in B01 to seven. These detection bands were chosen to allow
X-ray color analysis and the selection of sources in narrow X-ray
bands; the latter can be useful for detecting weak sources that
produce X-ray emission over a narrow energy band (e.g.,\ the
low-energy thermal emission from normal galaxies, or AGNs with large
equivalent width emission lines) and in comparing the rest-frame X-ray
emission from sources at different redshifts (e.g.,\ the rest-frame
1--2~keV emission from a $z=1$ source is observed at
0.5--1.0~keV). Two source-detection grade sets were used: the
``restricted ACIS grade set'', and the ``standard \asca\ grade set''
(see Table~2). All photometry is reported using the standard \asca\
grade set.

Point-source detection was performed in each band with {\sc wavdetect}
(Freeman et~al. 2002) using a ``$\sqrt{2}$~sequence'' of wavelet
scales (i.e.,\ 1, $\sqrt{2}$, 2, $2\sqrt{2}$, 4, $4\sqrt{2}$, and 8
pixels). Previous studies have shown that using larger scales can
detect a few additional sources at large off-axis angles (see
\S3.2.2 in B01). However, we have not searched for sources using 
larger scales here because the source properties and positions are too
poorly defined to give useful results. Our key criterion for source
detection is that a source must be found with a given false-positive
probability threshold in at least one of the seven standard bands
using either the standard \asca\ or restricted ACIS grade sets. The
chosen false-positive probability thresholds in each band are $1\times
10^{-7}$ and $1\times 10^{-5}$ for the main source catalog and the
supplementary optically bright source catalog, respectively.

If we treat the 14 images (i.e.,\ the seven \asca-grade images and the
seven restricted ACIS-grade images) searched as entirely independent,
$\approx 9$ false sources are expected in the main \chandra\ source
catalog for the case of a uniform background. As mentioned in B01,
due to the large variation in effective exposure time across the field
and the increase in background near bright point sources (due to the
PSF ``wings''), we might expect the number of false sources to be
increased by a factor of $\approx$~2--3. However, since {\sc wavdetect}
suppresses fluctuations on scales smaller than the PSF, a single pixel
is unlikely to be considered a source detection cell, particularly at
large off-axis angles. Hence, our false-source estimates are likely
to be conservative. Indeed, we argue in \S3.4.1 that the true number
of false sources is likely to be considerably lower than our
predictions. We provide an estimation of the number of false
sources in the supplementary optically bright
\chandra\ source catalog in \S3.4.2.

We also searched for sources in the 6--8~keV band. Although a few
intrinsically hard X-ray sources were detected in this very hard band
(12 sources had $\Gamma<1.0$), the bulk of the 34 detected sources
were found to be bright X-ray sources with comparatively steep X-ray
spectral slopes. Only one very faint source was detected in this band
and not detected in all of the other bands (the source position is
$\alpha_{2000}=$~12$^{\rm h}$ 36$^{\rm m}$ 30\fs 5,
$\delta_{2000}=$~$+62^\circ$19$^{\prime}$26$^{\prime\prime}$);
however, since $\approx$~1 false source is formally expected in this
band for a $1\times 10^{-7}$ false-positive probability threshold,
this source may not be real. We do not provide counts and flux
information for the sources detected in this band, but we do indicate
the detected sources in column~13 of Table~3a (see \S3.4.1).

\subsection{Astrometry}

We refined the absolute X-ray source positions by matching X-ray
sources from the main point-source catalog (see \S3.4.1) to 1.4~GHz
radio sources detected by Richards (2000). The 240 1.4~GHz sources
detected across the CDF-N field have accurate ($\simlt 0\farcs 3$)
individual positions, and the Richards (2000) radio image is tied to
the FK5 reference frame to an accuracy of $0\farcs 03$. The X-ray
sources were matched to the radio sources using a $2\farcs 5$ search
radius. One hundred (100) X-ray sources were found to have a radio
counterpart (these sources are indicated in column 13 of Table~3a; see
\S3.4.1). The vast majority of these matches are expected to be
correct, but $\approx 2$ are statistically expected to be false
matches. We also note that in some cases the X-ray source may be
offset from the radio source even though both are associated with the
same galaxy (e.g., a galaxy with a radio-emitting nuclear starburst
that also has an off-nuclear ultraluminous X-ray binary). From a
comparison of the X-ray and radio source positions, we found small
shift and plate-scale corrections. The plate-scale correction
[$\approx$~$0\farcs 35$ every 12$\arcmin$ in both right ascension (RA)
and declination (Dec)] implies a $\approx$~0.05\% adjustment to the
pixel size; we found the same correction was required for the CDF-S
observations (see the Appendix). These corrections have been applied
to the X-ray source positions.

We investigated the accuracy of the X-ray source positions using the
100 X-ray detected radio sources. In Figure~7 we show the positional
offset between the X-ray and radio sources versus off-axis angle. The
median offset is $\approx 0\farcs 3$; however, there are also clear
off-axis angle and source-count dependencies. The off-axis angle
dependence is due to the HRMA PSF becoming broad at large off-axis
angles, while the count dependency is due to the difficulty of
centroiding a faint X-ray source. The median offset of the bright
X-ray sources ($\ge$~200 full-band counts) is only $\approx 0\farcs
2$, and almost all sources have offsets within $0\farcs 5$. The median
offset of the faint X-ray sources ($<$~200 full-band counts) is
$\approx 0\farcs 4$. However, some faint X-ray sources can have
offsets as large as 1--2$\arcsec$ at large off-axis angles; for these
sources we verified that the radio emission is not extended by
visually inspecting the 1.4~GHz image. Since the positional
uncertainty of individual radio sources can be up to $\approx 0\farcs
3$, it is possible that the accuracy of the X-ray source positions is
partially limited by the radio data. The positional uncertainty of
each source is estimated following equations 1 \& 2 (see \S3.4.1).

\subsection{Point-source Catalogs}

\subsubsection{Main \chandra\ Source Catalog}

We ran {\sc wavdetect} with a false-positive probability threshold of
$1\times 10^{-7}$ on all of the 14 images. The resulting source lists
were then merged to create the point-source catalog given as Tables~3a
and 3b. For cross-band matching, a matching radius of $2\farcs 5$ was
used for sources within $6\arcmin$ of the average aim point. For
larger off-axis angles, a matching radius of $4\farcs 0$ was used.
These matching radii were chosen based on inspection of histograms
showing the number of matches obtained as a function of angular
separation (see \S2 of Boller et~al. 1998); with these radii the
mismatch probability is $\approx 1$\% over the entire field.

We also used {\sc wavdetect} to search the seven standard-band
\asca-grade images for lower-significance, cross-band counterparts to
the highly significant sources already detected at the $1\times
10^{-7}$ level in at least one of the seven standard bands; in these
runs we used a false-positive probability threshold of $1\times
10^{-5}$. We found 195 additional cross-band counterparts in this
manner. Since the spatial-matching requirement greatly reduces the
number of pixels being searched, only $\approx 1$ of these cross-band
counterparts is expected to be false.

We improved upon the {\sc wavdetect} source positions using a
matched-filter technique. This technique convolves the full-band image
of each source with a combined PSF. The combined PSF is automatically
generated as part of the {\sc acis\_extract} procedure (Broos \etal
2002) within {\sc tara} (see Footnote 4) and is produced by combining
the theoretical PSF of a source for each observation, weighted by the
number of detected counts.\footnote{{\sc acis\_extract} can be
accessed from
http://www.astro.psu.edu/xray/docs/TARA/ae\_users\_guide.html. The
theoretical PSFs are taken from the CXC PSF library; see
http://asc.harvard.edu/ciao2.2/documents\_dictionary.html\#psf.} This
technique takes into account of the fact that, due to the complex PSF
at large off-axis angles, the peak of the X-ray emission does not
always correspond to the X-ray source position. The matched-filter
technique provides a significant improvement ($\approx 0\farcs1$ on
average) in the positional accuracy for sources further than
5$\arcmin$ from the average aim-point. For smaller off-axis angles
there was no overall improvement, and the source positions were
determined using a simple centroiding algorithm.

Manual correction of the source properties and source positions were
required in some special cases. We manually separated five close
doubles ($2\farcs 5$--$3\farcs 5$) and determined the position of each
separated source. These sources incur larger photometric errors due to
the difficulty of the separation process. It was also necessary to
perform manual photometry for 75 sources that were located close to
bright sources, lay in regions of high background, or lay partially
outside of the field. Finally, visual inspection of the X-ray source
positions showed small offsets from the peak of the X-ray emission for
31 sources; we took into account of the fact that the peak of the
X-ray emission does not always correspond to the X-ray source position
for sources at large off-axis angles. The positions of these sources
were adjusted manually. We have flagged sources that required manual
correction in column~13 of Table~3a (see below).

The main \chandra\ source catalog is presented as Tables 3a and
3b. Table~3a provides the basic source properties: source position and
uncertainty, source counts in each band, and additional
notes. Details of the columns in Table~3a are given below.

\begin{itemize}

\item
Column~1 gives the source number. Sources are listed in order of
increasing RA.

\item
Columns~2 and 3 give the RA and Dec of the X-ray
source, respectively. These positions have been determined using the
procedure described above. To avoid truncation error, we quote the
positions to higher precision than in the International Astronomical
Union approved names beginning with the acronym
``CXO~HDFN.''\footnote{See http://cxc.harvard.edu/udocs/naming.html.}

\item
Column~4 gives the positional uncertainty. As shown in
\S3.3, the positional uncertainty is dependent on off-axis 
angle and the number of detected counts. For the brighter X-ray
sources ($\ge$~200 full-band counts) the positional uncertainties are
given by the empirically determined equation:

\begin{equation}
\Delta= \left\{\begin{array}{ll}
0.3 & \theta<5\arcmin \\
 & \\
0.3+\left({\theta-5\arcmin\over 25\arcmin}\right) & \theta\ge5\arcmin \\
\end{array}
\right.
\end{equation}

\noindent
where $\Delta$ is the positional uncertainty in arcsec and $\theta$ is
the off-axis angle in arcmin (compare with Figure~7). The stated
positional uncertainties are for $\approx$~80--90\% confidence (see
also Barger \etal 2003a for analyses using optical data), and the
accuracy of our astrometric solution is discussed in \S3.3.

For the fainter X-ray sources ($<$~200 full-band counts) the positional
uncertainties are given by the empirically determined equation:

\begin{equation}
\Delta= \left\{\begin{array}{ll}
0.6 & \theta<5\arcmin \\
 & \\
0.6+\left({\theta-5\arcmin\over 6.25\arcmin}\right) & \theta\ge5\arcmin \\
\end{array}
\right.
\end{equation}

\item
Column~5 gives the off-axis angle for each source in arcminutes. This
is calculated using the source position given in columns~2 and 3 and
the average aim point (see Table~1).

\item
Columns~6--12 give the counts in the seven standard bands. All values
are for the standard \asca\ grade set, and they have not been
corrected for vignetting. Source counts and $1\sigma$ statistical
errors (from Gehrels 1986) have been calculated using circular
aperture photometry; extensive testing showed this method to be more
reliable than the {\sc wavdetect} photometry. The circular aperture
was centered at the position given in columns 2 and 3 for all bands. 

The local background is determined in an annulus outside of the
source-extraction region. The mean number of background counts per
pixel is calculated from a Poisson model using ${n_1}\over{n_0}$,
where $n_0$ is the number of pixels with 0 counts and $n_1$ is the
number of pixels with 1 count. Although only the numbers of pixels with
0 and 1 counts are measured, this technique directly provides the mean
background even when ${n_1}\gg{n_0}$. This technique is more robust
than ${n_0}\over{{n_0}+{n_1}}$, which is only applicable when the mean
background is $\ll1$ count pixel$^{-1}$. Furthermore, by ignoring all
pixels with $>1$ count, this technique guards against background
contamination from sources. The principal requirement for using this
technique is that the background follows a Poisson distribution; we
show in \S4.2 that the \hbox{ACIS-I} background matches this
criterion. The total background for each source is calculated and
subtracted to give the net number of source counts.

For sources with fewer than 1000 full-band counts, we have chosen the
aperture radii based on the encircled-energy function of the \chandra\
PSF as determined using the CXC's {\sc mkpsf} software (Feigelson
et~al. 2000; Jerius et~al. 2000; M.~Karovska and P.~Zhao 2001, private
communication). In the softest bands (i.e.,\ SB, SB1, and SB2), where
the image quality is the best, the aperture radius was set to the 95\%
encircled-energy radius of the PSF. In the other bands, the 90\%
encircled-energy radius of the PSF was used. Appropriate aperture
corrections were applied to the source counts.

For sources with more than 1000 full-band counts, systematic errors in
the aperture corrections often exceed the expected errors from photon
statistics when the apertures described in the previous paragraph are
used. Therefore, for such sources we used larger apertures to minimize
the importance of the aperture corrections; this is appropriate since
these bright sources dominate over the background. We set the aperture
radii to be twice those used in the previous paragraph and manually
inspected these sources to verify that the measurements were not
contaminated by neighboring objects.

We have performed several consistency checks to verify the quality of
the photometry. For example, we have checked that the sum of the
counts measured in the soft and hard bands does not differ from the
counts measured in the full band by an amount larger than that
expected from measurement error (we also performed similar tests for
SB1, SB2, HB1, and HB2). Systematic errors in our photometry are
estimated to be $\simlt 4$\%.

We have verified that the cosmic ray afterglow removal procedure (see
\S2.2) has not led to significant systematic photometric errors. Due 
to the low count rates of our sources, incident X-ray photons are
almost never incorrectly flagged as afterglow events. To guard against
the possibility that some of the faintest X-ray sources are cosmic ray
afterglows missed by the cosmic ray afterglow removal procedure, we
inspected the photon arrival times of all of the X-ray sources with
$\le$10 counts in either the soft or hard band. None of the sources
appeared to be a cosmic ray afterglow.

When a source is not detected in a given band, an upper limit is
calculated.  All upper limits are determined using the circular
apertures described above.  When the number of counts in the aperture
is $\leq 10$, the upper limit is calculated using the Bayesian method
of Kraft, Burrows,
\& Nousek (1991) for 99\% confidence. The uniform prior used by these
authors results in fairly conservative upper limits (see Bickel 1992),
and other reasonable choices of priors do not materially change our
scientific results.  For larger numbers of counts in the aperture,
upper limits are calculated at the $3\sigma$ level for Gaussian
statistics.

\item
Column~13 gives notes on the sources.
``H'' denotes objects lying in the \hbox{HDF-N} (see Figure~2).
``O'' refers to objects that have large cross-band positional offsets
($>2\farcs 5$). All but one of these sources lie at off-axis angles of
$>6\arcmin$, and the flagged counterpart may not be associated with
the main counterpart in some cases. ``M'' refers to sources where the
photometry was performed manually. ``S'' refers to close-double
sources where manual separation was required.  ``R'' refers to
radio-detected sources that were used in determining the astrometric
accuracy (see \S3.3).  ``HB3'' refers to a detection in the 6--8~keV
band.  For further explanation of many of these notes, see the above
text in this section on manual correction of the {\sc wavdetect}
results.

\end{itemize}

\noindent Table~3b provides additional source properties: effective exposure,
band ratios, the effective photon index, and source fluxes. Details of
the columns in Table~3b are given below.

\begin{itemize}

\item
Column~1 gives the source number (see column 1 of Table~3a for details).

\item
Columns~2 and 3 give the RA and Dec of the X-ray
source, respectively (see columns~2 and 3 of Table~3a for details).

\item
Column~4 gives the effective exposure time derived from the full-band
exposure map (see \S3.1 for details on the exposure maps). Dividing
the counts listed in columns 6--12 of Table~3a by the effective
exposure will provide vignetting-corrected count rates.

\item
Column~5 gives the main band ratio, defined as the ratio of counts
between the hard and soft bands (see columns 14--16 for descriptions
of the other band ratios). Quoted band ratios have been corrected for
differential vignetting between the hard band and soft band using the
appropriate exposure maps. Errors for this quantity are calculated
following the ``numerical method'' described in \S1.7.3 of Lyons
(1991); this avoids the failure of the standard approximate variance
formula when the number of counts is small (see \S2.4.5 of Eadie
et~al. 1971). Note that the error distribution is not Gaussian when
the number of counts is small.

\item
Column~6 gives the effective photon index ($\Gamma$) for a power-law
model with the Galactic column density, taking into account the
continuous ACIS quantum efficiency degradation described in columns
7--13. The effective photon index has been calculated based on the
band ratio in column~5 when the number of counts is not low.

A source with a low number of counts is defined as being (1) detected
in the soft band with $<30$ counts and not detected in the hard band,
(2) detected in the hard band with $<15$ counts and not detected in
the soft band, (3) detected in both the soft and hard bands, but with
$<15$ counts in each, or (4) detected only in the full band.  When the
number of counts is low, the photon index is poorly constrained and
set to $\Gamma=1.4$, a representative value for faint sources that
should give reasonable fluxes.

\item
Columns~7--13 give observed-frame fluxes in the seven standard bands;
fluxes are in units of $10^{-15}$~erg~cm$^{-2}$~s$^{-1}$.
Fluxes have been computed using the counts in columns~6--12 of
Table~3a, the appropriate exposure maps, and the spectral slopes given
in column~6. The fluxes have not been corrected for absorption by the
Galaxy or material intrinsic to the source. For a power-law model with
$\Gamma=1.4$, the soft-band and hard-band Galactic absorption
corrections are $\approx 4.2$\% and $\approx 0.1$\%,
respectively.

X-ray spectral and variability analyses of the brighter X-ray sources
have shown that many have complex X-ray spectra and vary significantly
in X-ray count rate over the 20 observations (Bauer
\etal 2003). More accurate fluxes for these sources would require
direct fitting of the X-ray spectra for each observation, which is beyond
the scope of this paper.

It has been recently shown that there has been continuous degradation
of the ACIS quantum efficiency since the launch of {\it Chandra}. This
degradation is thought to be due to absorption caused by molecular
contamination of the ACIS optical blocking filters. We have corrected
the fluxes for this absorption using {\sc acisabs} version 1.1, taking
into account the dates of each of the 20 observations.\footnote{The
IDL code for {\sc acisabs} can be obtained from
http://www.astro.psu.edu/users/chartas/xcontdir/xcont.html. See
http://cxc.harvard.edu/cal/Acis/Cal\_prods/qeDeg/ for more information
on the ACIS quantum efficiency degradation.} For a $\Gamma=$~1.4 source,
the soft-band and hard-band corrections are $\approx$~10\% and
$\approx$~0\%, respectively.

\item
Columns~14--16 give additional band ratios. These are used to provide
X-ray color information (see \S4.1). Band ratio 2 is defined as the
ratio of counts between SB2 and SB1, band ratio 3 is defined as the
ratio of counts between HB2 and HB1, and band ratio 4 is defined as
the ratio of counts between HB1 and SB2. Errors for the band ratios
have been calculated following the procedure given in column~5, and
all quoted band ratios have been corrected for differential vignetting
using the appropriate exposure maps (see
\S3.1).

\end{itemize}

\noindent Many of the sources in Table~3 have been presented in earlier 
papers. The source properties given here supersede those presented in
earlier papers.

In Table~4 we summarize the source detections in the seven standard
bands. In total 503 independent point sources are detected; 20
($13600^{+3800}_{-3000}$ deg$^{-2}$) of these lie in the HDF-N. In
Table~5 we summarize the number of sources detected in one band but
not another. All but one of the sources are detected in either the
full band or the soft band; the full-band and soft-band undetected
source is detected in the hard band.

In Figure~8 we show the distributions of detected counts in the full,
soft and hard bands. The median number of full-band counts is
$\approx$~100 which is sufficiently large for basic X-ray spectral
analysis (see Bauer \etal 2003 for X-ray spectral analyses of the
brightest X-ray sources in this field, Vignali \etal 2002a,b for X-ray
spectral analyses of EROs and $z>4$ AGNs, and Alexander \etal 2003
for X-ray spectral analyses of X-ray detected bright submm
sources). In Figure~9 we show the distribution of full-band effective
exposure time. The median exposure time (1.7~Ms) shows that the
majority of the sources have deep X-ray coverage. The two ``bumps'' at
$\approx$~0.7~Ms and $\approx$~0.95~Ms are due to off-axis sources
detected in regions only covered by one of the two main roll angles
(see Table~1). In Figure~10 we show the distributions of X-ray flux in
the full, soft, and hard bands. A broad range of X-ray fluxes is
found, with $\approx$~50\% of the sources having soft-band and
hard-band fluxes of $< 2\times 10^{-16}$~erg~cm$^{-2}$~s$^{-1}$ and $<
2\times 10^{-15}$~erg~cm$^{-2}$~s$^{-1}$, respectively.

In Figure~11 we show ``postage-stamp'' images in the full band for all
detected sources. The broad range of source sizes is primarily due to
the broadening of the HRMA PSF with off-axis angle. A few of the
sources appear to be resolved; in most cases this is an artifact of
the complex \chandra\ off-axis PSF. In Figure~12a we plot the
positions of the sources. This format removes the illusory effect
produced by the changing PSF size across the field of view. The
majority of the sources lie in the most sensitive region of the field
(i.e.,\ close to the average aim point). One hundred and thirty-eight
(138) sources are newly detected in the 2~Ms \chandra\
exposure. Figure~12a shows that the new sources are detected over the
majority of the field.

Five of the 370 sources detected in the 1~Ms \chandra\ exposure are
undetected here (see Table~6). The five undetected sources were weakly
detected in the 1~Ms \chandra\ exposure: four sources were only
detected with the restricted ACIS grade set and with $<$~15 full-band
counts, while the other source (CXOHDFN J123813.4+621134) had 32.9
full-band counts but lay at a large off-axis angle
($\approx$~10$\arcmin$) and could have a significant contribution from
background. Only one of these five sources (CXOHDFN J123633.8+621327)
has an $R<27$ optical counterpart within $2\arcsec$ of the X-ray
position (Barger
\etal 2002; see Table~6). Such faint or absent optical counterparts
raise the possibility that some of these sources are false. We
examined the source regions in the full, soft, and hard bands and found
evidence for possible X-ray emission in two cases (see Table~6);
however, in both cases the emission is not clearly distinct from the
background. We also extracted the X-ray events within 1$\arcsec$ of
the position of each source. In four cases 0--2 counts were found for
the second $\approx$~1~Ms exposure, and in one case 4 counts were found
(CXOHDFN J123627.2+621308). With the exception of the latter case, the
number of counts is similar to that expected from the background;
however, we cannot rule out the possibility that these are variable or
transient sources. Since we conservatively predicted $\approx$~10--15
false X-ray sources in the 1~Ms
\chandra\ catalog, our results here may suggest that the true number
of false sources is $\approx$~2--3 times lower than this.

\subsubsection{Supplementary Optically Bright \chandra\ Source Catalog}

Since the density of optically bright sources on the sky is
comparatively low, we can search for X-ray counterparts to optically
bright sources at a lower significance threshold than that used in the
main \chandra\ source catalog without introducing many false sources
(see \S5.3 of Richards \etal 1998 for a similar technique at radio
frequencies). We ran {\sc wavdetect} with a false-positive probability
threshold of $1\times 10^{-5}$ on all of the 14 standard-band
images. A basic lower significance \chandra\ catalog was produced
containing 430 independent X-ray sources not present in the
main \chandra\ source catalog. Many of these sources are likely to be
false; however, we can extract a large number of real sources by
matching to optically bright counterparts.

We used the Liu et~al. (1999) $R$-band images in the construction of
our optical source catalog. These images are relatively shallow (a
3$\sigma$ magnitude limit of $R\approx$~23.1); however, they cover the
entire area of the CDF-N, and deep optical images are not required for
this catalog. An optical source list of $\approx$~3,400 sources was
generated using the {\sc sextractor} photometry tool (Bertin \&
Arnouts 1996), assuming the ``Best'' magnitude criteria. We searched
for X-ray counterparts to these optical sources using a matching
radius of $1\farcs 5$. A false-matching probability was calculated
based on the density of optical sources down to the magnitude of each
source. The $R$-band number counts used in determining this
false-matching probability were derived directly from the Liu
et~al. (1999) data. In the limited region of overlap, the number
counts were found to be consistent with those of Hogg et~al. (1997). A
``boosted'' X-ray significance was then determined by multiplying this
false-matching probability with the $1\times 10^{-5}$ false-positive
threshold used to define the lower significance catalog. Only sources
with a boosted X-ray significance of $\le1.5\times 10^{-7}$ are
included in the supplementary optically bright \chandra\ source
catalog; this significance level threshold was empirically chosen to
provide a good balance between the number of detected sources and the
number of false sources.

In total 74 optically bright X-ray sources met this criterion. We
estimated the expected number of false matches by offsetting the X-ray
source coordinates in RA and Dec by $5\arcsec$ (both positive and
negative shifts) and re-correlating with the optical sources. On
average $\approx$~8 matches were found to the $\approx$~3,400 optical
sources with these tests, demonstrating that the vast majority of the
74 X-ray matches are real; considering the optical source density,
the search radius, and the total number of X-ray sources used in the
matching, $\approx$~6 false matches are expected. We also included
five $R<21$ sources where the X-ray source lay $1\farcs 5$--$10\farcs
0$ from the centroid of the optical source but was still within the
extent of the optical emission. The X-ray counterparts for these
sources may be off-nuclear X-ray binaries or star forming
regions. Since these five sources were identified in a somewhat
subjective manner, it is not meaningful to determine a false-matching
probability. These sources are indicated in column 18 of
Table~7a. Thus the supplementary optically bright
\chandra\ source catalog contains 79 sources.

As for the main \chandra\ source catalog, the supplementary
optically bright \chandra\ source catalog is divided into two
tables (Tables~7a and 7b). The format of Table~7a (basic source
properties) is similar to that of Table~3a, with the properties
of the optical counterparts and the probabilities of a false match
also included. Details of the columns in Table~7a are given below.

\begin{itemize}

\item
Column~1 gives the source number (see column~1 of Table~3a for details).

\item
Columns~2 and 3 give the RA and Dec of the X-ray
source, respectively. The {\sc wavdetect} positions are given for
these faint X-ray sources. Whenever possible, we quote the position
determined in the full band; when a source is not detected in the full
band we use, in order of priority, the soft-band position, hard-band
position, SB1 position, SB2 position, HB1 position, or HB2
position. The priority ordering of position choices above was designed
to maximize the signal-to-noise of the data being used for
positional determination. To avoid truncation error, we quote the
positions to higher precision than in the International Astronomical
Union approved names beginning with the acronym ``CXO~HDFN'' (see
Footnote 11).

\item
Column~4 gives the positional uncertainty in arcseconds. For these faint
X-ray sources the positional uncertainty corresponds to $1\farcs2$,
the 90th percentile of the average optical-X-ray positional offsets
given in column 15.

\item
Column~5 gives the off-axis angle for each source in arcminutes (see
column~5 of Table~3a for details).

\item
Columns~6--12 give the counts in the seven standard bands. The
photometry is taken directly from {\sc wavdetect} for these faint
X-ray sources.

\item
Columns~13 and 14 give the RA and Dec of the
optical source centroid, respectively.

\item
Column~15 gives the offset between the optical and X-ray sources in arcseconds.

\item
Column~16 gives the $R$-band magnitude of the optical source
(Vega-based photometric system).

\item
Column~17 gives the false-match probability described above. This
should be multiplied by 10$^{-5}$ to calculate the boosted X-ray
significance.

\item
Column~18 gives notes on the sources. With the exception of the
additional note given below, the key for these notes is given in
column 13 of Table~3a.
``L'' refers to objects where the X-ray source lies $>1\farcs 5$ from
the centroid of the optical source but is still within the extent of
the optical emission (see the text above for further discussion).

\end{itemize}

\noindent The format of Table~7b (additional source properties) is similar
to that of Table~3b. However, since these are very faint X-ray
sources, the band ratios and photon indices are not calculated
($\Gamma=2.0$ is assumed when calculating the source fluxes
below). Details of the columns in Table~7b are given below.

\begin{itemize}

\item
Column~1 gives the source number (see column 1 of Table~3a for details).

\item
Columns~2 and 3 give the RA and Dec of the X-ray source, respectively
(see columns~2 and 3 of Table~7a for details).

\item
Column~4 gives the effective exposure time derived from the full-band
exposure map (see \S3.1 for details of the exposure maps). Dividing by
the counts listed in columns 6--12 of Table~7a by the effective
exposure will provide vignetting-corrected count rates.

\item
Columns~5--11 give observed-frame fluxes in the seven standard bands;
fluxes are in units of $10^{-15}$~erg~cm$^{-2}$~s$^{-1}$ and have been
calculated assuming $\Gamma=2.0$ (see the stacking results below).
The fluxes have not been corrected for absorption by the Galaxy or
material intrinsic to the source; however, the fluxes have been
adjusted to take into account the continuous ACIS quantum efficiency
degradation. See columns~7--13 of Table~3b for further details.

\end{itemize}

\noindent The properties of the 79 sources in Table~7 are more homogeneous 
than those of the sources in Table~3, probably due to the optical
selection criteria. As expected, all of the 79 sources are X-ray faint (all
sources have $<$~30 soft-band counts, and all but two sources have
soft-band fluxes of $< 10^{-16}$~erg~cm$^{-2}$~s$^{-1}$), and the
majority lie in the most sensitive region of the field (i.e.,\ close
to the average aim point; see Figure~12b). Sixty eight
($\approx$~86\%) of the sources are detected in at least one of the
softest X-ray bands (SB, SB1, and SB2) while, by comparison, only four
($\approx$~5\%) of the sources are detected in the hardest bands (HB,
HB1, and HB2); all but one of the 11 sources not detected in the
softest bands are detected in the full band.

The $R$-band magnitudes of the supplementary sources cover
$R=$~17.4--23.1, and the majority ($\approx$~66\%) have
$R=$~21--23. In Figure~13 we show the $R$-band magnitude versus
soft-band flux. All of the sources reside in the region expected for
starburst and normal galaxies. A fraction of these sources may be
low-luminosity AGNs; however, the small number of hard-band detected
sources suggests that very few are luminous absorbed AGNs. We can
place an average constraint on the X-ray spectral slopes of these
sources by stacking the individual sources together. Following \S3.3
of Alexander \etal (2001) we find an average band ratio of
$0.28\pm0.02$, corresponding to an average effective photon index of
$\Gamma=2.0\pm0.1$. This result is consistent with those found by
Alexander \etal (2002a) and Hornschemeier \etal (2003) from stacking
X-ray-infrared detected emission-line galaxies and optically bright,
X-ray faint sources, respectively. This supports the idea that few
absorbed AGNs are present in these optically bright sources.

Due to their optical selection, the optically bright supplementary
sources may not be representative of the faintest X-ray
sources. However, we cannot easily extract a reliable list of
$R\simgt$~23 lower significance sources as a comparison to these
optically bright sources. Although clearly an optically bright X-ray
faint selection will favor the identification of non-AGNs (e.g.,\
Hornschemeier \etal 2003), the observed steepening of the
stacked-average photon index at the lowest count rates found in
\S4.1 provides evidence that normal and starburst galaxies are 
comparitively common at faint X-ray fluxes (see also Alexander \etal
2002a and Bauer \etal 2002b).

%
\section{Data and Background Analysis}
%

\subsection{X-ray Band Ratio and X-ray Color-Color Analysis}

In Figure~14 we show the band ratio as a function of full-band count
rate for the sources in the main \chandra\ catalog. A trend toward
larger band ratios for decreasing count rates is seen. This hardening
of the X-ray spectral slope has been reported in other studies (e.g.,\
della Ceca \etal 1999; Ueda \etal 1999; Mushotzky \etal 2000; Tozzi
\etal 2001; B01) and appears to be due to an increase in the number 
of absorbed AGNs detected. For the sources in this study the
stacked-average photon index flattens from $\Gamma\approx$~1.8 to
$\Gamma\approx$~0.8 for full-band count rates of $\approx10^{-2}$ to
$\approx 3\times10^{-5}$ counts~s$^{-1}$. At the lowest full-band
count rates ($\simlt 3\times10^{-5}$ counts~s$^{-1}$) the
stacked-average photon index steepens ($\Gamma\approx$~1.2), possibly
due to the increasingly large contribution from normal and starburst
galaxies (e.g.,\ Hornschemeier \etal 2003); the difference is
significant at the 6.4$\sigma$ level. Absorbed AGNs are undoubtedly
detected at these low count rates; however, they are likely to
represent a smaller fraction of the X-ray source population than that
found at higher count rates.

In Figure~15 we show X-ray color-color diagrams. These diagrams
provide diagnostics on the nature of X-ray detected sources and they
are particularly useful for investigating X-ray sources too faint for
X-ray spectral analysis. In our comparisons we have simplistically
assumed that AGNs can be characterized by power-law emission with
varying amounts of neutral intrinsic absorption. Starburst and normal
galaxies are also often characterized by power-law emission in the
rest-frame \hbox{$\approx$~2--8~keV} band due to the presence of X-ray
binaries (e.g.,\ Kim, Fabbiano, \& Trinchieri 1992a,b; Ptak \etal
1999). However, at soft X-ray energies (i.e.,\ $\simlt$~2~keV), the
emission from starburst and normal galaxies is often better modeled
with a soft thermal component (e.g.,\ Ptak \etal 1999).

Figure~15a compares the SB2/SB1 and HB1/SB2 band ratios. This choice
of band ratios is useful for identifying AGNs with low-to-moderate
absorption, starburst galaxies, and normal galaxies. The bulk of the
sources fall in the region where AGNs are expected to lie (i.e.,\ an
average spectral slope of $\Gamma\approx$~1.8 with varying amounts of
intrinsic absorption); similar conclusions were drawn by Barger \etal
(2002) and Mainieri \etal (2002) from X-ray color-color analyses of
bright X-ray sources. However, a considerable fraction
($\approx$~20\%, 103 sources) have very flat X-ray spectra
(HB1/SB2$>$1.0) and hence properties inconsistent with simple AGN
models. The majority of these sources also show flat X-ray spectra
from other diagnostics (e.g.,\ all but two of the sources have
HB/SB$>$1.0, and 75 of the 87 sources with HB/SB$>$2.0 have
HB1/SB2$>$1.0). At high column densities the direct AGN emission is
weak, and other components such as reflection, scattering, and line
emission can become important. In such cases the X-ray spectrum of a
highly absorbed AGN is often poorly characterized by a simple absorbed
power-law model.

Figure~15b compares the HB1/SB2 and HB2/HB1 band ratios. This choice
of band ratios is useful for identifying highly absorbed
AGNs. Approximately 70--90\% of the AGNs that showed very flat X-ray
spectra in Figure~15a (i.e.,\ HB1/SB2$>$1.0) fall in the region where
highly absorbed AGN are expected to lie in Figure~15b; the very flat
X-ray spectra that these sources showed in Figure~15a may be due to
other components dominating over the absorbed power-law emission at
softer X-ray energies (e.g.,\ scattering, reflection, and line
emission). The X-ray emission from the other AGNs with very flat X-ray
spectra is still not well characterized by a simple absorbed power-law
model, and these sources may be Compton thick AGNs (i.e.,\ $N_{\rm H}>
1.5\times 10^{24}$~cm$^{-2}$; e.g.,\ Matt \etal 2000).

Based on these simple models, there appears to be few
($\approx$~10--30) Compton-thick AGNs; see also Bauer \etal 2003 for
direct X-ray spectral analysis constraints of the X-ray brightest
sources. Given that Compton-thick AGNs account for $\approx$~50\% of
nearby AGNs (e.g.,\ Maiolino \etal 1998; Risaliti \etal 1999), this
result may seem surprising. However, since the observed luminosities
of nearby Compton-thick AGNs are typically $L_{\rm 2-10~keV}\approx
10^{41}$~erg~s$^{-1}$ (see Figure~4 in Maiolino \etal 1998), a similar
source at $z=$~0.5--2.0 would have an X-ray flux of
\hbox{$\approx$~(0.1--1.4)}$\times 10^{-16}$~erg~cm$^{-2}$~s$^{-1}$
(calculated assuming a typical observed spectral slope for
Compton-thick AGNs of $\Gamma=$~1.0; e.g.,\ Maiolino \etal 1998,
Bassani \etal 1999). Thus, even in a 2~Ms \chandra\ exposure, few
typical Compton-thick AGNs are expected.

Finally, we note that very few sources lie along the tracks expected
for soft thermal emission. This is because we have not yet reached the
sensitivity to detect such emission in the HB1 band (see Figure~15a).

\subsection{Background and Sensitivity Limits}

The faintest sources in the main \chandra\ catalog have $\approx$~5
counts (soft band) and $\approx$~7 counts (hard band), corresponding
to count rates of $\approx$~1 count every 4.5 days (soft band) and 3.2
days (hard band). For a $\Gamma=$~1.4 power law with Galactic
absorption, the corresponding soft-band and hard-band fluxes at the
aim point are $\approx 1.3\times 10^{-17}$~erg~cm$^{-2}$~s$^{-1}$ and
$\approx 8.3\times 10^{-17}$~erg~cm$^{-2}$~s$^{-1}$,
respectively. This gives a measure of the ultimate sensitivity of this
survey; however, these numbers are only relevant for a small region
close to the aim point. In order to determine the sensitivity across
the field it is necessary to take into account the broadening of the
PSF with off-axis angle, as well as changes in the effective exposure
and background rate across the field. Under the simplifying assumption
of $\sqrt{N}$ uncertainties, we can determine the sensitivity across
the field following Muno \etal (2003) as

\begin{equation}
S~=~{\frac{n^2_{\sigma}}{2}~(1+[1+\frac{8b}{n^2_{\sigma}}]^{\frac{1}{2}})}
\end{equation}

\noindent where $S$ is the number of source counts for a given 
signal-to-noise ratio ($n_\sigma$) and the number of background counts
($b$) in a source cell. The only component within this equation that
we need to measure is the background. For the sensitivity calculations
here we measured the background in a source cell using the background
maps described below and assuming an aperture size of 70\% of the
encircled-energy radius of the PSF; the 70\% encircled energy-radius
was chosen as a compromise between having too few source counts and
too many background counts. The total background includes
contributions from the unresolved cosmic background, particle
background, and instrumental background (e.g.,\ Markevitch 2001;
Markevitch \etal 2003). For our analyses we are only interested in the
total background and do not distinguish between these different
components.

To create background maps in all of the seven \asca-grade images, we
first masked out the point-sources from the main \chandra\ catalog
using apertures with radii twice that of the $\approx$~90\% PSF
encircled-energy radii. The resultant images should include minimal
contributions from detected point sources. They will include
contributions from extended sources (e.g.,\ Bauer \etal 2002a), which
will cause a slight overestimation of the measured background close to
extended sources. In Figure~16 we show the distributions of background
counts per pixel in the full, soft, and hard bands, and in Table~8 we
provide the mean backgrounds in the seven standard bands. Even with a
2~Ms \chandra\ exposure, most of the pixels have no background counts
(e.g.,\ in the full band 81.6\% of the pixels are zero). For a small
number of detected counts, the expected distribution is Poissonian. We
compared the background count distributions to Poisson distributions
and found them to be very similar (the probability as derived from the
Kolmogorov-Smirnov test is $>$99.99\% in all of the seven standard
bands; see Figure~16). Since the effective exposure varies across the
field, we must be careful that this result is not an effect due to the
combination of many different Poisson distributions. To guard against
this, we performed the same analyses on many different small
($\approx$~4~arcmin$^2$) regions of similar exposure across the
field. In all cases the Kolmogorov-Smirnov test probabilities were
$>$99.99\% in all of the seven standard bands. We filled in the masked
regions for each source with a local background estimated by making a
probability distribution of counts using an annulus with inner and
outer radii of 2 and 4 times the $\approx$~90\% PSF encircled-energy
radius, respectively. The full-band background map is shown in
Figure~17. The total number of background counts over the whole image
dominates the counts in the detected sources (see Table~8). Since
these detected sources contribute the bulk of the cosmic X-ray
background, this shows that the unresolved cosmic X-ray background
component is a small fraction of the total background counts. Detailed
modeling would be required in order to measure the unresolved cosmic
X-ray background (see Markevitch \etal 2003 for modeling of the total
background using ACIS-S observations).

Following equation 3, we generated sensitivity maps [for a
signal-to-noise ratio (S/N) of 3] using the background and exposure
maps; we assumed a $\Gamma=$~1.4 power-law model with Galactic
absorption. In Figure~18 we show the full-band sensitivity map, and in
Figure~19 we show plots of flux limit versus solid angle for the full,
soft, and hard bands. The $\approx1$~arcmin$^2$ region at the aim
point has 0.5--2.0~keV and \hbox{2--8~keV} sensitivity limits of
$\approx 2.5\times10^{-17}$~erg~cm$^{-2}$~s$^{-1}$ and $\approx
1.4\times10^{-16}$~erg~cm$^{-2}$~s$^{-1}$, respectively (see Table~9
for the aim point sensitivities in all of the seven standard
bands). Since we do not filter out detected sources with S/N$<3$, a
few sources have fluxes below these sensitivity limits (35 sources in
the soft band and 22 sources in the hard band). Approximately
190~arcmin$^2$ of the field [including almost all of the expected area
of the IRAC observations for the Great Observatories Origins Deep
Survey (GOODS; e.g.,\ Dickinson \& Giavalisco 2003)] has a full-band
sensitivity limit of $\approx 3\times10^{-16}$~erg~cm$^{-2}$~s$^{-1}$,
and $\approx$~360~arcmin$^2$ (i.e.,\ $\approx$~1.3 times the size of a
single ACIS-I field) has a full-band sensitivity limit of $\approx
1\times10^{-15}$~erg~cm$^{-2}$~s$^{-1}$. The latter flux limit is
equivalent to the faintest X-ray sources detected at the aim point in
moderately deep ($\approx$~100~ks) \chandra\ observations (see
Figure~1).

\subsection{Prospects for Longer Chandra Exposures}

Doubling the exposure of a \chandra\ observation leads to an increase
in sensitivity between a factor of $\sqrt2$ (for the case of a large
background; background limited) and 2 (for the case of no background;
photon limited). The total number of background counts is larger than
the total number of source counts in the CDF-N field (see
Table~8). However, the number of background counts in a detection cell
is often negligible; in our analyses here, we will again assume a
circular detection cell with a radius equal to the 70\%
encircled-energy radius of the PSF. For example, the average numbers
of background counts in a detection cell $\approx$~2$\arcmin$ from the
aim point in the soft and hard bands are $\approx$~0.8 and
$\approx$~2.5, respectively.

We can examine the transition from photon to background limited using
equation 3. For S/N$=3$, $\approx$~3.3 background counts per detection
cell marks the point when an observation moves toward being background
limited; for 3.3 background counts, the effective increase in
sensitivity when doubling the exposure is 1.71 (i.e.,\ between
$\sqrt2$ and 2). For our analyses here we will take photon limited as
being when there are $<$3.3 background counts in a detection cell.

In Figure~20 we show the approximate photon-limited regions for the
full, soft, and hard bands, and in Table~9 we provide the approximate
photon-limited solid angles in the seven standard bands. The
photon-limited region is the largest in the softest bands due to their
lower backgrounds and smaller PSFs; however, even in the full band
$\approx$~25~arcmin$^2$ of the field is photon limited. We can
estimate the longest \chandra\ exposure for a photon-limited
observation by increasing the background in the current background
images (assuming a Poisson distribution with mean count rates taken
from Table~8) and re-calculating the photon-limited area for the new
exposure. In Table~9 we provide the longest photon-limited exposures
in the seven standard bands; for our analyses here the longest
exposure corresponds to that found when $\simlt$~1~arcmin$^2$ of the
field remains photon limited. The longest photon-limited exposures
(25--50~Ms) are in the softest bands (i.e.,\ $\le$~2~keV). The shorter
photon-limited exposures in the hard bands are largely due to the
large background contribution from 4--8~keV (see column 3 in Table~8),
although photon-limited exposures up to $\approx$~15~Ms should be
possible in the HB1 band (2--4~keV). Since 489 ($\approx$~97\%) of the
503 sources are detected in either the soft band or the HB1 band,
significantly deeper observations should be able to efficiently detect
sources out to $\approx$~4~keV. \chandra\ clearly cannot detect hard
sources as efficiently as soft sources; however, due to the negative
$K$-correction of absorbed AGN emission, significantly longer
\chandra\ exposures should be an effective tool in the identification
of high-redshift heavily absorbed AGNs.

In Table~9 we provide predicted S/N$=3$ sensitivities at the aim point
for 4 and 8~Ms \chandra\ exposures in all of the seven X-ray
bands.\footnote{The future effects of the optical blocking filter
contamination on the ACIS quantum efficiency is uncertain (see columns
7--13 of Table~3b). In our calculations here, we have assumed that
ACIS has been ``baked'' out to remove the contaminants and restore the
quantum efficiency to pre-contamination values.} The soft-band
limiting fluxes (down to $\approx
7\times10^{-18}$~erg~cm$^{-2}$~s$^{-1}$) are faint enough to start
individually detecting sources from the source populations
statistically detected in stacking analyses (e.g.,\ Brandt
\etal 2001c; Alexander \etal 2002b; Brusa \etal 2002; Hornschemeier
et~al. 2002; Nandra \etal 2002). Based on the fluctuation analyses of
the 1~Ms CDF-N observation, X-ray source densities up to
$\approx$~70,000 deg$^{-2}$ are possible down to soft-band fluxes of
$\approx 7\times10^{-18}$~erg~cm$^{-2}$~s$^{-1}$ (e.g.,\ Miyaji \&
Griffiths 2002). At such large source densities, the effects of source
confusion upon source positional determination, source photometry, and
source density must be considered.

The source-confusion limit is based on the number of ``beams'' per
source. A beam is typically taken to be a circle with a radius equal
to the FWHM of the PSF (e.g.,\ Hogg 2001). Since the shape of the HRMA
PSF is complex at large off-axis angles, we conservatively assume here
a ``beam'' as a circle with a radius equal to the 70\%
encircled-energy radius of the PSF. The minimum required number of
beams per source is determined from the number of sources lying below
the detection threshold and depends upon the slope of the number
counts for these undetected sources. Based on the predicted number
count slopes from the fluctuation analysis of the 1~Ms CDF-N
observation (Miyaji \& Griffiths 2002), the minimum required number of
beams per source for source confusion is likely to be 30--50 (e.g.,\
Hogg 2001). In Figure~21 we show the density of the detected sources
versus off-axis angle and compare this to our calculated
source-confusion limit. Within 6$\arcmin$ of the aim point source
confusion is unlikely to be a problem even in the most pessimistic
sources-per-beam scenario; however, we may suffer source confusion at
larger off-axis angles. Within 3$\arcmin$ of the aim point, source
confusion is unlikely to be present even with extremely deep
observations (i.e.,\ where source densities exceed 100,000
deg$^{-2}$). These calculations do not take into account source
clustering, which can also effect the determination of source
properties.

A 4--8~Ms \chandra\ exposure can reach the flux limits being discussed
for the next generation of X-ray observatories such as
\xeus\ (not to be launched for at least 12 years).\footnote{Details of
the XEUS mission can be found at
http://astro.esa.int/SA-general/Projects/XEUS/main/xeus\_main.html.}
Although the prime scientific focus of \xeus\ is X-ray spectral
analysis rather than X-ray imaging (e.g.,\ Bleeker \& Mendez 2003), an
extremely deep \chandra\ survey can explore the likely source
populations to be detected in deep \xeus\ surveys. In addition to
exploring new discovery space, such a survey would also provide firm
constraints on the X-ray source density while being free of source
confusion. This information could prove important for the design of
\xeus\ and the next generation of X-ray observatories.

%
\section{Conclusions}
%

We have presented source catalogs and basic analyses for point sources
detected in the 2~Ms \chandra\ Deep Field-North survey down to on-axis
flux limits of \hbox{$\approx 2.5\times
10^{-17}$~erg~cm$^{-2}$~s$^{-1}$} (0.5--2.0~keV) and \hbox{$\approx
1.4\times 10^{-16}$~erg~cm$^{-2}$~s$^{-1}$} (2--8~keV). We have
provided two point-source catalogs: a main \chandra\ catalog of 503
\hbox{X-ray} sources detected with a {\sc wavdetect} false-positive 
probability threshold of 10$^{-7}$ and a supplementary optically
bright \chandra\ catalog of 79 \hbox{X-ray} sources detected with a
{\sc wavdetect} false-positive probability threshold of 10$^{-5}$ but
matched to optically bright ($R\simlt$~23) counterparts. The
\hbox{X-ray} colors of the sources in the main \chandra\ catalog imply
a broad variety of source types; however, absorbed AGNs (including a
small number of possible Compton-thick sources) are clearly the
dominant source type. The X-ray and optical properties of the sources
in the supplementary optically bright \chandra\ catalog are mostly
consisent with those expected from starburst and normal galaxies. Due
to their optical selection, these sources may not be representative of
the faintest X-ray sources.

We have provided analyses of the image backgrounds and X-ray
sensitivity across the field. We found that the background count
distributions are very similar to Poisson distributions for all seven
X-ray bands. We showed that this \chandra\ observation is nearly
photon limited for regions close to the aim point in all of the seven
\hbox{X-ray} bands and predicted that exposures up to $\approx$~25~Ms
(0.5--2.0~keV) and $\approx$~4~Ms (2--8~keV) should remain nearly
photon limited. We also demonstrated that this observation does not
suffer from source confusion within $\approx$~6$\arcmin$ of the aim
point and showed that future observations are unlikely to be
source-confusion limited within $\approx$~3$\arcmin$ of the aim point
even for source densities exceeding \hbox{100,000 deg$^{-2}$}. These analyses
directly show that \chandra\ can achieve significantly fainter
\hbox{X-ray} fluxes in an efficient nearly photon-limited manner and
be largely free of source confusion.

%
\acknowledgments
%

This work would not have been possible without the enormous efforts of
the entire \chandra\ and ACIS teams. We are grateful to E.D.~Feigelson
and J.A.~Nousek for valuable contributions early in the CDF-N
program. We thank D.N.~Burrows, Y.~Butt, P.E.~Freeman, J.~Gaffney,
K. Getman, and C.~Liu, and members of the GOODS team (particularly
N.~Grogin, A.~Koekemoer, and M.~Nonino) for interesting discussions
and providing data.
We gratefully acknowledge the anonymous referee for an efficient and
thoughtful report.
We acknowledge the financial support of NASA grants NAS~8-38252 and
NAS~8-01128 (GPG, PI), NSF CAREER award AST-9983783 (DMA, FEB, WNB,
CV), CXC grant GO2-3187A (DMA, FEB, WNB, CV), Chandra fellowship grant
PF2-30021 (AEH), NSF grant AST-9900703 (DPS), NSF grant AST-0084847
(AJB), and NSF grant AST-0084816 (LLC).

%
%

\appendix

\section{The 1 Ms Chandra Deep Field-South}

The $\approx$~1~Ms CDF-S observation provides the second deepest view
of the Universe in the 0.5--8.0~keV band. X-ray catalogs have already
been published in G02. However, the source searching and data
processing procedures used to create the G02 catalogs were, in many
respects, quite different from those applied to the CDF-N
observations. In order to allow consistent comparisons between the
CDF-N and CDF-S, we have produced catalogs for the CDF-S using the
methods described in \S2 and \S3. We only provide details regarding
the differences required for the production of the CDF-S catalogs
here. The CDF-S catalogs, images, and exposure maps can be accessed
from the World Wide Web site listed in Footnote 8.

\subsection{Observations and Conditions}

The CDF-S consists of 12 separate observations taken between 1999 Oct
14 and 2000 Dec 23. Observation 581 (1999 Oct 14) is not included in
the reduction due to telemetry saturation and CCD problems; see
Table~A1 for the details of the 11 observations used in this
analysis. The four
\hbox{ACIS-I} CCDs and the ACIS-S CCD S2 were operated in all of the
observations; however, due to its large off-axis angle, and
consequently its low sensitivity, CCD S2 is not used here. The first
observation was taken in Very Faint mode while the other 10
observations were taken in Faint mode; the focal-plane temperature was
$-110^\circ$\,C for the first two observations and $-120^\circ$\,C for
the other nine observations. The background light curves were
inspected using {\sc event browser}. The background was found to be
high during the first observation, and 8.3~ks of exposure was
removed. The total exposure time for the 11 observations is 939.4~ks;
see Table~A1. The observing strategy for the CDF-S observations was to
keep the aim point approximately constant throughout the
observations. The average aim point weighted by exposure time is
$\alpha_{2000}=$~3$^{\rm h}$ 32$^{\rm m}$ $28\fs 2$,
$\delta_{2000}=$~$-27^\circ$48$^{\prime}$36$^{\prime\prime}$, and the
individual aim points are offset by 5--30$^{\prime\prime}$ from the
average aim point; see Table~A1. Due to the different roll angles, the
total region covered by the 11 observations is 391.3~arcmin$^2$.

\subsection{Production of the Point-Source Catalogs}

The main \chandra\ catalog and supplementary optically bright
\chandra\ catalog were both produced following \S3. The images and 
exposure maps were created following \S3.1; the images were registered
to observation 582. Point-source detection was performed following
\S3.2. Since the CDF-S region does not have deep published radio coverage,
absolute source positions were determined by matching X-ray sources to
optical counterparts, following the method given in \S3.3. The optical
data used to create the optical source list was the $R$-band
Wide-Field Imager (WFI) observations taken as part of the ESO Imaging
Survey (Arnouts \etal 2001). The optical source list was generated
using {\sc sextractor}, assuming the ``Best'' magnitude criteria. We
matched X-ray sources to point-like (FWHM $<$1.4, as measured by {\sc
sextractor}) $R<$~25 sources with a $2\farcs 5$ search radius. From a
comparison of the X-ray and optical source positions we found a small
shift in the X-ray source positions. We also found a plate-scale
correction identical to that found in the CDF-N (i.e.,\
$\approx$~$0\farcs 35$ every 12$\arcmin$ in both RA and Dec; see
\S3.3). These corrections have been applied to the X-ray source
positions.

The main \chandra\ catalog was produced following \S3.4.1 and is
presented as Tables~A2a and A2b. The format of these tables are the
same as those of Tables~3a and 3b. In total 326 independent sources
are detected. In 18 cases we had to perform manual photometry, in 52
cases we had to make positional changes, and in two cases we had to
separate manually a close double and determine the position of each
separated source. The fluxes and photon indices have been determined
taking into account the continuous ACIS quantum efficiency degradation
described in columns 7--13 of Table~3b; for a $\Gamma=$~1.4 source,
the soft-band and hard-band corrections are $\approx$~2\% and
$\approx$~0\%, respectively.

The supplementary optically bright \chandra\ catalog was produced
following \S3.4.2 and is presented as Tables~A3a and A3b. The format
of these tables are the same as those of Tables~7a and 7b. A basic
lower significance \chandra\ catalog of 363 sources was
constructed. The optical data used to identify the optically bright
counterparts to these \chandra\ sources were produced from the
$R$-band WFI observations (see above). We matched optical counterparts
to the X-ray sources with a search radius of $1\farcs 5$, and selected
sources with a ``boosted'' X-ray significance of $\le1.5\times
10^{-7}$ (see \S3.4.2). In total 40 optically bright sources were
selected; by offsetting the source coordinates in RA and Dec we
estimate $\approx$~5 of these matches are false. We also included two
$R<21$ sources where the X-ray source lay $1\farcs 5$--$10\farcs 0$
from the centroid of the optical source but was still within the
extent of the optical emission. Thus, 42 sources are included in the
supplementary optically bright \chandra\ catalog.

We determined S/N$=3$ sensitivity limits following \S4.2. The
$\approx1$~arcmin$^2$ region at the aim point has 0.5--8.0~keV,
0.5--2.0~keV, and 2--8~keV sensitivity limits of
$\approx 1.3\times10^{-16}$~erg~cm$^{-2}$~s$^{-1}$,
$\approx 5.2\times10^{-17}$~erg~cm$^{-2}$~s$^{-1}$, and
$\approx 2.8\times10^{-16}$~erg~cm$^{-2}$~s$^{-1}$, respectively.

\subsection{Comparisons of the Main CDF-S Catalogs}

In this final section we compare our main \chandra\ catalog to the
catalogs of G02. The source-searching strategy of G02 had four main
differences from that employed by us: (1) they searched for X-ray
sources using both {\sc wavdetect} and a modified version of {\sc
sextractor}, (2) they only performed source searching in the
0.5--7.0~keV band (detections in the 0.5--2.0~keV and 2--10~keV bands
were determined based on the signal-to-noise ratio of the 0.5--7.0~keV
detected sources in those bands), (3) when searching for sources with
{\sc wavdetect} they used a false-positive probability threshold of
$1\times 10^{-6}$, and (4) they performed additional processing of
their results, removing all sources with S/N$<2.1$. Since we only used
{\sc wavdetect} for source searching, we will generally only make
comparisons to the G02 {\sc wavdetect} sources.

We matched the sources from both studies. G02 detected 346 independent
sources: 318 with {\sc wavdetect}, and 332 with {\sc sextractor}. The
published G02 source positions are offset from the true source
positions by $-1\farcs 2$ in RA and $0\farcs 8$ in Dec; see the
accompanying notes with the Centre de Donnees astronomiques de
Strasbourg (CDS) on-line version of the catalogs.\footnote{The CDS
World Wide Web site is at http://cdsweb.u-strasbg.fr/.} We removed
this $\approx 1\farcs 4$ offset from the G02 source positions when
performing our source matching. Using a $4\arcsec$ search radius we
matched the X-ray sources from the main
\chandra\ catalog to the G02 {\sc wavdetect} sources, finding 284 in
common; a further five {\sc sextractor}-only sources were also
matched. In Figure~A1 we show the results of this source matching. The
sources unique to the G02 {\sc wavdetect} catalog tend to lie at large
off-axis angles. We visually inspected all of these sources, and the
majority are plausibly real. If we match the G02 {\sc wavdetect}
sources to both our main
\chandra\ catalog and the lower significance
\chandra\ catalog of 363 sources we find 308 of the {\sc wavdetect}
sources in common; a further 18 {\sc sextractor}-only sources are also
matched. These results suggest that a lower false-positive probability
threshold than that adopted by us for the main \chandra\ catalog is
probably required to identify all of the sources at large off-axis
angles. By comparison, many of the sources unique to our catalog lie
close to the aim point; see Figure~A1. Visual inspection of these
sources also suggests that the majority are real. Many of these
sources are not included in the G02 catalogs because they are either
detected with S/N$<2.1$ or they are not detected in the full band
(i.e.,\ they would not have been detected in the 0.5--7.0~keV band
used by G02). From these basic analyses it therefore seems that a
source-searching strategy that combines our approach with that of G02
would optimize the number of detected sources.

We compared the overall positional accuracy of the sources in the {\sc
wavdetect} catalogs. In this comparison we matched X-ray sources to
$R<24$ sources detected in the WFI observations; we removed the
$\approx 1\farcs 4$ offset from the G02 source positions before
performing the source matching. Using a $2\farcs 5$ search radius, we
found $R<24$ counterparts for 176 of the 284 {\sc wavdetect} sources
common to both studies. In Figure~A2 we show the X-ray-optical
positional offset distributions. The distribution is considerably
tighter for our sources and the median positional offset is better by
$\approx 0\farcs 15$ ($0\farcs 37$ versus $0\farcs 52$); the median
positional offset for the 176 sources is also $0\farcs 52$ if the
published G02 X-ray-optical offsets are used (based on FORS1 $R$-band
observations). Our improvements in positional accuracy are mostly due
to the plate-scale correction (see \S3.3) and our refined source
position technique (see \S3.4.1).

We compared the photometry between the sources in the {\sc wavdetect}
catalogs. G02 performed source photometry in the \hbox{0.5--2.0~keV}
and 2--10~keV bands. Although we did not perform source photometry in
the 2--10~keV band, few source counts are detected at \hbox{8--10~keV}
so a comparison to our 2--8~keV band photometry is valid. In Figure~A3
we show the results of the photometry comparisons. Our source
photometry is systematically higher than that of G02 by $\approx$~2\%
in the 0.5--2.0~keV band and $\approx$~4\% in the 2--8~keV band. Since
the results of extensive simulations by G02 showed that their
photometric technique underestimated the true number of counts by
$\approx$~4\%, the agreement between the catalogs is very good. There
are large disagreements (i.e.,\ $>$50\%) in the measured counts for a
small number of faint sources (i.e.,\ $\simlt$~150 counts). Visual
inspection of these sources showed that they generally lay in regions
where the estimation of the background is problematic (i.e.,\ the
background was high or variable due to diffuse emission, a bright
nearby source, or the edge of a CCD). 

We compared the fluxes between the sources in the {\sc wavdetect}
catalogs. The source fluxes in both studies were calculated using the
measured counts, an effective exposure, and a spectral slope. In their
flux determination, G02 assumed a spectral slope of $\Gamma=$~1.4 for
all sources while we determined spectral slopes on a source-by-source
basis (see column 6 of Table~3a). The average fluxes in the
0.5--2.0~keV and 2--8~keV bands are higher in our catalog by
$\approx$~9\% and $\approx$~18\%, respectively (see Figure~A4); we
corrected the 2--10~keV fluxes of the G02 sources to the 2--8~keV band
assuming $\Gamma=$~1.4. These small differences are mostly due to the
measured counts and effective exposures for the sources. However,
there is some scatter, particularly in the 2--8~keV band, due to
differences in the adopted spectral slopes; see Figure~A4.

%

\clearpage

%
%

\begin{deluxetable}{lccclccl}
\tablenum{1}
\tabletypesize{\small}
\tablewidth{0pt}
\tablecaption {Journal of \chandra\ Deep Field-North Observations}
\small
%
\tablehead{
\colhead{Obs.}                                 &
\colhead{Obs.}                                 &
\colhead{Exposure}                             &
\multicolumn{2}{c}{Aim Point$^{\rm b}$}        &
\colhead{Roll}                                 & 
\colhead{Obs.}                                 &
\colhead{Pipeline}                             \\
\colhead{ID}                                 &
\colhead{Start (UT)}                         &
\colhead{Time (ks)$^{\rm a}$ }               &
\colhead{$\alpha_{\rm 2000}$}                &
\colhead{$\delta_{\rm 2000}$}                &
\colhead{Angle ($^\circ$)$^{\rm c}$}         &
\colhead{Mode$^{\rm d}$}                     &
\colhead{Version$^{\rm e}$}         
}
\startdata

580  & 1999 Nov 13, 01:14 &  49.4 & 12 37 16.3 & +62 12 37   & 36.6  & F & R4CU5UPD9 \\
\\   
967  & 1999 Nov 14, 19:47 &  57.4 & 12 37 04.3 & +62 12 53   & 38.2  & F & R4CU5UPD9 \\
\\  
966  & 1999 Nov 21, 04:03 &  57.4 & 12 37 04.3 & +62 12 55   & 39.2  & F & R4CU5UPD9 \\
\\  
957  & 2000 Feb 23, 06:31 &  57.4 & 12 36 36.4 & +62 14 44   & 134.3  & F & R4CU5UPD12.1 \\
\\
\enddata
\tablecomments{The focal-plane temperature was 
$-110^\circ$\,C during the first three observations and 
$-120^\circ$\,C during the others.}
\tablenotetext{a}{All observations were continuous. These times have been corrected for 
the lost exposure due to the read-out time per CCD frame and also reflect
the total exposure after the data affected by high background have been
removed. The short time intervals with bad satellite aspect are
negligible and have not been removed. The total exposure time for the
20 observations is 1.945~Ms.}
\tablenotetext{b}{The average aim point, weighted by exposure time, is
$\alpha_{2000}=$~12$^{\rm h}$ 36$^{\rm m}$ $45\fs 7$,
$\delta_{2000}=$~$+62^\circ$13$^{\prime}$58$^{\prime\prime}$.} 
\tablenotetext{c}{Roll angle describes the orientation of the \chandra\ instruments
on the sky.  The angle is between 0--360$^{\circ}$, and it increases to the West
of North (opposite to the sense of traditional position angle).}
\tablenotetext{d}{The observing mode (F=Faint mode; VF=Very Faint mode).}
\tablenotetext{e}{The version of the CXC pipeline software used for basic 
processing of the data.}
\end{deluxetable}

\clearpage

%
%
\begin{deluxetable}{lcll}
\tablenum{2}
\tabletypesize{\small}
\tablewidth{0pt}
\tablecaption {Source Detection Bands}
\scriptsize
\tablehead{
\colhead{}                  &
\colhead{Energy}            &
\colhead{Standard {\it ASCA}} &
\colhead{Restricted ACIS} \\
\colhead{Band}                  &
\colhead{(keV)}                 &
\colhead{Grade Set}             &
\colhead{Grade Set}
}
\startdata
Full & 0.5--8.0 & {\it ASCA} grades 0, 2, 3, 4, 6 & ACIS grades 0, 2, 8, 16, 64\\
Soft & 0.5--2.0 & {\it ASCA} grades 0, 2, 3, 4, 6 & ACIS grades 0, 64\\
Hard & 2--8     & {\it ASCA} grades 0, 2, 3, 4, 6 & ACIS grades 0, 2, 8, 16\\
SB1  & 0.5--1.0 & {\it ASCA} grades 0, 2, 3, 4, 6 & ACIS grades 0, 64\\
SB2  & 1--2     & {\it ASCA} grades 0, 2, 3, 4, 6 & ACIS grades 0, 64\\
HB1  & 2--4     & {\it ASCA} grades 0, 2, 3, 4, 6 & ACIS grades 0, 2, 8, 16\\
HB2  & 4--8     & {\it ASCA} grades 0, 2, 3, 4, 6 & ACIS grades 0, 2, 8, 16\\
\enddata
\end{deluxetable}
\pagebreak

%

%
%

\begin{deluxetable}{lllcccccccccl}
\tablenum{3a}
\rotate
\tabletypesize{\scriptsize}
\tablewidth{0pt}
\tablecaption{Main {\it Chandra} Catalog: Basic Source Properties}
\tablehead{
\colhead{} &
\multicolumn{2}{c}{X-ray Coordinates} &
\colhead{Pos}                   &
\colhead{Off-Axis}                   &
\multicolumn{7}{c}{Counts}      &
\colhead{}  \\
\colhead{\#}                    &
\colhead{$\alpha_{2000}$}       &
\colhead{$\delta_{2000}$}       &
\colhead{Error}       &
\colhead{Angle}       &
\colhead{FB}          &
\colhead{SB}          &
\colhead{HB}          &
\colhead{SB1}         &
\colhead{SB2}         &
\colhead{HB1}         &
\colhead{HB2}         &
\colhead{Notes}  \\
\colhead{(1)}         &
\colhead{(2)}         &
\colhead{(3)}         &
\colhead{(4)}         &
\colhead{(5)}         &
\colhead{(6)}         &
\colhead{(7)}         &
\colhead{(8)}         &
\colhead{(9)}         &
\colhead{(10)}        &
\colhead{(11)}        &
\colhead{(12)}        &
\colhead{(13)}  \\
}
\startdata
  1&12 35 15.04&+62 14 07.0& 1.5&10.57&                       79.6$^{+16.8}_{-14.6}$&                       68.0$^{+14.4}_{-12.2}$&                                      $<$19.6&               14.2$^{+8.2}_{-6.0}$$^{\rm b}$&                       53.8$^{+13.0}_{-10.8}$&                                      $<$13.0&                                      $<$14.8&                   M\\
  2&12 35 16.76&+62 15 37.8& 1.5&10.50&                       26.7$^{+17.6}_{-16.6}$&             14.3$^{+11.5}_{-10.4}$$^{\rm b}$&                                      $<$32.9&                                      $<$16.8&               13.3$^{+9.4}_{-8.3}$$^{\rm b}$&                                      $<$21.5&                                      $<$24.9&                    \\
  3&12 35 18.82&+62 15 51.5& 1.4&10.30&                      112.8$^{+21.1}_{-20.1}$&                       68.7$^{+14.4}_{-13.4}$&                                      $<$39.5&                                      $<$19.6&                       56.0$^{+12.1}_{-10.9}$&                                      $<$26.0&                                      $<$29.9&                    \\
  4&12 35 19.47&+62 13 39.7& 1.4&10.05&                       71.1$^{+14.7}_{-13.6}$&                        40.2$^{+10.1}_{-9.1}$&                                      $<$26.6&                                      $<$13.1&                         33.2$^{+8.8}_{-7.7}$&                                      $<$18.9&                                      $<$18.8&                    \\
  5&12 35 21.32&+62 16 28.1& 0.5&10.15&                      596.8$^{+32.6}_{-31.4}$&                      410.0$^{+24.4}_{-23.4}$&                      169.6$^{+21.7}_{-20.6}$&                      124.8$^{+14.3}_{-13.3}$&                      286.0$^{+20.3}_{-19.2}$&                      120.9$^{+15.8}_{-14.7}$&                       46.7$^{+15.4}_{-14.3}$&                    \\
\\
\enddata

\tablecomments{Table 3a is available in its entirety from the World Wide Web site listed in Footnote 8.}

\tablenotetext{a}{Source was detected with the restricted ACIS grade set with a probability threshold of $10^{-7}$.}
\tablenotetext{b}{Source was detected with the standard {\it ASCA} grade set with a probability threshold of $10^{-5}$.}
\tablenotetext{c}{Source positional offset $>$2.5{\ifmmode ^{\prime\prime}\else$^{\prime\prime}$\fi} from the primary source position.}

\end{deluxetable}

%
%

\begin{deluxetable}{lllcccrrrrrrrccc}
\tablenum{3b}
\rotate
\tabletypesize{\scriptsize}
\tablewidth{0pt}
\tablecaption{Main {\it Chandra} Catalog: Additional Source Properties}
\tablehead{
\colhead{} &
\multicolumn{2}{c}{X-ray Coordinates} &
\colhead{Effective}             &
\colhead{Band}                  &
\colhead{Effective}             &
\multicolumn{7}{c}{Flux}        &
\colhead{Band}                  &
\colhead{Band}                  &
\colhead{Band}               \\
\colhead{\#}                    &
\colhead{$\alpha_{2000}$}       &
\colhead{$\delta_{2000}$}       &
\colhead{Exposure}     &
\colhead{Ratio1}       &
\colhead{$\Gamma$}     &
\colhead{FB}           &
\colhead{SB}           &
\colhead{HB}           &
\colhead{SB1}          &
\colhead{SB2}          &
\colhead{HB1}          &
\colhead{HB2}          &
\colhead{Ratio2}       &
\colhead{Ratio3}       &
\colhead{Ratio4}       \\
\colhead{(1)}         &
\colhead{(2)}         &
\colhead{(3)}         &
\colhead{(4)}         &
\colhead{(5)}         &
\colhead{(6)}         &
\colhead{(7)}         &
\colhead{(8)}         &
\colhead{(9)}         &
\colhead{(10)}         &
\colhead{(11)}         &
\colhead{(12)}         &
\colhead{(13)}         &
\colhead{(14)}         &
\colhead{(15)}         &
\colhead{(16)} \\
}
\startdata
  1&12 35 15.04&+62 14 07.0& 165.7&                                      $<$0.31&                                      $>$1.87&                                4.41&                                2.09&                             $<$2.59&                                0.56&                                1.39&                             $<$1.16&                             $<$3.57&                       $3.91^{+3.01}_{-1.63}$&                                          $-$&                                      $<$0.24\\
  2&12 35 16.76&+62 15 37.8& 492.8&                                      $<$2.44&                               1.40$^{\rm a}$&                                0.63&                                0.14&                             $<$1.60&                             $<$0.23&                                0.13&                             $<$0.70&                             $<$2.17&                                      $>$0.83&                                          $-$&                                      $<$1.60\\
  3&12 35 18.82&+62 15 51.5& 891.1&                                      $<$0.60&                                      $>$1.28&                                1.57&                                0.38&                             $<$1.10&                             $<$0.15&                                0.32&                             $<$0.47&                             $<$1.55&                                      $>$3.19&                                          $-$&                                      $<$0.49\\
\enddata

\tablecomments{Table 3b is available in its entirety from the World Wide Web site listed in Footnote 8. Band Ratio1=HB/SB; Band Ratio2=SB2/SB1; Band Ratio3=HB2/HB1; Band Ratio4=HB1/SB2.}
\tablenotetext{a}{Photon index set to given value due to low number of counts (see \S3.4.1).}

\end{deluxetable}

%
%
\begin{deluxetable}{lccccc}
\tablenum{4}
\tabletypesize{\small}
\tablewidth{0pt}
\tablecaption {Summary of {\it Chandra} Source Detections}
\scriptsize
\tablehead{
\colhead{}                                 &
\colhead{Number of}                              &
\multicolumn{4}{c}{Detected Counts Per Source}   \\
\colhead{Band (keV)}                                      &
\colhead{Sources}                        &
\colhead{Maximum}                                   &
\colhead{Minimum}                                   &
\colhead{Median}                                    &
\colhead{Mean}                                      
}
\startdata
Full (0.5--8.0) & 479 & 20994.8 &  11.4 &  99.3 & 438.7 \\
Soft (0.5--2.0) & 451 & 15233.9 &   4.3 &  54.3 & 307.8 \\
Hard  (2--8) & 332 &  5793.7 &   6.6 &  76.7 & 209.2 \\
SB1 (0.5--1.0) & 249 &  4696.3 &   4.1 &  33.8 & 161.3 \\
SB2 (1--2) & 413 & 10517.2 &   4.8 &  48.6 & 234.5 \\
HB1 (2--4) & 310 &  3933.0 &   4.1 &  49.1 & 146.7 \\
HB2 (4--8) & 183 &  1851.1 &   6.7 &  53.3 & 114.8 \\
\enddata
\tablecomments{There are 503 independent X-ray sources detected in total with a false-positive probability threshold of $1\times 10^{-7}$. We have included cross-band counterparts from the {\sc wavdetect} runs with a false-positive probability threshold of $1\times 10^{-5}$ (see \S3.4.1).}
\end{deluxetable}
\pagebreak

%
%
\begin{deluxetable}{lccccccc}
\tablenum{5}
\tabletypesize{\small}
\tablewidth{0pt}
\tablecaption {Sources Detected in One Band but Not Another}
\scriptsize
\tablehead{
\colhead{Detection}                              &
\multicolumn{7}{c}{Non-Detection Energy Band}    \\
\colhead{Band (keV)}                            &
\colhead{Full}                    &
\colhead{Soft}                    &
\colhead{Hard}                        &
\colhead{SB1}                       &
\colhead{SB2}                         &
\colhead{HB1}                         &
\colhead{HB2} 
}
\startdata
Full (0.5--8.0)& -- & 51 &148 &230 & 84 &169 &296\\
Soft (0.5--2.0)& 23 & -- &166 &203 & 43 &179 &289\\
Hard (2--8)    &  1 & 47 & -- &162 & 56 & 34 &149\\
SB1 (0.5--1.0) &  0 &  1 & 79 & -- & 18 & 82 &142\\
SB2 (1--2)     & 18 &  5 &137 &182 & -- &145 &254\\
HB1 (2--4)     &  0 & 38 & 12 &143 & 42 & -- &131\\
HB2 (4--8)     &  0 & 21 &  0 & 76 & 24 &  0 & --\\
\enddata
\tablecomments{For example, there were 51 sources detected in the full band that were not detected in the soft band.}
\end{deluxetable}
%

%
%
\begin{deluxetable}{llcl}
\tablenum{6}
\tabletypesize{\small}
\tablewidth{0pt}
\tablecaption {1 Ms Chandra sources not detected in the 2 Ms Chandra exposure}
\scriptsize
\tablehead{
\multicolumn{2}{c}{X-ray Coordinates} &
\colhead{$R$ mag}  &
\colhead{Notes} 
}
\startdata
12 36 27.28 & +62 13 08.3 & 27.2    & Possible very weak full and hard-band emission \\
12 36 33.86 & +62 13 27.7 & 24.5    & No evidence for X-ray emission \\
12 37 12.66 & +62 13 42.5 & $>$27.5 & No evidence for X-ray emission  \\
12 37 44.53 & +62 06 41.4 & $>$27.3 & Possible very weak full and soft-band emission \\
12 38 13.46 & +62 11 34.1 & $>$28.4 & No evidence for X-ray emission  \\
\enddata
\end{deluxetable}
%

%

%
%

\begin{deluxetable}{lllcccccccccllcccl}
\tablenum{7a}
\rotate
\tabletypesize{\scriptsize}
\tablewidth{0pt}
\tablecaption{Supplementary Optically Bright {\it Chandra} Catalog: Basic Source Properties}
\tablehead{
\colhead{} &
\multicolumn{2}{c}{X-ray Coordinates} &
\colhead{Pos}                   &
\colhead{Off-Axis}                   &
\multicolumn{7}{c}{Counts}      &
\multicolumn{2}{c}{Optical Coordinates} &
\colhead{Offset}                &
\colhead{}                      &
\colhead{False}                 &
\colhead{}  \\
\colhead{\#}                    &
\colhead{$\alpha_{2000}$}       &
\colhead{$\delta_{2000}$}       &
\colhead{Error}       &
\colhead{Angle}       &
\colhead{FB}          &
\colhead{SB}          &
\colhead{HB}          &
\colhead{SB1}         &
\colhead{SB2}         &
\colhead{HB1}         &
\colhead{HB2}         &
\colhead{$\alpha_{2000}$}       &
\colhead{$\delta_{2000}$}       &
\colhead{$O-X$}       &
\colhead{$R$ mag}     &
\colhead{Prob}      &
\colhead{Notes}  \\
\colhead{(1)}         &
\colhead{(2)}         &
\colhead{(3)}         &
\colhead{(4)}         &
\colhead{(5)}         &
\colhead{(6)}         &
\colhead{(7)}         &
\colhead{(8)}         &
\colhead{(9)}         &
\colhead{(10)}        &
\colhead{(11)}        &
\colhead{(12)}        &
\colhead{(13)}        &
\colhead{(14)}        &
\colhead{(15)}        &
\colhead{(16)}        &
\colhead{(17)}        &
\colhead{(18)}  \\
}
\startdata
  1&12 35 48.31&+62 15 05.7& 1.2& 6.78&                       25.2$\pm8.2$$^{\rm b}$&                       16.9$\pm5.6$$^{\rm b}$&                                      $<$27.3&                                      $<$14.4&                                      $<$21.9&                                      $<$16.1&                                      $<$23.1&12 35 48.26&+62 15 06.6& 1.0&20.9&   0.00270&                    \\
  2&12 35 54.31&+62 15 33.3& 1.2& 6.20&                                      $<$29.3&                                      $<$20.2&                                      $<$23.1&                                      $<$11.8&                        9.1$\pm3.6$$^{\rm a}$&                                      $<$13.9&                                      $<$18.7&12 35 54.28&+62 15 32.0& 1.3&22.5&   0.00902&                    \\
  3&12 35 55.47&+62 10 51.9& 1.2& 6.62&                                      $<$26.3&                       10.7$\pm4.2$$^{\rm b}$&                                      $<$22.0&                                      $<$11.5&                        9.5$\pm3.7$$^{\rm a}$&                                      $<$13.7&                                      $<$17.0&12 35 55.40&+62 10 51.1& 1.0&20.1&   0.00150&                   \\
  4&12 35 57.88&+62 15 36.5& 1.2& 5.81&                                      $<$28.9&                       17.2$\pm5.3$$^{\rm b}$&                                      $<$21.6&                                      $<$16.1&                                      $<$15.8&                                      $<$12.9&                                      $<$17.9&12 35 57.97&+62 15 36.7& 0.7&19.6&   0.00104&                    \\
\enddata

\tablecomments{Table 7a is available in its entirety from the World Wide Web site listed in Footnote 8.}

\tablenotetext{a}{Source was detected with the restricted ACIS grade set with a probability threshold of $10^{-5}$.}
\tablenotetext{b}{Source positional offset $>$2.5{\ifmmode ^{\prime\prime}\else$^{\prime\prime}$\fi} from the primary source position.}

\end{deluxetable}

%
%

\begin{deluxetable}{lllcrrrrrrr}
\tablenum{7b}
\rotate
\tabletypesize{\scriptsize}
\tablewidth{0pt}
\tablecaption{Supplementary Optically Bright {\it Chandra} Catalog: Additional Source Properties}
\tablehead{
\colhead{} &
\multicolumn{2}{c}{X-ray Coordinates} &
\colhead{Effective}             &
\multicolumn{7}{c}{Flux}   \\
\colhead{\#}                    &
\colhead{$\alpha_{2000}$}       &
\colhead{$\delta_{2000}$}       &
\colhead{Exposure}     &
\colhead{FB}           &
\colhead{SB}           &
\colhead{HB}           &
\colhead{SB1}          &
\colhead{SB2}          &
\colhead{HB1}          &
\colhead{HB2}         \\
\colhead{(1)}         &
\colhead{(2)}         &
\colhead{(3)}         &
\colhead{(4)}         &
\colhead{(5)}         &
\colhead{(6)}         &
\colhead{(7)}         &
\colhead{(8)}         &
\colhead{(9)}         &
\colhead{(10)}         &
\colhead{(11)} \\
}
\startdata
  1&12 35 48.31&+62 15 05.7&1689.3&                                0.13&                                0.05&                             $<$0.34&                             $<$0.06&                             $<$0.06&                             $<$0.15&                             $<$0.54\\
  2&12 35 54.31&+62 15 33.3&1694.9&                             $<$0.15&                             $<$0.06&                             $<$0.28&                             $<$0.05&                                0.02&                             $<$0.13&                             $<$0.43\\
  3&12 35 55.47&+62 10 51.9&1063.7&                             $<$0.21&                                0.05&                             $<$0.43&                             $<$0.07&                                0.04&                             $<$0.19&                             $<$0.60\\
  4&12 35 57.88&+62 15 36.5&1739.4&                             $<$0.14&                                0.05&                             $<$0.26&                             $<$0.06&                             $<$0.04&                             $<$0.11&                             $<$0.40\\
  5&12 36 00.07&+62 14 23.0&1756.4&                                0.11&                             $<$0.04&                                0.26&                             $<$0.04&                             $<$0.03&                                0.08&                             $<$0.41\\
\\
  6&12 36 01.61&+62 17 13.2&1530.7&                             $<$0.15&                                0.05&                             $<$0.30&                             $<$0.03&                             $<$0.05&                             $<$0.13&                             $<$0.44\\
\enddata

\tablecomments{Table 7b is available in its entirety from the World Wide Web site listed in Footnote 8.}

\end{deluxetable}

%
%
\begin{deluxetable}{lcccc}
\tablenum{8}
\tabletypesize{\small}
\tablewidth{0pt}
\tablecaption {Background Parameters for the {\it ASCA}-grade images}
\scriptsize
\tablehead{
\colhead{}                  &
\multicolumn{2}{c}{Mean Background} &
\colhead{Total Background$^{\rm c}$}   &
\colhead{Count Ratio$^{\rm d}$} \\
\colhead{Band (keV)}                  &
\colhead{(Counts pixel$^{-1}$)$^{\rm a}$} &
\colhead{(Counts Ms$^{-1}$ pixel$^{-1}$)$^{\rm b}$} &
\colhead{($10^{5}$ counts)}              &
\colhead{(Background/Source)} 
}
\startdata
Full (0.5--8.0)& 0.211 & 0.191 & 14.1 &  6.7 \\
Soft (0.5--2.0)& 0.062 & 0.056 &  4.1 &  2.9 \\
Hard (2--8)    & 0.150 & 0.135 & 10.0 & 14.0 \\
SB1 (0.5--1.0) & 0.027 & 0.025 &  1.8 &  4.4 \\
SB2 (1--2)     & 0.035 & 0.031 &  2.3 &  2.3 \\
HB1 (2--4)     & 0.051 & 0.046 &  3.4 &  7.2 \\
HB2 (4--8)     & 0.099 & 0.089 &  6.6 & 27.6 \\
\enddata
\tablenotetext{a}{The mean numbers of counts per pixel. These are measured from the masked background images described in \S4.2.}
\tablenotetext{b}{The mean numbers of counts per pixel divided by the mean effective exposure. These are measured from the exposure maps and masked background images described in \S4.2.}
\tablenotetext{c}{Total number of background counts (calculated from the total number of pixels and the mean background).}
\tablenotetext{d}{Ratio of the total number of background counts to the total number of source counts.} 
\end{deluxetable}

%
%
\begin{deluxetable}{lccccc}
\tablenum{9}
\tabletypesize{\small}
\tablewidth{0pt}
\tablecaption {Sensitivity parameters}
\scriptsize
\tablehead{
\colhead{}                         &
\multicolumn{2}{c}{Photon Limited}    &
\multicolumn{3}{c}{Flux Limit}     \\
\colhead{Band (keV)}                  &
\colhead{Area$^{\rm a}$}      &
\colhead{Exposure$^{\rm b}$}              &
\colhead{Current$^{\rm c}$}           &
\colhead{4~Ms$^{\rm d}$}           &
\colhead{8~Ms$^{\rm d}$}
}
\startdata
Full (0.5--8.0)& 25  &  4 &  7.1 &  4.2 &  2.6 \\
Soft (0.5--2.0)& 83  & 25 &  2.5 &  1.3 &  0.7 \\
Hard (2--8)    & 34  &  4 & 14.1 &  8.1 &  5.1 \\
SB1 (0.5--1.0) & 134 & 50 &  3.2 &  1.6 &  0.8 \\
SB2 (1--2)     & 122 & 40 &  2.3 &  1.2 &  0.6 \\
HB1 (2--4)     & 92  & 15 &  7.4 &  3.9 &  2.2 \\
HB2 (4--8)     & 47  &  6 & 22.4 & 12.6 &  7.8 \\
\enddata
\tablenotetext{a}{Current area of the field that is photon limited (arcmin$^2$); see \S4.3.}
\tablenotetext{b}{Predicted maximum photon-limited exposure for an area of $\approx$~1~arcmin$^2$ (Ms); see \S4.3.}
\tablenotetext{c}{Current S/N$=3$ flux limit (10$^{-17}$~erg~cm$^{-2}$~s$^{-1}$) at the aim point (corresponding to the deepest $\approx$~1~arcmin$^2$ region); see \S4.2.}
\tablenotetext{d}{Predicted S/N$=3$ flux limit (10$^{-17}$~erg~cm$^{-2}$~s$^{-1}$) at the aim point (corresponding to the deepest $\approx$~1~arcmin$^2$ region) for 4~Ms and 8~Ms exposures; see footnote 13 and \S4.3 for assumptions.}
\end{deluxetable}

\clearpage

%
%

\clearpage

\begin{deluxetable}{lccclccl}
\tablenum{A1}
\tabletypesize{\small}
\tablewidth{0pt}
\tablecaption{Journal of \chandra\ Deep Field-South Observations}
\small
%
\tablehead{
\colhead{Obs.}                                 &
\colhead{Obs.}                                 &
\colhead{Exposure}                             &
\multicolumn{2}{c}{Aim Point$^{\rm b}$}        &
\colhead{Roll}                                 & 
\colhead{Obs.}                                 &
\colhead{Pipeline}                             \\
\colhead{ID}                                 &
\colhead{Start (UT)}                              &
\colhead{Time (ks)$^{\rm a}$ }               &
\colhead{$\alpha_{\rm 2000}$}                &
\colhead{$\delta_{\rm 2000}$}                &
\colhead{Angle ($^\circ$)$^{\rm c}$}         &
\colhead{Mode$^{\rm d}$}                     &
\colhead{Version$^{\rm e}$}     
}
\startdata

1431-0 & 1999 Oct 15, 17:38 & 21.1  & 3 32 29.44 & $-$27 48 21.8 & 47.3  & VF & R4CU5UPD9\\
\\
1431-1 & 1999 Nov 23, 02:30 & 93.6  & 3 32 29.35 & $-$27 48 40.4 & 353.9 & F & R4CU5UPD9\\
\\
 441 & 2000 May 27, 01:18   & 56.0  & 3 32 26.75 & $-$27 48 17.4 & 166.7 & F & R4CU5UPD13.3\\
\\
 582 & 2000 June 03, 02:38  & 130.5 & 3 32 26.82 & $-$27 48 16.3 & 162.9 & F & R4CU5UPD13.3\\
\\
\enddata
\tablecomments{The focal-plane temperature was 
$-110^\circ$\,C during the first two observations and 
$-120^\circ$\,C during the others.}
\tablenotetext{a}{All observations were continuous. These times have been corrected for 
the lost exposure due to the read-out time per CCD frame and also reflect
the total exposure after the data affected by high background have been
removed. The short time intervals with bad satellite aspect are
negligible and have not been removed. The total exposure time for the
11 observations is 939.4~ks.}
\tablenotetext{b}{The average aim point, weighted by exposure time, is
$\alpha_{2000}=$~3$^{\rm h}$ 32$^{\rm m}$ $28\fs 2$,
$\delta_{2000}=$~$-27^\circ$48$^{\prime}$36$^{\prime\prime}$; individual aim points are offset by 5--30$^{\prime\prime}$ from the average aim point.} 
\tablenotetext{c}{Roll angle describes the orientation of the \chandra\ instruments
on the sky.  The angle is between 0--360$^{\circ}$, and it increases to the West
of North (opposite to the sense of traditional position angle).}
\tablenotetext{d}{The observing mode (F=Faint mode; VF=Very Faint mode).}
\tablenotetext{e}{The version of the CXC pipeline software used for basic 
processing of the data.}
\end{deluxetable}

%

%
%

\begin{deluxetable}{lllcccccccccl}
\tablenum{A2a}
\rotate
\tabletypesize{\scriptsize}
\tablewidth{0pt}
\tablecaption{Main {\it Chandra} Catalog: Basic Source Properties}
\tablehead{
\colhead{} &
\multicolumn{2}{c}{X-ray Coordinates} &
\colhead{Pos}                   &
\colhead{Off-Axis}                   &
\multicolumn{7}{c}{Counts}      &
\colhead{}  \\
\colhead{\#}                    &
\colhead{$\alpha_{2000}$}       &
\colhead{$\delta_{2000}$}       &
\colhead{Error}       &
\colhead{Angle}       &
\colhead{FB}          &
\colhead{SB}          &
\colhead{HB}          &
\colhead{SB1}         &
\colhead{SB2}         &
\colhead{HB1}         &
\colhead{HB2}         &
\colhead{Notes}  \\
\colhead{(1)}         &
\colhead{(2)}         &
\colhead{(3)}         &
\colhead{(4)}         &
\colhead{(5)}         &
\colhead{(6)}         &
\colhead{(7)}         &
\colhead{(8)}         &
\colhead{(9)}         &
\colhead{(10)}        &
\colhead{(11)}        &
\colhead{(12)}        &
\colhead{(13)}  \\
}
\startdata
  1&3 31 44.21 &$-$27 49 26.1& 1.4& 9.76&                       76.0$^{+15.7}_{-14.6}$&                        38.1$^{+10.5}_{-9.4}$&                                      $<$32.0&                                      $<$15.1&                                      $<$22.3&                                      $<$19.4&                                      $<$25.5&                    \\
  2&3 31 44.72 &$-$27 51 56.3& 1.4&10.18&                      102.7$^{+18.6}_{-17.4}$&             85.8$^{+13.9}_{-12.7}$$^{\rm c}$&                                      $<$34.4&                                      $<$19.0&                       71.9$^{+12.0}_{-10.8}$&                                      $<$22.4&                                      $<$26.2&                   O\\
  3&3 31 44.85 &$-$27 51 39.2& 1.4&10.06&                       87.9$^{+18.3}_{-17.2}$&                       59.3$^{+12.7}_{-11.7}$&                                      $<$36.2&                                      $<$20.4&                        39.8$^{+10.3}_{-9.3}$&                                      $<$22.6&                                      $<$28.4&                    \\
  4&3 31 47.30 &$-$27 53 13.3& 0.5&10.15&                      485.9$^{+26.9}_{-25.8}$&                      325.5$^{+20.7}_{-19.7}$&                      158.9$^{+17.4}_{-16.3}$&                        71.7$^{+10.7}_{-9.7}$&                      253.8$^{+18.2}_{-17.2}$&                      110.3$^{+13.6}_{-12.3}$&                       47.1$^{+11.8}_{-10.7}$&                    \\
  5&3 31 47.98 &$-$27 50 45.5& 0.5& 9.15&                      438.2$^{+26.4}_{-25.3}$&                      338.7$^{+21.5}_{-20.3}$&                      102.3$^{+16.1}_{-14.9}$&                      125.3$^{+13.6}_{-12.5}$&                      213.1$^{+17.2}_{-16.1}$&                       76.2$^{+12.1}_{-11.0}$&                                      $<$28.2&                    \\
\\
\enddata

\tablecomments{Table A2a is available in its entirety from the World Wide Web site listed in Footnote 8.}

\tablenotetext{a}{Source was detected with the restricted ACIS grade set with a probability threshold of $10^{-7}$.}
\tablenotetext{b}{Source was detected with the standard {\it ASCA} grade set with a probability threshold of $10^{-5}$.}
\tablenotetext{c}{Source positional offset $>$2.5{\ifmmode ^{\prime\prime}\else$^{\prime\prime}$\fi} from the primary source position.}

\end{deluxetable}

%
%

\begin{deluxetable}{lllcccrrrrrrrccc}
\tablenum{A2b}
\rotate
\tabletypesize{\scriptsize}
\tablewidth{0pt}
\tablecaption{Main {\it Chandra} Catalog: Additional Source Properties}
\tablehead{
\colhead{} &
\multicolumn{2}{c}{X-ray Coordinates} &
\colhead{Effective}             &
\colhead{Band}                  &
\colhead{Effective}             &
\multicolumn{7}{c}{Flux}        &
\colhead{Band}                  &
\colhead{Band}                  &
\colhead{Band}               \\
\colhead{\#}                    &
\colhead{$\alpha_{2000}$}       &
\colhead{$\delta_{2000}$}       &
\colhead{Exposure}     &
\colhead{Ratio1}       &
\colhead{$\Gamma$}     &
\colhead{FB}           &
\colhead{SB}           &
\colhead{HB}           &
\colhead{SB1}          &
\colhead{SB2}          &
\colhead{HB1}          &
\colhead{HB2}          &
\colhead{Ratio2}       &
\colhead{Ratio3}       &
\colhead{Ratio4}       \\
\colhead{(1)}         &
\colhead{(2)}         &
\colhead{(3)}         &
\colhead{(4)}         &
\colhead{(5)}         &
\colhead{(6)}         &
\colhead{(7)}         &
\colhead{(8)}         &
\colhead{(9)}         &
\colhead{(10)}         &
\colhead{(11)}         &
\colhead{(12)}         &
\colhead{(13)}         &
\colhead{(14)}         &
\colhead{(15)}         &
\colhead{(16)} \\
}
\startdata
  1&3 31 44.21 &$-$27 49 26.1& 567.5&                                      $<$0.87&                                      $>$0.88&                                2.18&                                0.34&                             $<$1.68&                             $<$0.16&                             $<$0.19&                             $<$0.56&                             $<$2.29&                                          $-$&                                          $-$&                                          $-$\\
  2&3 31 44.72 &$-$27 51 56.3& 690.6&                                      $<$0.42&                                      $>$1.54&                                1.68&                                0.65&                             $<$1.28&                             $<$0.17&                                0.48&                             $<$0.52&                             $<$1.79&                                      $>$3.89&                                          $-$&                                      $<$0.31\\
  3&3 31 44.85 &$-$27 51 39.2& 715.4&                                      $<$0.64&                                      $>$1.16&                                1.72&                                0.42&                             $<$1.42&                             $<$0.17&                                0.26&                             $<$0.51&                             $<$1.96&                                      $>$2.00&                                          $-$&                                      $<$0.56\\
\enddata

\tablecomments{Table A2b is available in its entirety from the World Wide Web site listed in Footnote 8. Band Ratio1=HB/SB; Band Ratio2=SB2/SB1; Band Ratio3=HB2/HB1; Band Ratio4=HB1/SB2.}
\tablenotetext{a}{Photon index set to given value due to low number of counts.}

\end{deluxetable}

%

%
%

\begin{deluxetable}{lllcccccccccllcccl}
\tablenum{A3a}
\rotate
\tabletypesize{\scriptsize}
\tablewidth{0pt}
\tablecaption{Supplementary Optically Bright {\it Chandra} Catalog: Basic Source Properties}
\tablehead{
\colhead{} &
\multicolumn{2}{c}{X-ray Coordinates} &
\colhead{Pos}                   &
\colhead{Off-Axis}                   &
\multicolumn{7}{c}{Counts}      &
\multicolumn{2}{c}{Optical Coordinates} &
\colhead{Offset}                &
\colhead{}                      &
\colhead{False}                 &
\colhead{}  \\
\colhead{\#}                    &
\colhead{$\alpha_{2000}$}       &
\colhead{$\delta_{2000}$}       &
\colhead{Error}       &
\colhead{Angle}       &
\colhead{FB}          &
\colhead{SB}          &
\colhead{HB}          &
\colhead{SB1}         &
\colhead{SB2}         &
\colhead{HB1}         &
\colhead{HB2}         &
\colhead{$\alpha_{2000}$}       &
\colhead{$\delta_{2000}$}       &
\colhead{$O-X$}       &
\colhead{$R$ mag}     &
\colhead{Prob}      &
\colhead{Notes}  \\
\colhead{(1)}         &
\colhead{(2)}         &
\colhead{(3)}         &
\colhead{(4)}         &
\colhead{(5)}         &
\colhead{(6)}         &
\colhead{(7)}         &
\colhead{(8)}         &
\colhead{(9)}         &
\colhead{(10)}        &
\colhead{(11)}        &
\colhead{(12)}        &
\colhead{(13)}        &
\colhead{(14)}        &
\colhead{(15)}        &
\colhead{(16)}        &
\colhead{(17)}        &
\colhead{(18)}  \\
}
\startdata
  1&3 31 59.36 &$-$27 46 50.8& 1.2& 6.62&                                 22.8$\pm6.8$&                                      $<$16.6&                                      $<$19.7&                                      $<$11.0&                                      $<$13.8&                                      $<$13.1&                                      $<$15.0&3 31 59.31 &$-$27 46 51.3& 0.8&22.8&   0.01200&                    \\
  2&3 32 10.87 &$-$27 47 21.9& 1.2& 4.03&                                  5.8$\pm2.8$&                                  4.1$\pm2.2$&                                      $<$12.1&                                      $<$15.7&                                      $<$13.5&                                       $<$9.3&                                      $<$10.5&3 32 10.91 &$-$27 47 22.7& 0.9&20.5&   0.00193&                   \\
  3&3 32 12.33 &$-$27 49 12.1& 1.2& 3.56&                        8.9$\pm3.7$$^{\rm a}$&                        5.1$\pm2.4$$^{\rm a}$&                                      $<$10.3&                        2.8$\pm1.7$$^{\rm a}$&                                       $<$9.3&                                       $<$7.9&                                       $<$8.3&3 32 12.38 &$-$27 49 12.6& 0.9&22.7&   0.01120&                   \\
  4&3 32 12.46 &$-$27 47 53.6& 1.2& 3.56&                                  5.9$\pm2.8$&                                  4.3$\pm2.2$&                                      $<$11.7&                                       $<$9.7&                                      $<$10.7&                                       $<$7.9&                                       $<$9.4&3 32 12.52 &$-$27 47 53.7& 0.7&22.6&   0.00990&                   \\
\enddata

\tablecomments{Table A3a is available in its entirety from the World Wide Web site listed in Footnote 8.}

\tablenotetext{a}{Source was detected with the restricted ACIS grade set with a probability threshold of $10^{-5}$.}
\tablenotetext{b}{Source positional offset $>$2.5{\ifmmode ^{\prime\prime}\else$^{\prime\prime}$\fi} from the primary source position.}

\end{deluxetable}

%
%

\begin{deluxetable}{lllcrrrrrrr}
\tablenum{A3b}
\rotate
\tabletypesize{\scriptsize}
\tablewidth{0pt}
\tablecaption{Supplementary Optically Bright {\it Chandra} Catalog: Additional Source Properties}
\tablehead{
\colhead{} &
\multicolumn{2}{c}{X-ray Coordinates} &
\colhead{Effective}             &
\multicolumn{7}{c}{Flux}   \\
\colhead{\#}                    &
\colhead{$\alpha_{2000}$}       &
\colhead{$\delta_{2000}$}       &
\colhead{Exposure}     &
\colhead{FB}           &
\colhead{SB}           &
\colhead{HB}           &
\colhead{SB1}          &
\colhead{SB2}          &
\colhead{HB1}          &
\colhead{HB2}         \\
\colhead{(1)}         &
\colhead{(2)}         &
\colhead{(3)}         &
\colhead{(4)}         &
\colhead{(5)}         &
\colhead{(6)}         &
\colhead{(7)}         &
\colhead{(8)}         &
\colhead{(9)}         &
\colhead{(10)}         &
\colhead{(11)} \\
}
\startdata
  1&3 31 59.36 &$-$27 46 50.8& 855.6&                                0.24&                             $<$0.10&                             $<$0.53&                             $<$0.08&                             $<$0.07&                             $<$0.24&                             $<$0.73\\
  2&3 32 10.87 &$-$27 47 21.9& 842.6&                                0.06&                                0.03&                             $<$0.32&                             $<$0.12&                             $<$0.07&                             $<$0.17&                             $<$0.50\\
  3&3 32 12.33 &$-$27 49 12.1& 909.2&                                0.09&                                0.03&                             $<$0.26&                                0.02&                             $<$0.05&                             $<$0.14&                             $<$0.37\\
  4&3 32 12.46 &$-$27 47 53.6& 887.0&                                0.06&                                0.03&                             $<$0.30&                             $<$0.07&                             $<$0.05&                             $<$0.14&                             $<$0.42\\
  5&3 32 15.86 &$-$27 47 49.9& 815.0&                                0.06&                             $<$0.10&                             $<$0.37&                             $<$0.06&                                0.02&                             $<$0.17&                             $<$0.45\\
\\
  6&3 32 16.10 &$-$27 49 41.1& 938.8&                                0.08&                                0.02&                             $<$0.24&                                0.03&                             $<$0.06&                             $<$0.10&                             $<$0.44\\
\enddata

\tablecomments{Table A3b is available in its entirety from the World Wide Web site listed in Footnote 8.}

\end{deluxetable}


%
%

\begin{figure}
\vspace{-0.5truein}
\epsscale{0.9}
\figurenum{1}
\centerline{\includegraphics[width=15.0cm]{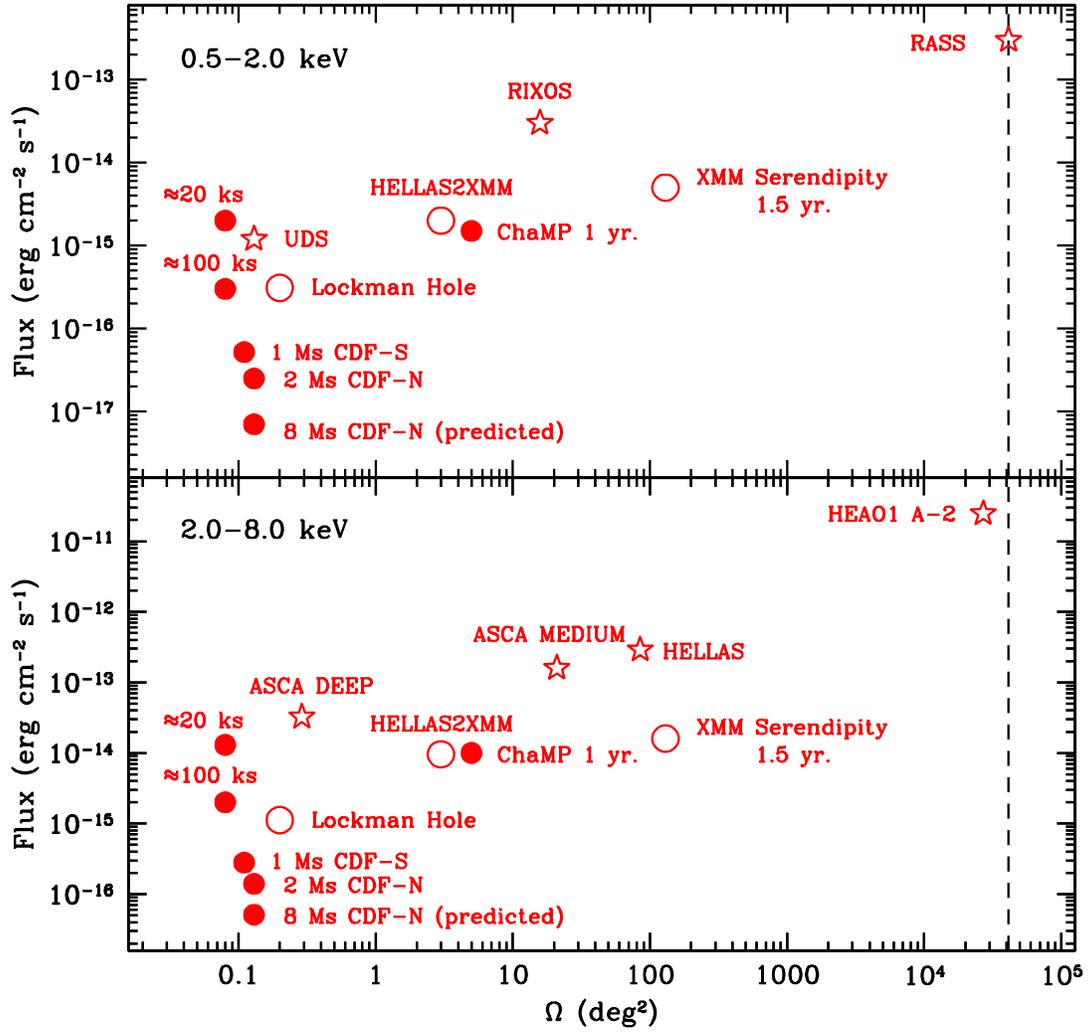}}
\caption{Flux versus solid angle for a variety of
X-ray surveys in the soft (top) and hard (bottom) bands. The vertical
dashed lines indicate the solid angle of the entire sky. The
pre-\chandra\ and \xmm\ surveys are shown as stars, the \chandra\
surveys are shown as filled circles, and the \xmm\ surveys are shown
as open circles. In general, only surveys with results in refereed
journals are shown. The pre-\chandra\ surveys shown are \rosat\ (RASS,
Voges \etal 1999; RIXOS, Mason \etal 2000; UDS, Lehmann \etal 2001),
\heao\ A-2 (Piccinotti \etal 1982), \asca\ (ASCA Medium, 
Cagnoni \etal 1998; ASCA Deep, Ogasaka \etal 1998),
\sax\ (HELLAS, Fiore \etal 2001). The \chandra\ surveys shown are 
$\approx$~20~ks \chandra\ surveys (e.g.,\ Brandt \etal 2000, Fiore
\etal 2000, Crawford \etal 2001, McHardy \etal 2003), $\approx$~100~ks \chandra\ surveys 
(e.g.,\ Mushotzky \etal 2000, Barger \etal 2001, Cowie \etal 2002,
Stern \etal 2002, Manners \etal 2003), ChaMP (Wilkes \etal 2001), the
1~Ms CDF-S survey (G02), and the 2~Ms and 8~Ms CDF-N surveys (see
\S4.2, \S4.3, and Table~9 for the sensitivity parameters of the CDF-N
surveys). The \xmm\ surveys shown are the Lockman Hole (Hasinger \etal
2001), HELLAS2XMM (Baldi \etal 2002), and the XMM Serendipity survey
(Watson \etal 2001, 2003, Barcons \etal 2002). For the small-area
surveys, where depth is usually important, we have shown the
approximate on-axis flux limits while for the large-area surveys
(i.e.,\ $>1$~deg$^2$), where areal coverage is usually imporant, we
have shown the flux limits corresponding to $\approx$~30\% of the
total area. When appropriate, we converted the flux limits into the
bands shown here assuming $\Gamma=$~1.4.}
\end{figure}

\clearpage

%
%

\begin{figure}
\vspace{-0.5truein}
\figurenum{2}
\centerline{\includegraphics[width=15.0cm]{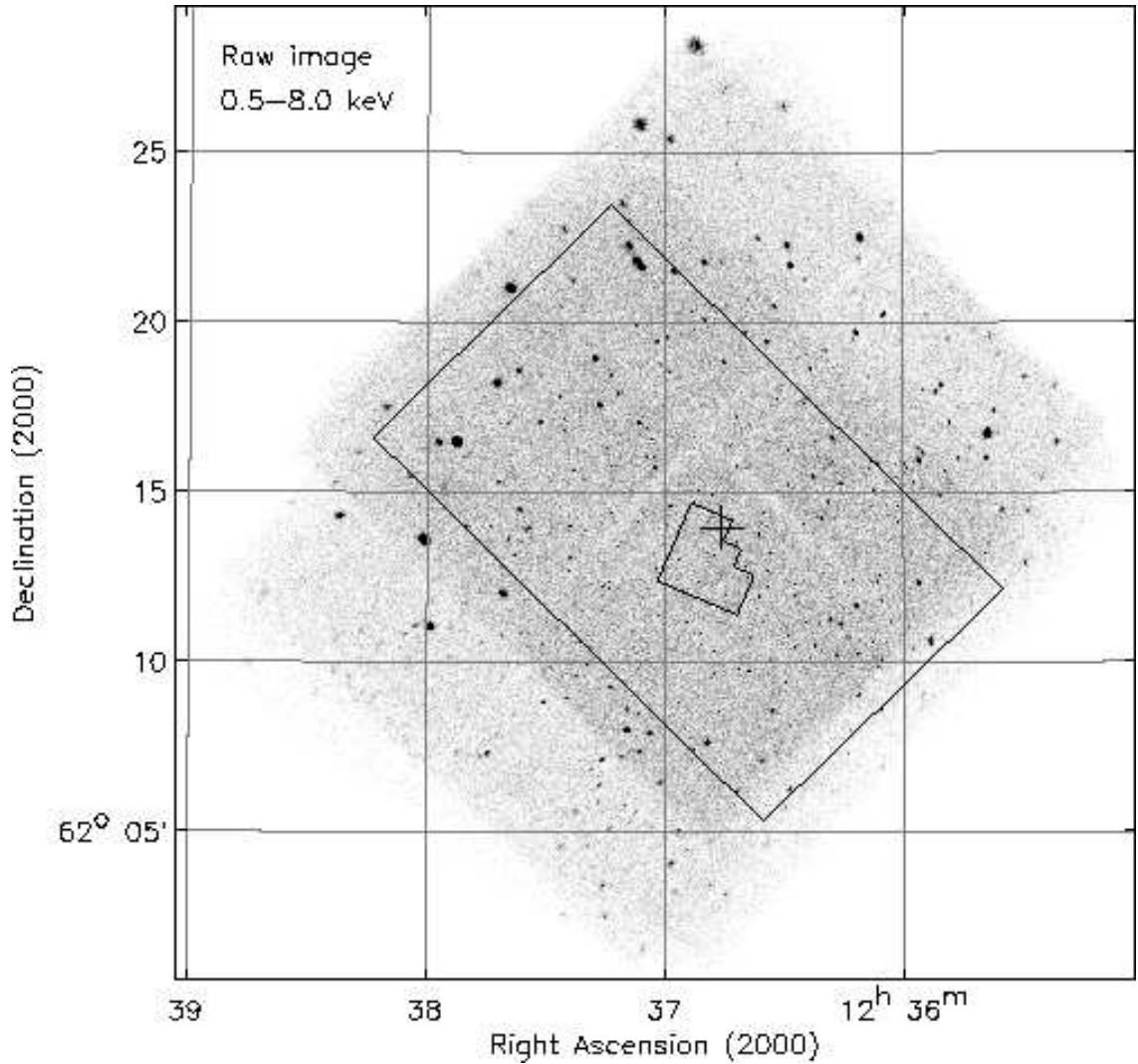}}
\vspace{-0.0truein}
\caption{Full-band raw image of the 2~Ms CDF-N. This 
image has been made using the standard \asca\ grade set (see Table~2)
and is binned by a factor of four in both right ascension (RA) and
$\approx$~$0\farcs 1$ in declination (Dec). The light grooves running
through the image correspond to the gaps between the CCDs. The small
polygon indicates the HDF-N, and the large rectangle indicates the
expected area of the IRAC observations for the Great Observatories
Origins Deep Survey (GOODS; e.g.,\ Dickinson \& Giavalisco 2003). The
cross near the center of the image indicates the average aim point,
weighted by exposure time (see Table~1).
Only one of the raw images is included here; please see the World Wide
Web site listed in Footnote~8 for raw images in the other bands listed
in Table~2.}

\end{figure}

\clearpage

%
%

\begin{figure}
\vspace{-0.5truein}
\figurenum{3}
\centerline{\includegraphics[width=15.0cm]{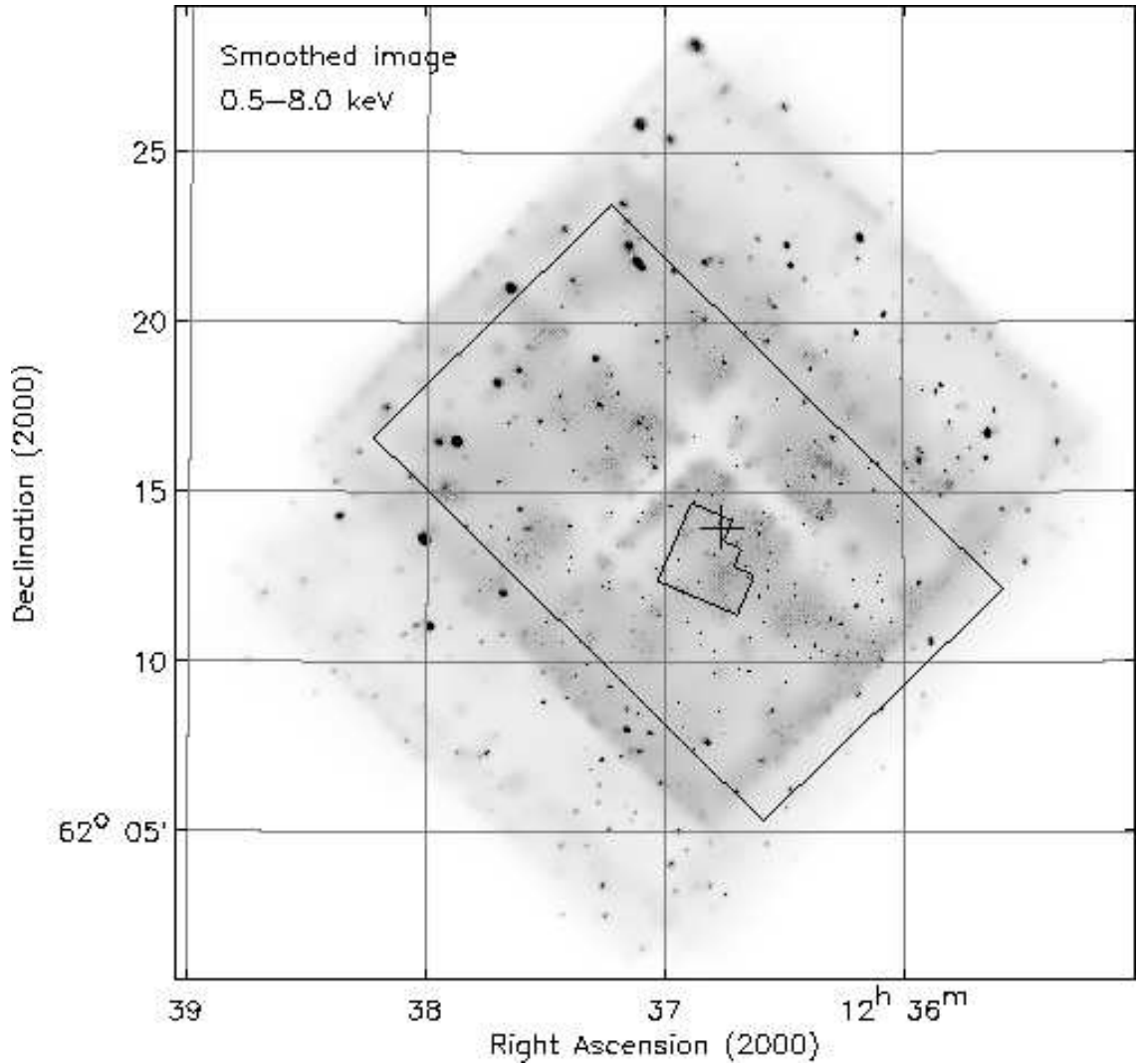}}
\vspace{-0.0truein}

\caption{Full-band adaptively smoothed image of the 2~Ms CDF-N. This 
image has been made using the standard \asca\ grade set (see Table~2)
and is binned by a factor of four in both RA and
Dec. The adaptive smoothing has been performed using the code
of Ebeling, White, \& Rangarajan (2003) at the $2.5\sigma$ level, and
the grayscales are linear; much of the apparent diffuse emission is
just instrumental background. The symbols and regions have the same
meaning as those given in Figure~2.
Only one of the adaptively smoothed images is included here;
please see the World Wide Web site listed in Footnote~8 for adaptively
smoothed images in the other bands listed in Table~2.}

\end{figure}

\clearpage

%
%

\begin{figure}
\figurenum{4}
\centerline{\includegraphics[width=15.0cm]{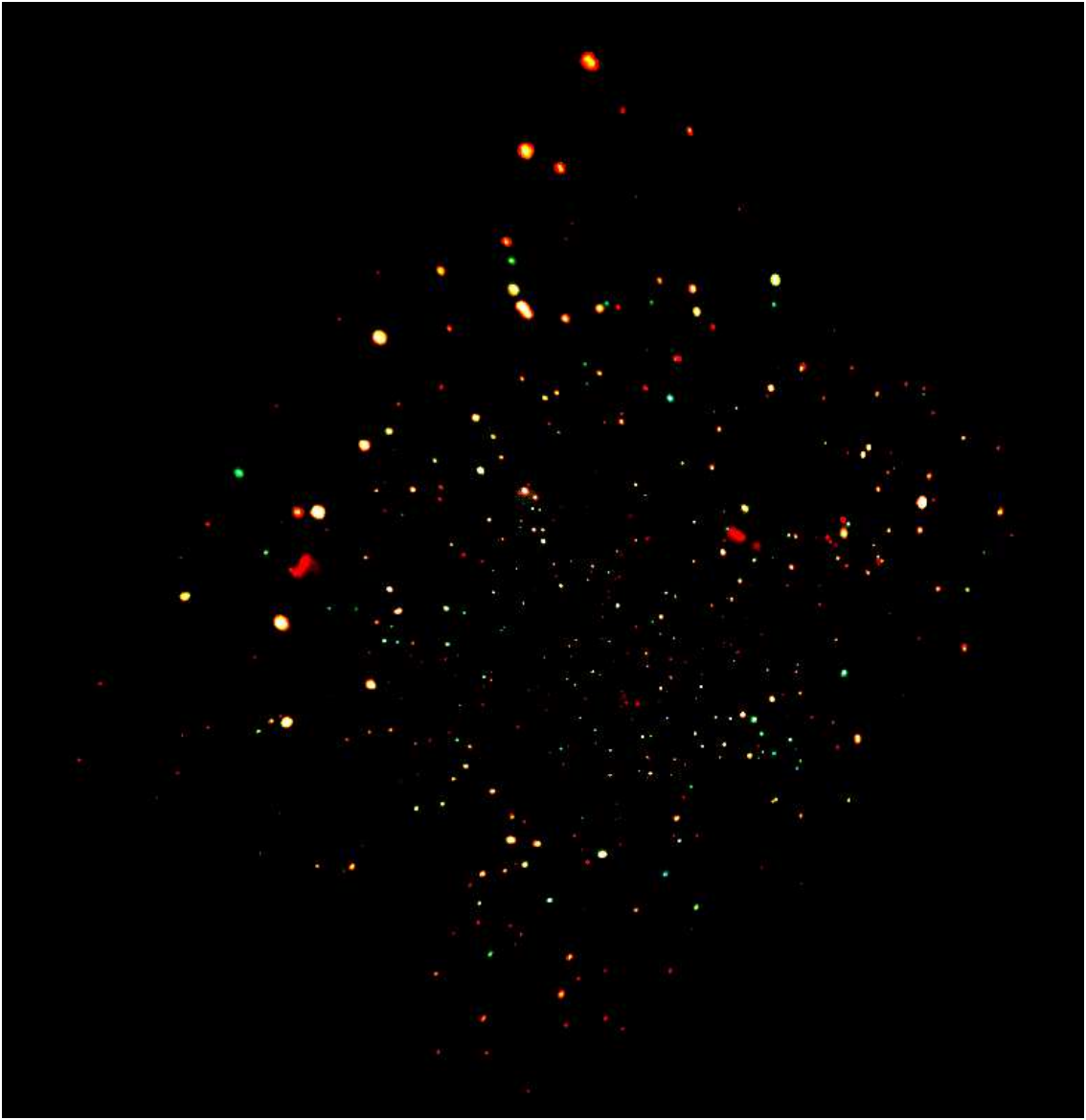}}
\vspace{0.2truein}
\caption{\chandra\ ``true-color'' image of the 2~Ms CDF-N. 
This image has been constructed from the 0.5--2.0~keV (red), 2--4~keV
(green), and 4--8~keV (blue) adaptively smoothed images available at
the World Wide Web site listed in Footnote~8. The two most prominent
red diffuse patches are galaxy groups/clusters (see Bauer
et~al. 2002a).}
\end{figure}

%
%

\begin{figure}
\vspace{-0.5truein}
\figurenum{5}
\centerline{\includegraphics[width=15.0cm]{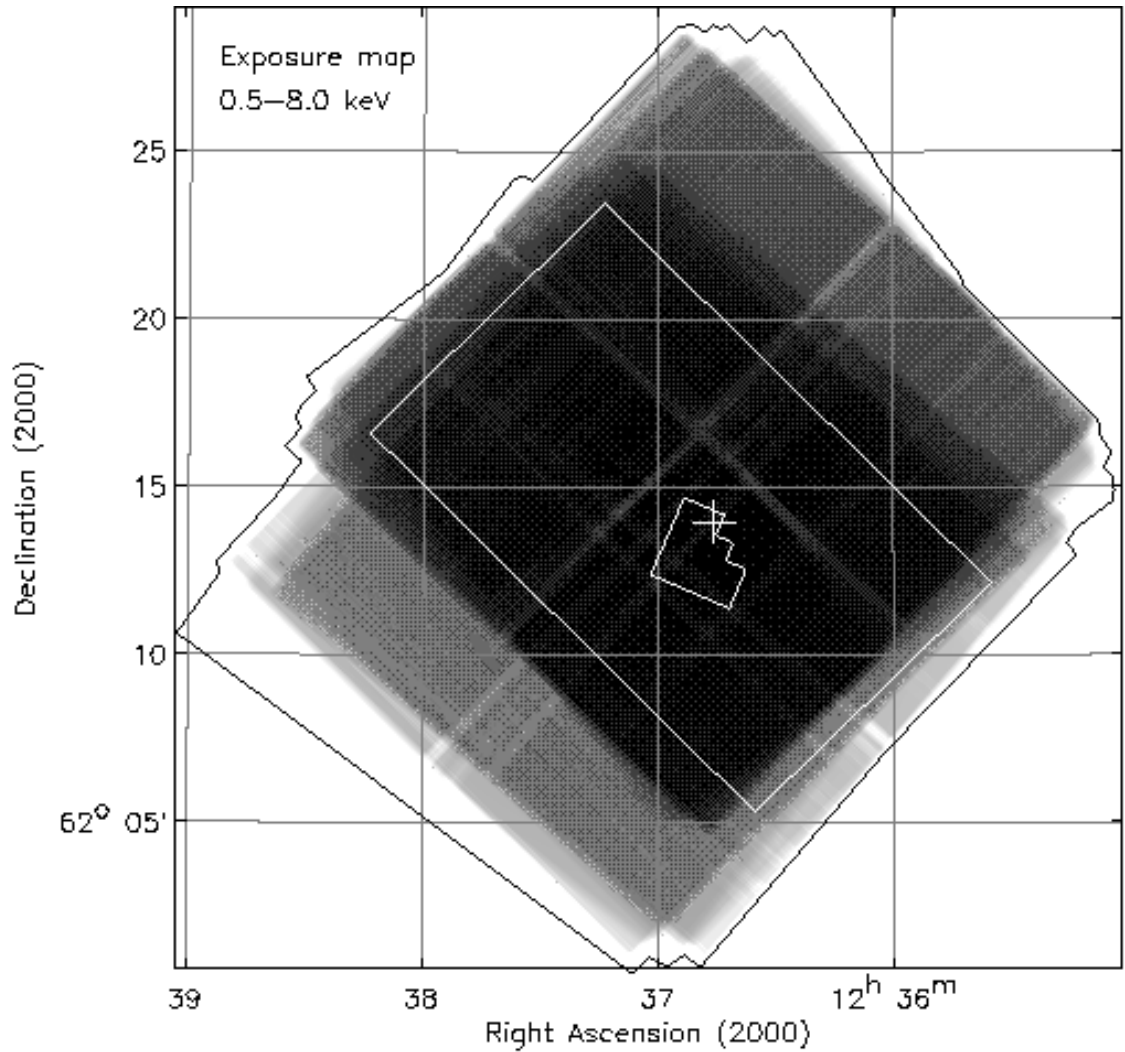}}
\vspace{-0.0truein}
\caption{Full-band exposure map of the 2~Ms CDF-N. This image has 
been binned by a factor of four in both RA and
Dec. The darkest areas correspond to the highest effective
exposure times (the maximum value is 1.945~Ms), and the grayscales are
logarithmic. The symbols and regions have the same meaning as those
given in Figure~2. The black outline surrounding the exposure map
indicates the extent of all the ACIS-I observations; the regions of
the exposure map near the outline appear white due to low exposure
($<$200~ks).
Only one of the exposure maps is included here; please see the World
Wide Web site listed in Footnote~8 for exposure maps in the other
bands listed in Table~2.}
\end{figure}

\clearpage

%
%

\begin{figure}
\vspace{-0.5truein}
\epsscale{0.9}
\figurenum{6}
\centerline{\includegraphics[width=12.0cm]{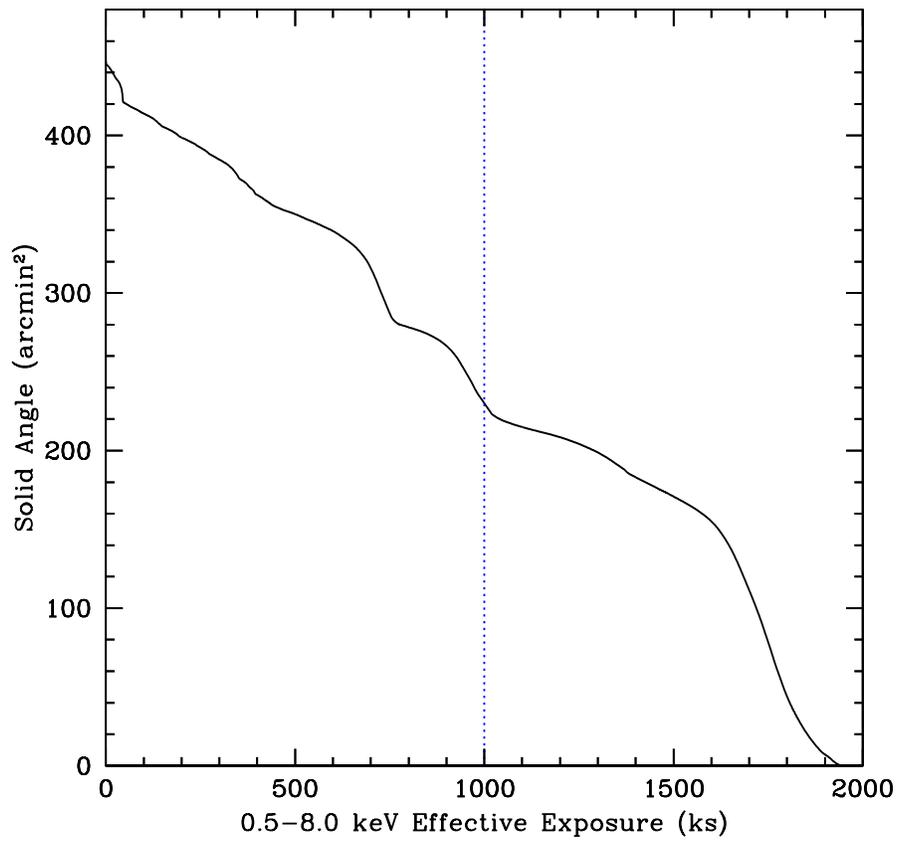}}
\vspace{-0.0truein}
\caption{Amount of survey solid angle having at least a given amount of 
effective exposure in the full-band exposure map. The vertical dotted
line indicates an effective exposure of $1$~Ms; $\approx$~51\%
($\approx$~230 arcmin$^2$) of the 2~Ms CDF-N field has $>1$~Ms of
exposure. Compare with Figure~5.}
\end{figure}

\clearpage

%
%

\begin{figure}
\epsscale{0.9}
\figurenum{7}
\centerline{\includegraphics[width=12.0cm]{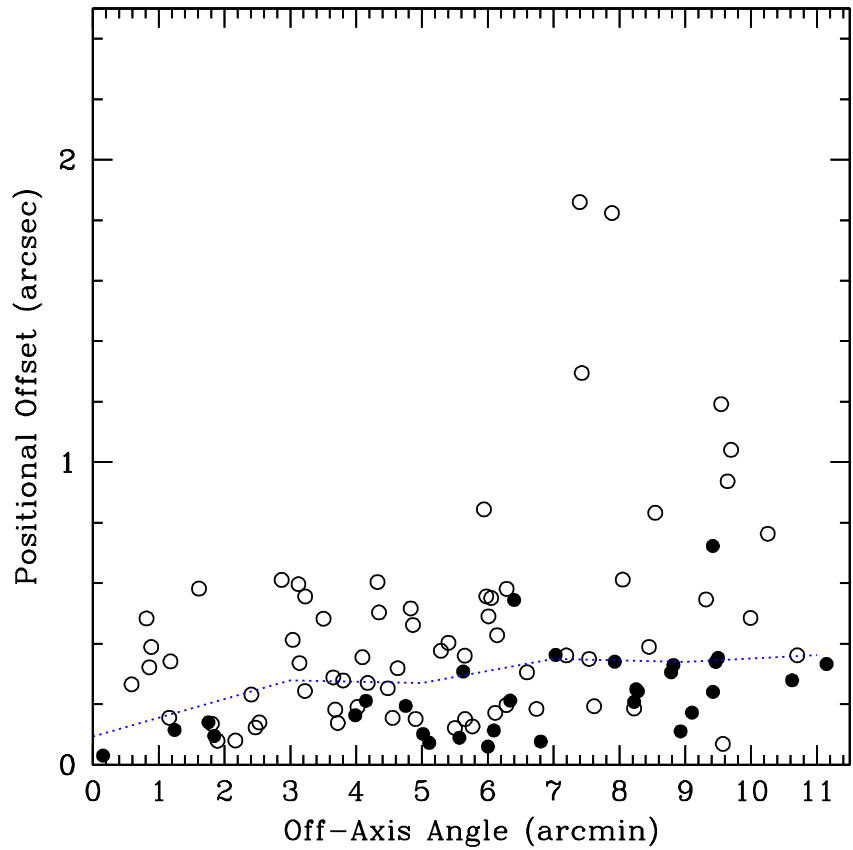}}
\caption{Positional offset versus off-axis angle for sources in the main 
\chandra\ catalog that match with 1.4~GHz sources (from Richards 2000) to within
$2\farcs 5$. The open circles are \chandra\ sources with $<$200
full-band counts, and the solid dots are \chandra\ sources with
$\ge200$ full-band counts. The dotted line shows the running
median. These data have been used to determine the positional
uncertainties of the \chandra\ sources; see \S3.3 and \S3.4.1.}
\end{figure}

\clearpage

%
%

\begin{figure}
\epsscale{0.9}
\figurenum{8}
\centerline{\includegraphics[width=12.0cm]{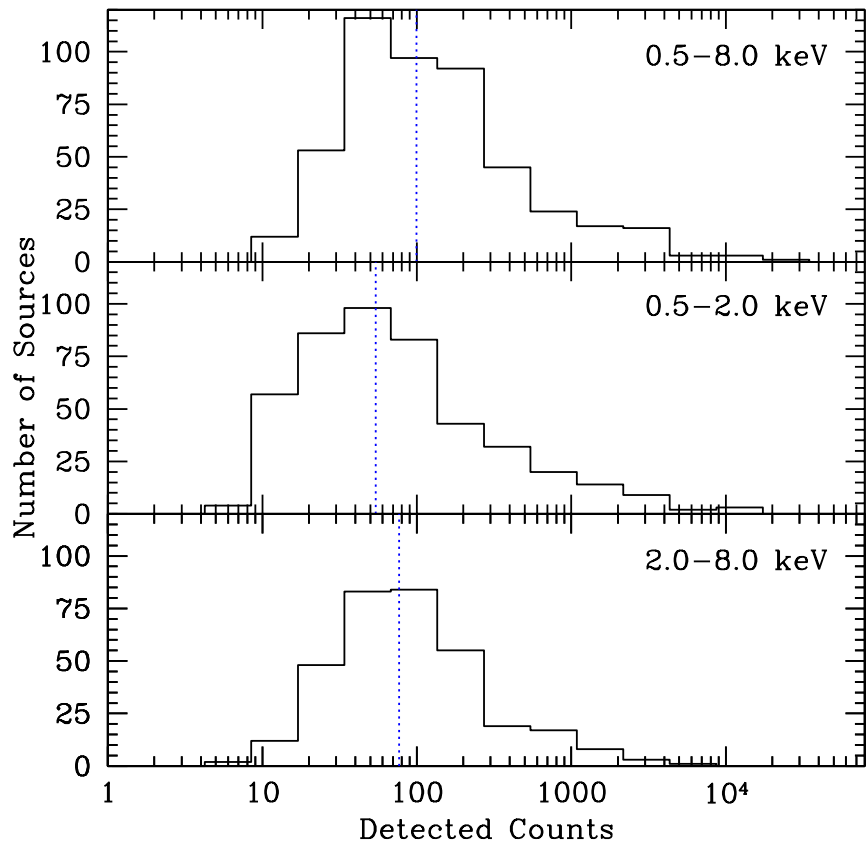}}
\caption{Histograms showing the distributions of detected counts in the full (top), soft (middle), and hard (bottom) bands. Sources with upper limits have not been plotted. The dotted lines indicate the median numbers of counts (see Table~4); the median number of full-band counts ($\approx$~100) is large enough for basic X-ray spectral analysis.}
\end{figure}

\clearpage

%
%

\begin{figure}
\epsscale{0.9}
\figurenum{9}
\centerline{\includegraphics[width=12.0cm]{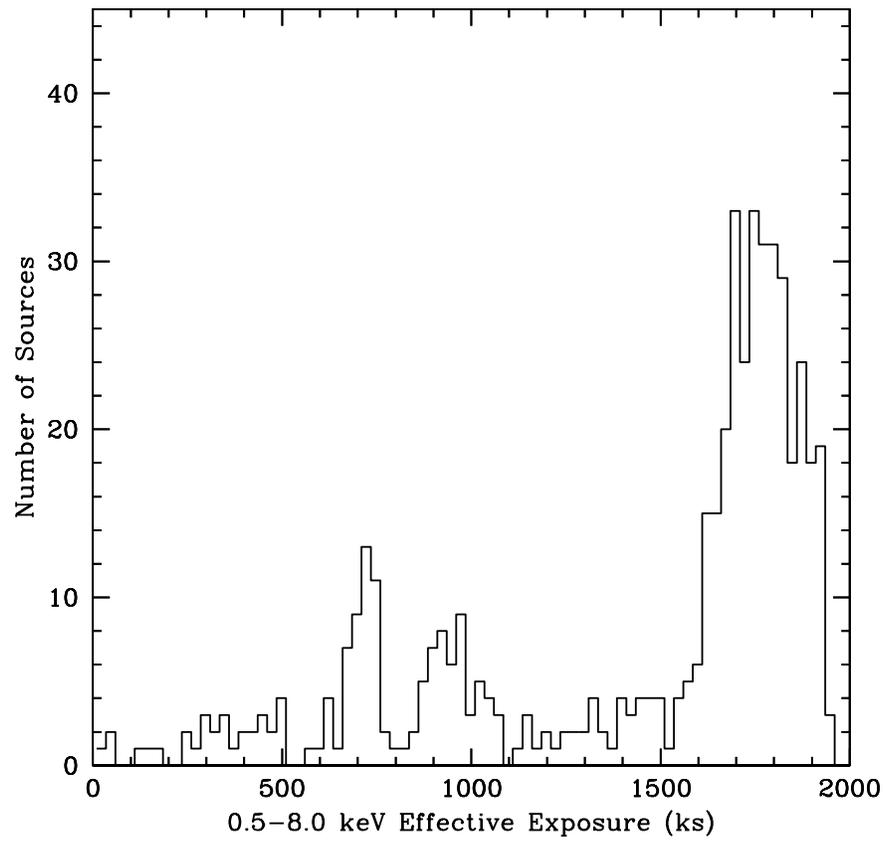}}
\caption{Histogram showing the distribution of full-band effective exposure time 
for the sources in the main \chandra\ catalog. A broad range of
exposure times is found due to the different roll angles and aim
points of the 20 observations. The median exposure time is 1.7~Ms.}
\end{figure}

\clearpage

%
%

\begin{figure}
\epsscale{0.9}
\figurenum{10}
\centerline{\includegraphics[width=12.0cm]{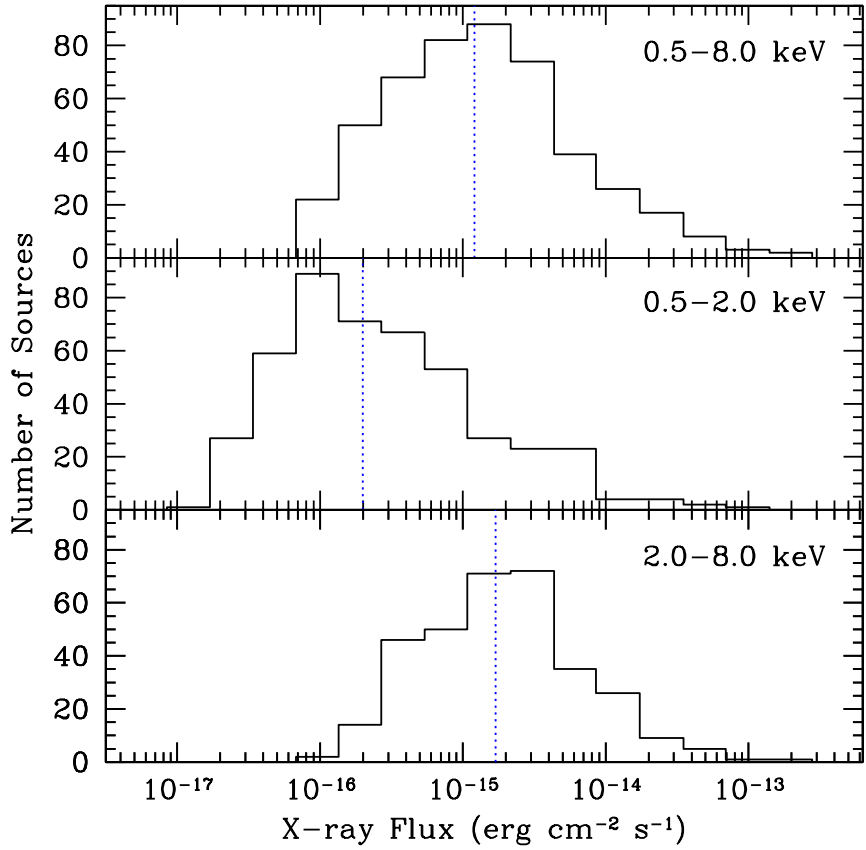}}
\caption{Histograms showing the distributions of X-ray fluxes in the full (top), soft (middle), and hard (bottom) bands. Sources with upper limits have not been plotted. The dotted lines indicate the median fluxes. Approximately 50\% of the sources have soft-band and hard-band fluxes of $< 2\times 10^{-16}$~erg~cm$^{-2}$~s$^{-1}$ and $< 2\times 10^{-15}$~erg~cm$^{-2}$~s$^{-1}$, respectively.}
\end{figure}

\clearpage

%
%

\begin{figure}
\figurenum{11}
\centerline{\includegraphics[width=12.0cm]{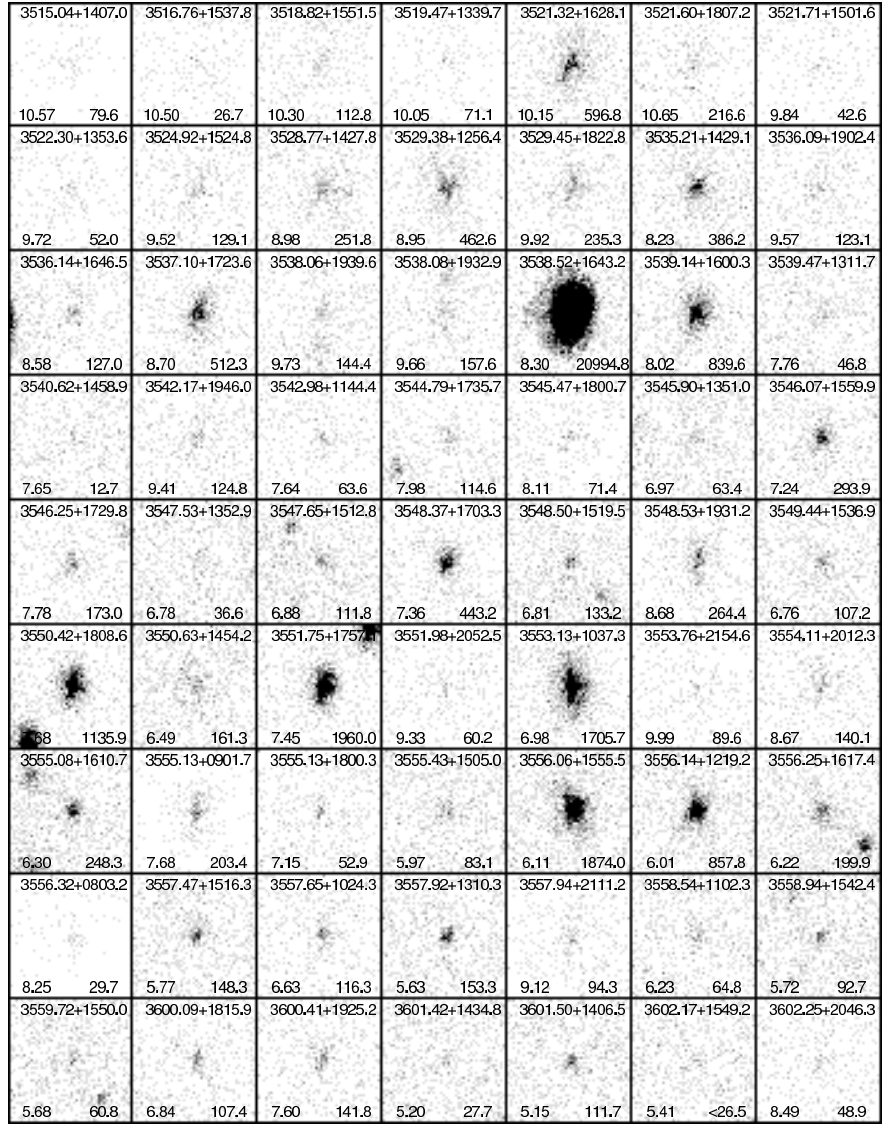}}
\caption{Full-band ``postage-stamp'' images for the sources in the 
main \chandra\ catalog. The label at the top of each image gives the
source coordinates (the hour and degree components are not included)
while the numbers at the bottom left and bottom right corners give the
off-axis angle (in arcminutes) and number of full-band counts,
respectively. Each image is oriented so that North is up and East is
to the left, and each is 50 pixels ($\approx 24\farcs 6$) on a
side. The source of interest is always at the center of the image. The
background varies significantly from image to image due to the varying
effective exposure time (see Figure~5 and Table~8). In a few cases no
source is apparent; often these sources were not detected in the full
band but sometimes the lack of an apparent source is due to the broad
PSF at large off-axis angles. A few of the sources appear to show
extent; in most cases this is an artifact of the complex \chandra\
off-axis PSF. Only one of the eight pages of cutouts could be included
here; please see the World Wide Web site listed in Footnote~8 for the
version with all the cutouts.}
\end{figure}

\clearpage

%
%

\begin{figure}
\figurenum{12a}
\centerline{\includegraphics[width=15.0cm]{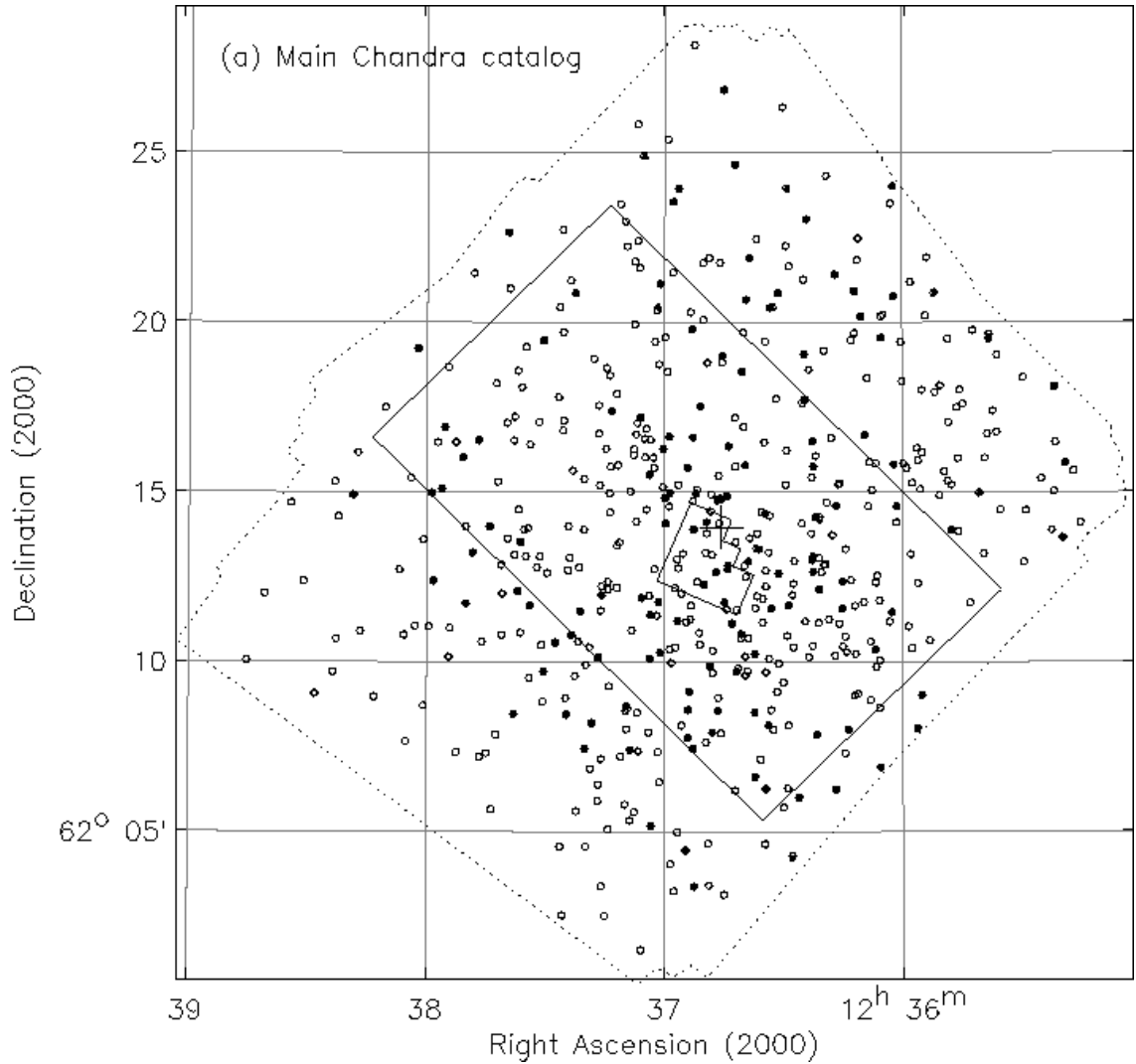}}
\caption{Positions of the sources in (a) the main \chandra\ 
catalog, and (b) the supplementary optically bright \chandra\ catalog.
The symbols and regions have the same meaning as those given in
Figure~2; the dotted outline indicates the extent of the 2~Ms CDF-N
field. The new sources in the main \chandra\ catalog 
(i.e.,\ those not detected in the 1~Ms exposure presented in B01) 
are indicated with solid dots.}
\end{figure}

\begin{figure}
\figurenum{12b}
\centerline{\includegraphics[width=15.0cm]{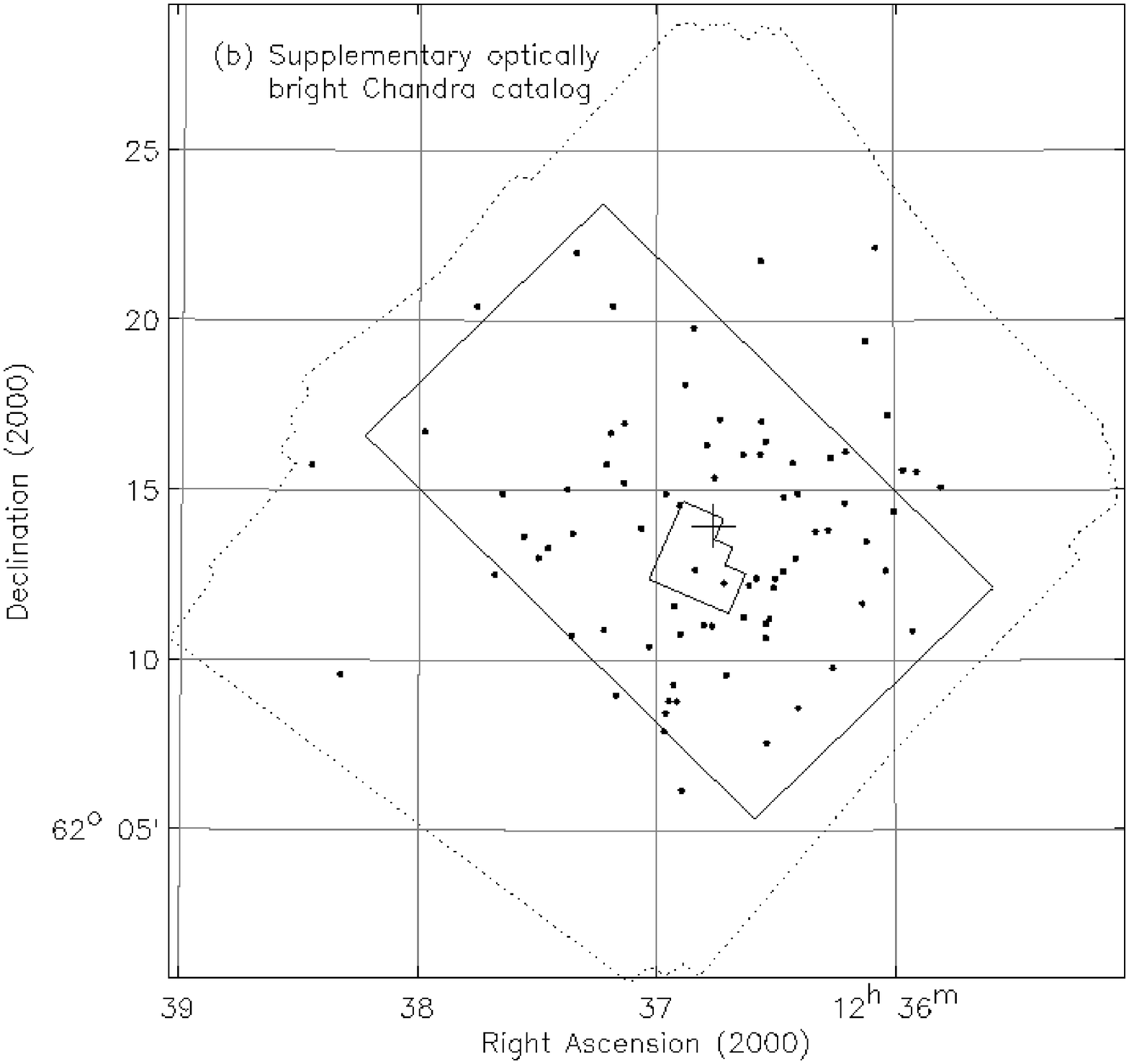}}
\caption{}
\end{figure}

\clearpage

%
%

\begin{figure}
\epsscale{0.9}
\figurenum{13}
\centerline{\includegraphics[width=12.0cm]{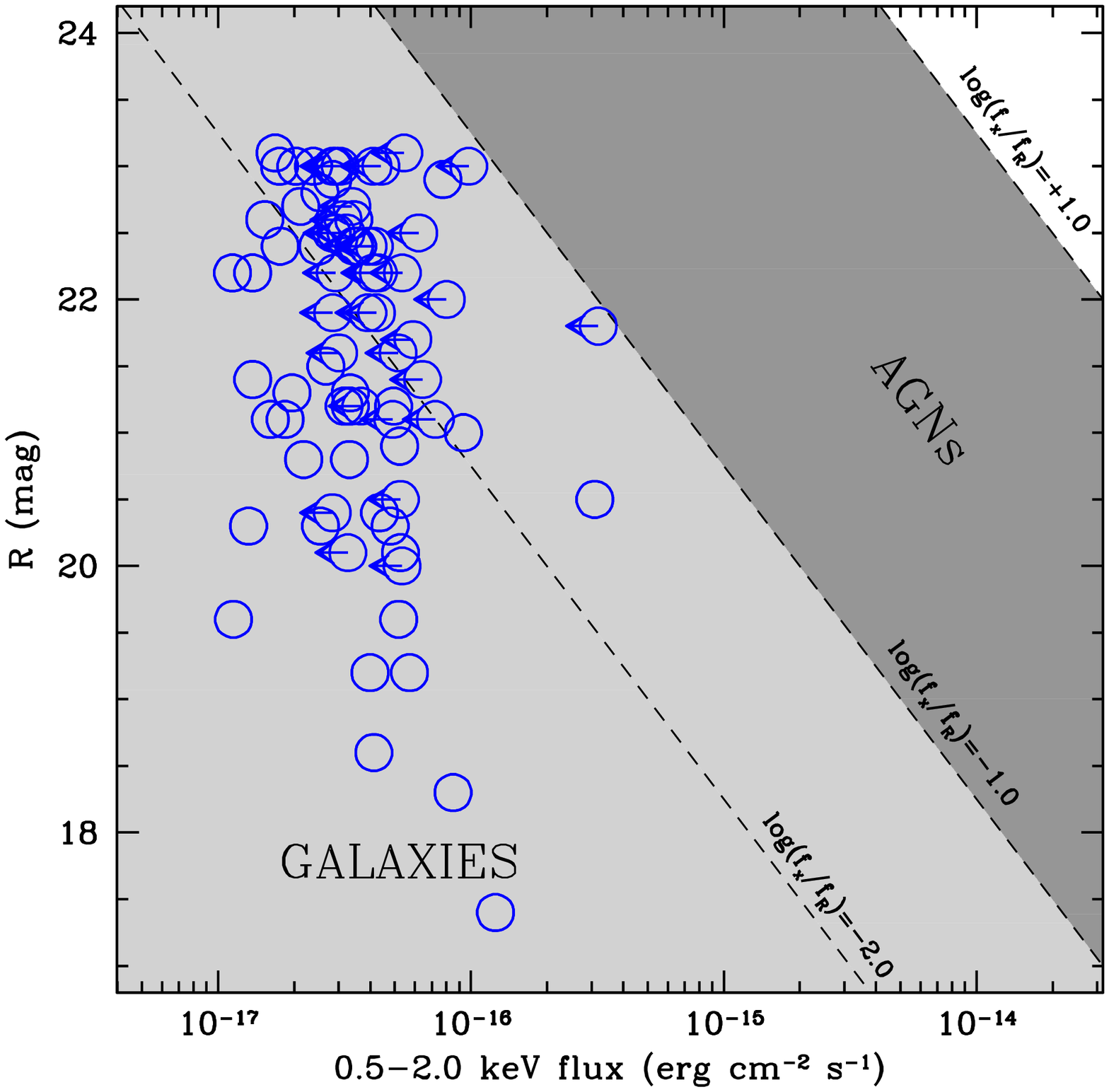}}
\caption{$R$-band magnitude versus soft-band flux for the 79 optically bright lower significance X-ray sources. The diagonal lines indicate constant flux ratios. The shaded regions show the approximate flux ratios for AGNs and galaxies (e.g.,\ Maccaccaro \etal 1988; Stocke \etal 1991; Hornschemeier \etal 2001).}
\end{figure}

\clearpage

%
%

\begin{figure}
\epsscale{0.90}
\figurenum{14}
\centerline{\includegraphics[width=17.0cm]{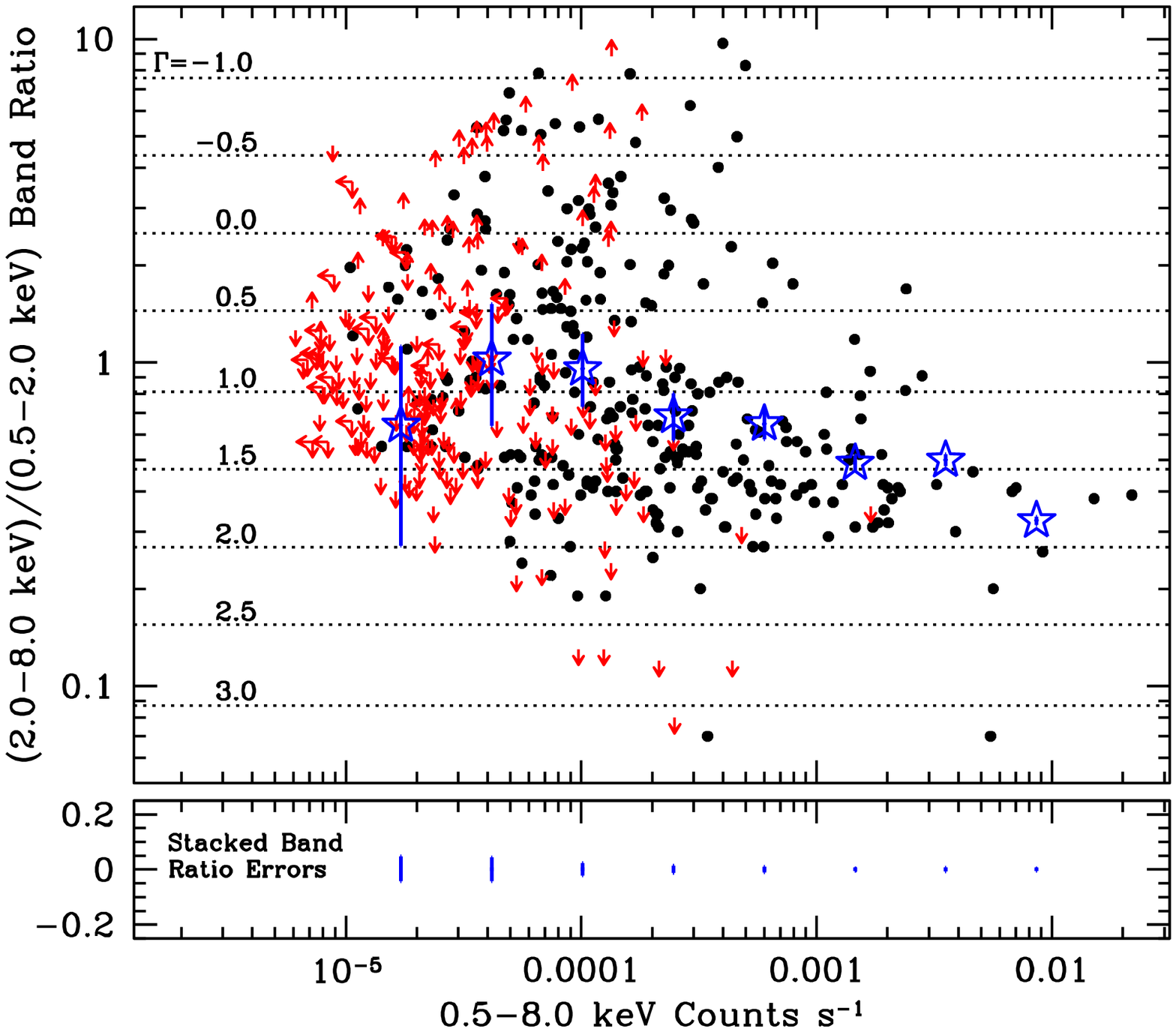}}
\vspace{-0.4truein}
\caption{Band ratio as a function of full-band count rate for the sources
in the main \chandra\ catalog.
Small solid dots show sources detected in both the soft and hard
bands. Plain arrows show sources detected in only one of these two
bands with the arrows indicating upper and lower limits; sources
detected in only the full band cannot be plotted. The open stars show
average band ratios as a function of full-band count rate derived from
stacking analyses (following \S3.3 of Alexander
\etal 2001). The vertical lines represent the average $1\sigma$ uncertainties on the
band ratio for a typical source at the shown count rate. These error
bars are not representative of the $1\sigma$ uncertainties on the
stacked band ratio, which are typically only $<0.04$ (see bottom
panel); note that the $y$-axis scaling in the bottom panel is
different to the $y$-axis scaling in the top panel.
Horizontal dotted lines are labeled with the photon indices that
correspond to a given band ratio assuming only Galactic absorption
(these were determined using the CXC's Portable, Interactive,
Multi-Mission Simulator; PIMMS).}
\end{figure}

\clearpage

%
%

\begin{figure}
\epsscale{0.90}
\figurenum{15a}
\centerline{\includegraphics[width=12.0cm]{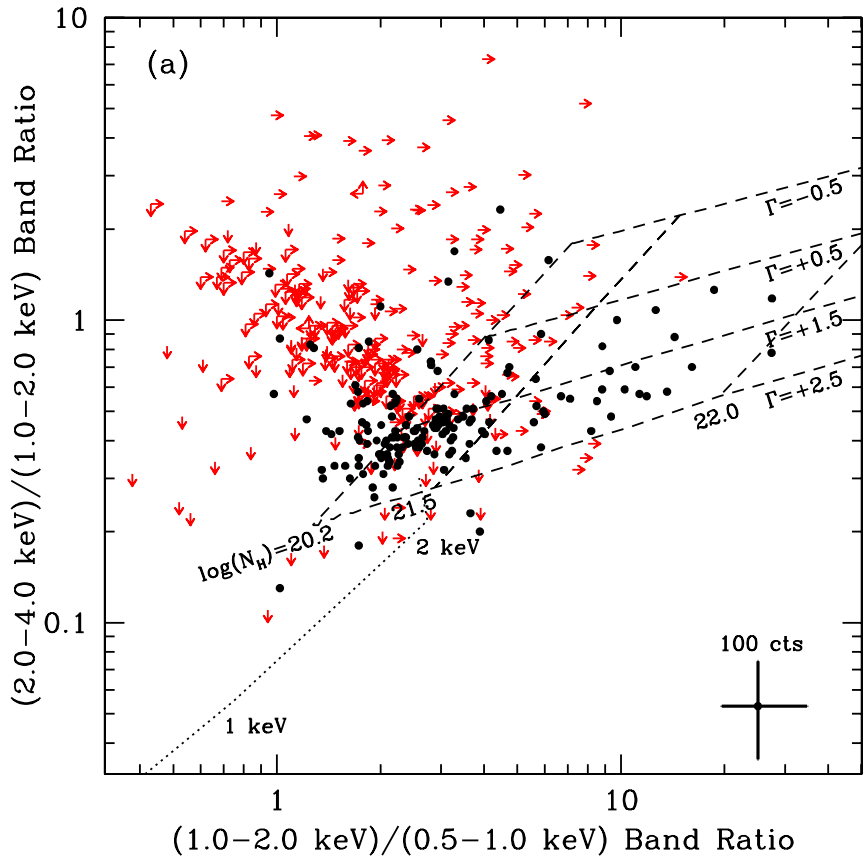}}
\caption{X-ray color-color diagrams for the sources in the main \chandra\ 
catalog showing (a) HB1/SB2 versus SB2/SB1 and (b) HB2/HB1 versus
HB1/SB2. Small solid dots show sources detected in all the plotted
bands, and plain arrows show sources with band ratio limits. To reduce
symbol crowding, we do not show error bars for each of the small solid
dots; instead the errors for a typical source with $\approx$~100
full-band counts is shown. The dashed lines show the expected band
ratios for a variety of absorbed power-law models, and the dotted
lines show the expected band ratios for a Raymond-Smith model with
different input temperatures (these were determined using PIMMS and
were calculated including the Galactic absorption). All of the models
have been calculated for $z=0$; higher redshifts for the Raymond-Smith
model will cause the tracks to move toward lower SB2/SB1 and HB1/SB2
band ratios, while higher redshifts for the absorbed power-law models
will lead to higher intrinsic absorption for a given set of band
ratios [for $N_{\rm H,z}\simlt 1.5\times10^{24}$ cm$^2$, $N_{\rm H,z}$
is related to the column density at $z=0$ by $N_{\rm
H,z}\approx(1+z)^{2.6}~N_{\rm H}$].}
\end{figure}

\begin{figure}
\epsscale{0.90}
\figurenum{15b}
\centerline{\includegraphics[width=12.0cm]{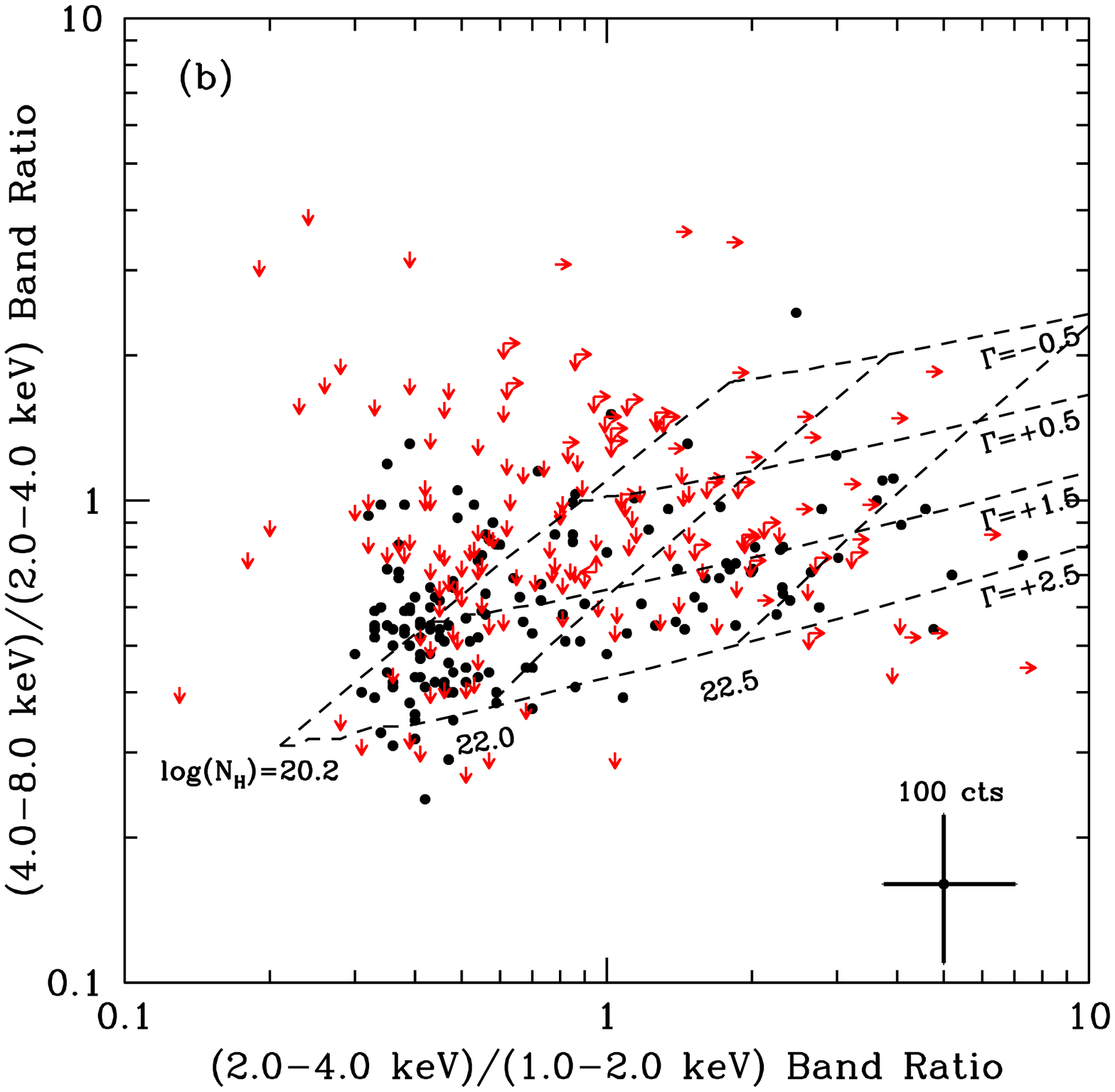}}
\caption{}
\end{figure}

\clearpage

%
%

\begin{figure}
\vspace{-0.5truein}
\epsscale{0.9}
\figurenum{16}
\centerline{\includegraphics[width=12.0cm]{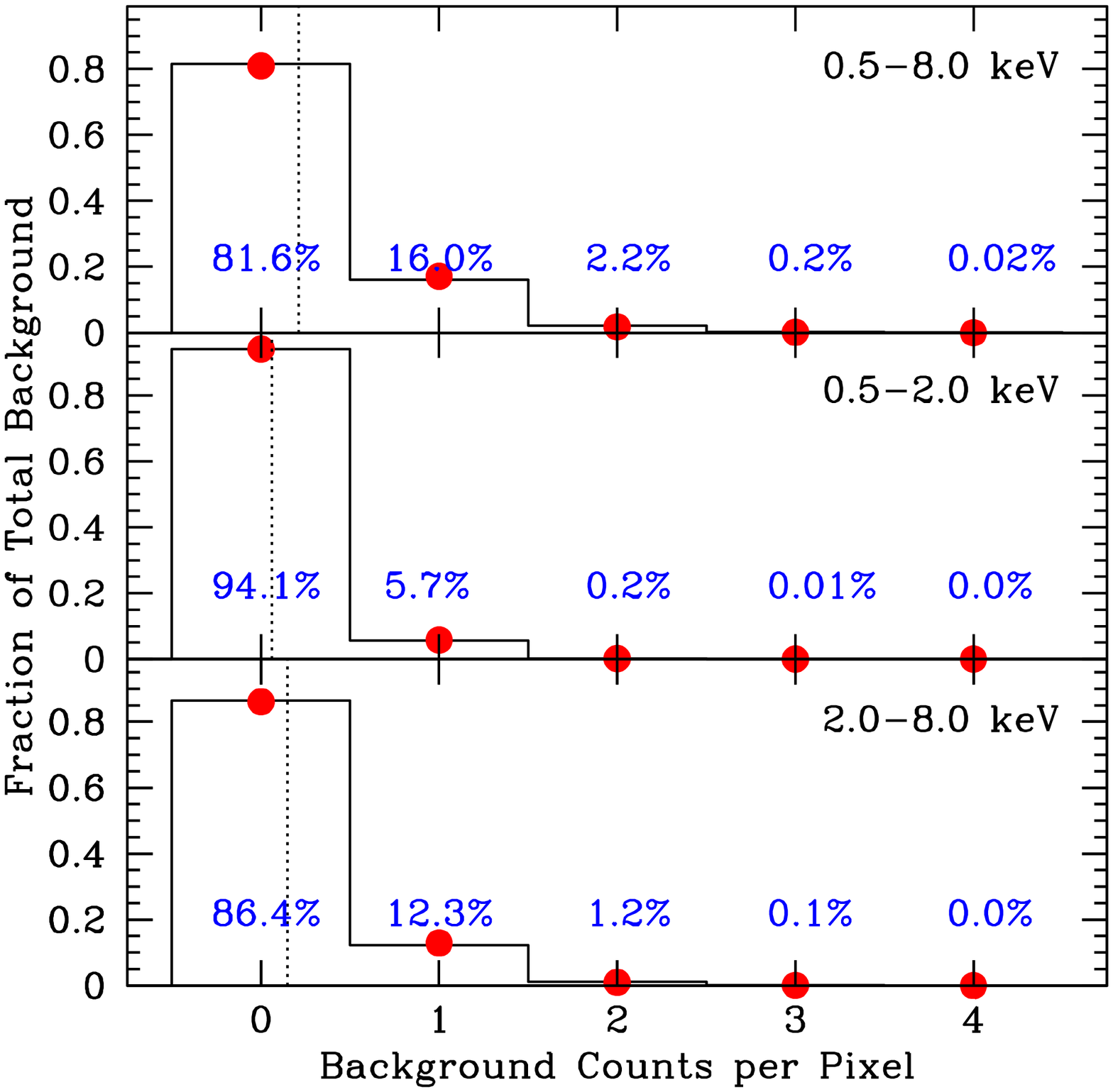}}
\vspace{-0.0truein}
\caption{Histograms showing the fractions of background counts per pixel in the full (top), soft (middle), and hard (bottom) bands. The vertical dotted lines indicate the mean numbers of background counts per pixel (see Table~8), and the numbers plotted show the fractions of the total numbers of background counts present in each bin. The solid dots indicate the expected numbers of background counts for Poisson distributions with the mean numbers of background counts per pixel. The background count distributions are very similar to Poisson distributions for all seven bands; the probability as derived from the Kolmogorov-Smirnov test is $>$99.99\%. We also performed these analyses on many different small ($\approx$~4~arcmin$^2$) regions of similar exposure across the field. In all cases the Kolmogorov-Smirnov test probabilites were $>$99.99\% in all of the seven bands (see \S4.2).}
\end{figure}

\clearpage

%
%

\begin{figure}
\vspace{-0.5truein}
\figurenum{17}
\centerline{\includegraphics[width=15.0cm]{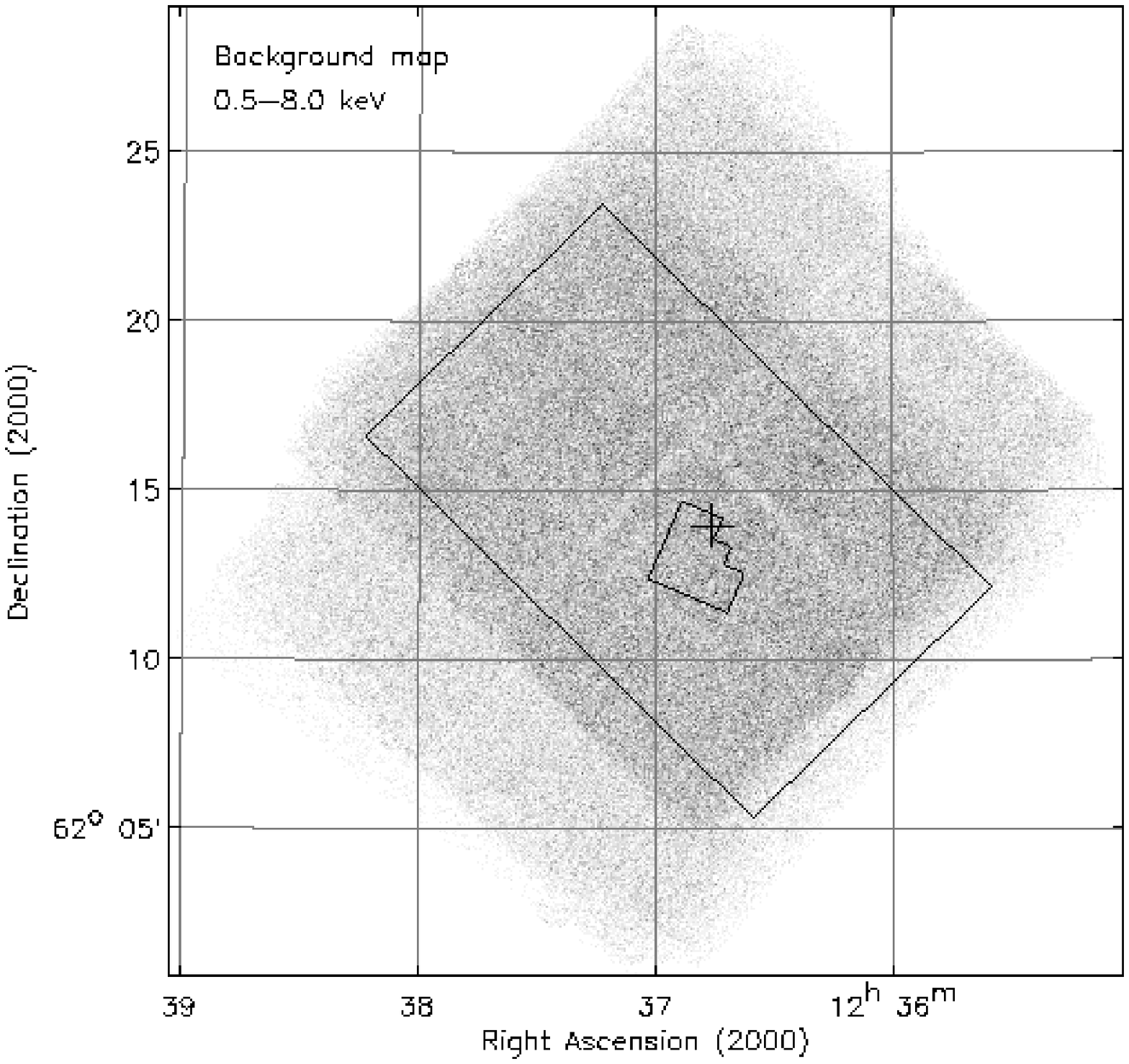}}
\vspace{-0.0truein}
\caption{Full-band background map of the 2~Ms CDF-N. This background map has been created
following \S4.2 using the standard \asca\ grade set (see Table~2) and
is binned by a factor of four in both RA and
Dec. The symbols and regions have the same meaning as those
given in Figure~2. The higher background around the GOODS IRAC region is
due to the higher effective exposure.  Diffuse emission from the
extended source \hbox{CXOHDFN J123620.0+621554} (Bauer \etal 2002a) is
faintly visible but is unlikely to increase the background
dramatically.}

\end{figure}

\clearpage

%
%

\begin{figure}
\vspace{-0.5truein}
\figurenum{18}
\centerline{\includegraphics[width=15.0cm]{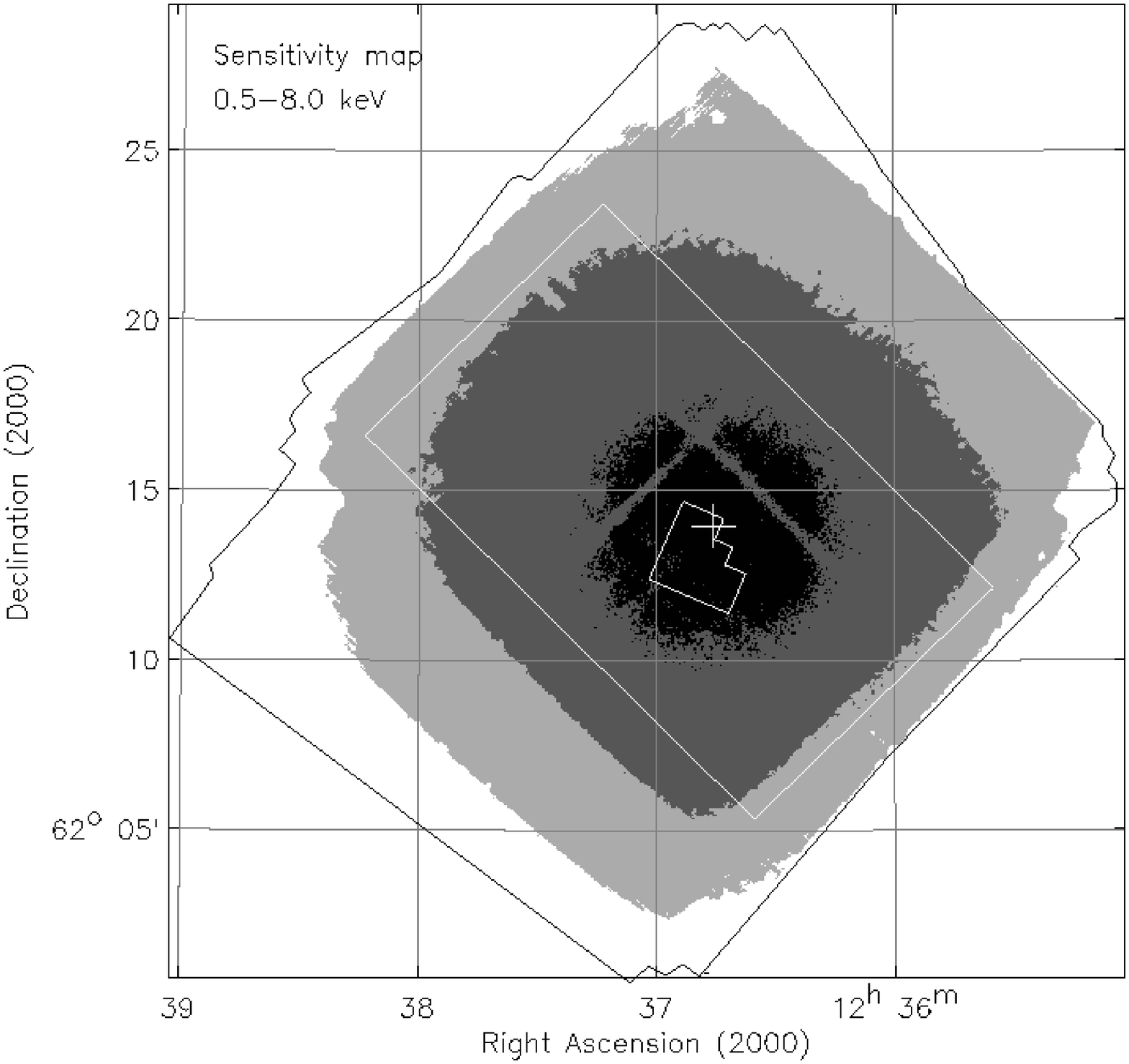}}
\vspace{-0.0truein}

\caption{Full-band S/N$=3$ sensitivity map of the 2~Ms CDF-N. This 
sensitivity map has been created following \S4.2 using the standard
\asca\ grade set (see Table~2) and is binned by a factor of four in
both RA and Dec. The symbols and regions have the
same meaning as those given in Figure~2; the solid outline indicates
the extent of the CDF-N field. The black, dark gray, light gray, and
white areas correspond to S/N$=3$ sensitivities of
$<10^{-16}$~erg~cm$^{-2}$~s$^{-1}$,
$10^{-16}$--$3.3\times10^{-16}$~erg~cm$^{-2}$~s$^{-1}$,
$3.3\times10^{-16}$--$10^{-15}$~erg~cm$^{-2}$~s$^{-1}$, and
$>10^{-15}$~erg~cm$^{-2}$~s$^{-1}$, respectively.}

\end{figure}

\clearpage

%
%

\begin{figure}
\vspace{-0.5truein}
\epsscale{0.9}
\figurenum{19}
\centerline{\includegraphics[width=12.0cm]{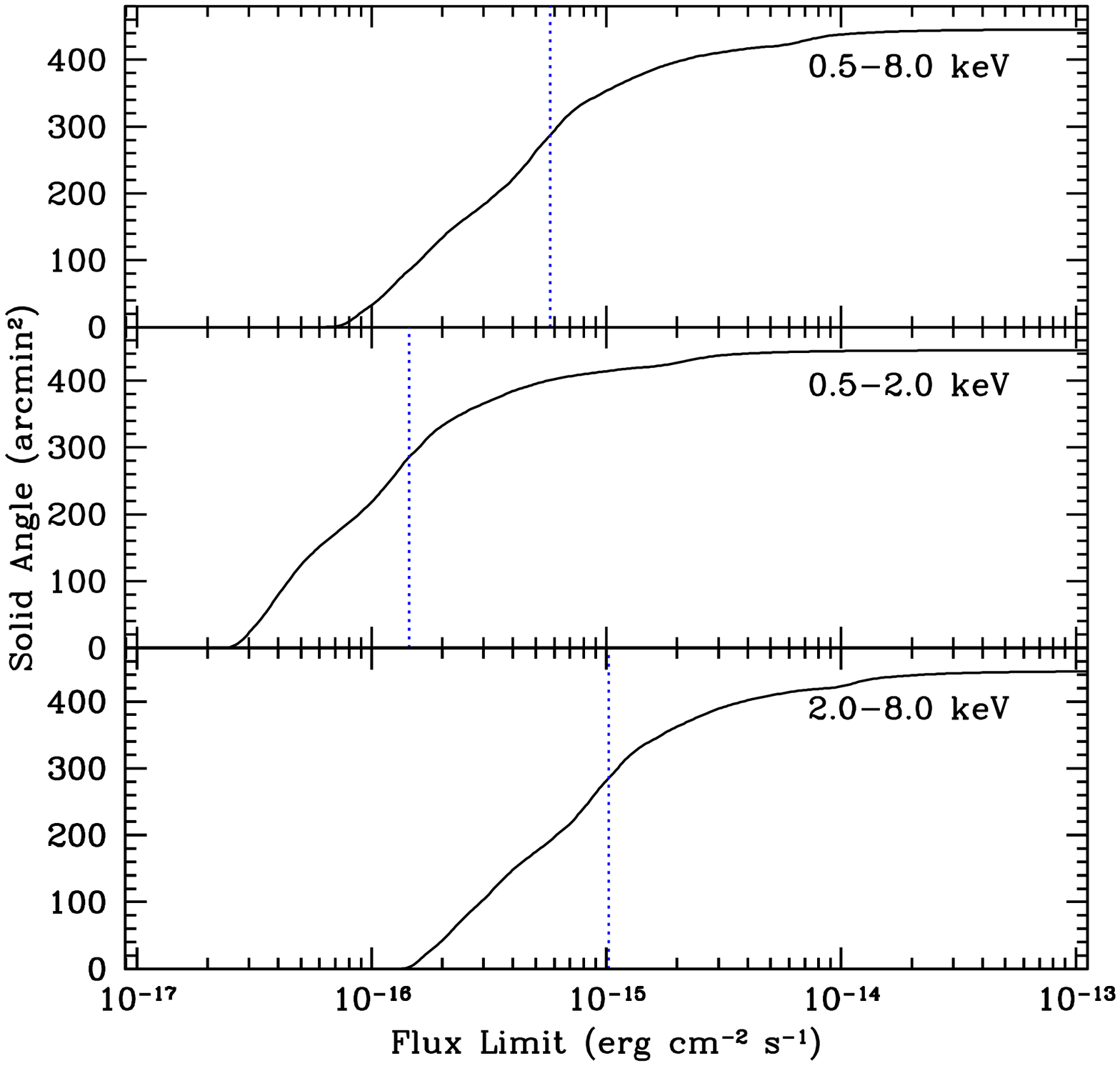}}
\vspace{-0.0truein}
\caption{Solid angle versus flux limit (S/N$=3$) for the full
(top), soft (middle), and hard (bottom) bands, determined following
\S4.2.  The flux limits at the aim point are $\approx
7.1\times10^{-17}$~erg~cm$^{-2}$~s$^{-1}$ (full band), $\approx
2.5\times10^{-17}$~erg~cm$^{-2}$~s$^{-1}$ (soft band), and $\approx
1.4\times10^{-16}$~erg~cm$^{-2}$~s$^{-1}$ (hard band). The vertical
dotted lines indicate the flux limits for an area of $\approx$~285
arcmin$^2$, equivalent to a single ACIS-I field of view; these
correspond to limiting fluxes of
$\approx 5.8\times10^{-16}$~erg~cm$^{-2}$~s$^{-1}$ (full band),
$\approx 1.4\times10^{-16}$~erg~cm$^{-2}$~s$^{-1}$ (soft band), and
$\approx 1.0\times10^{-15}$~erg~cm$^{-2}$~s$^{-1}$ (hard band).}
\end{figure}

\clearpage

%
%

\begin{figure}
\vspace{-0.5truein}
\figurenum{20}
\centerline{\includegraphics[width=15.0cm]{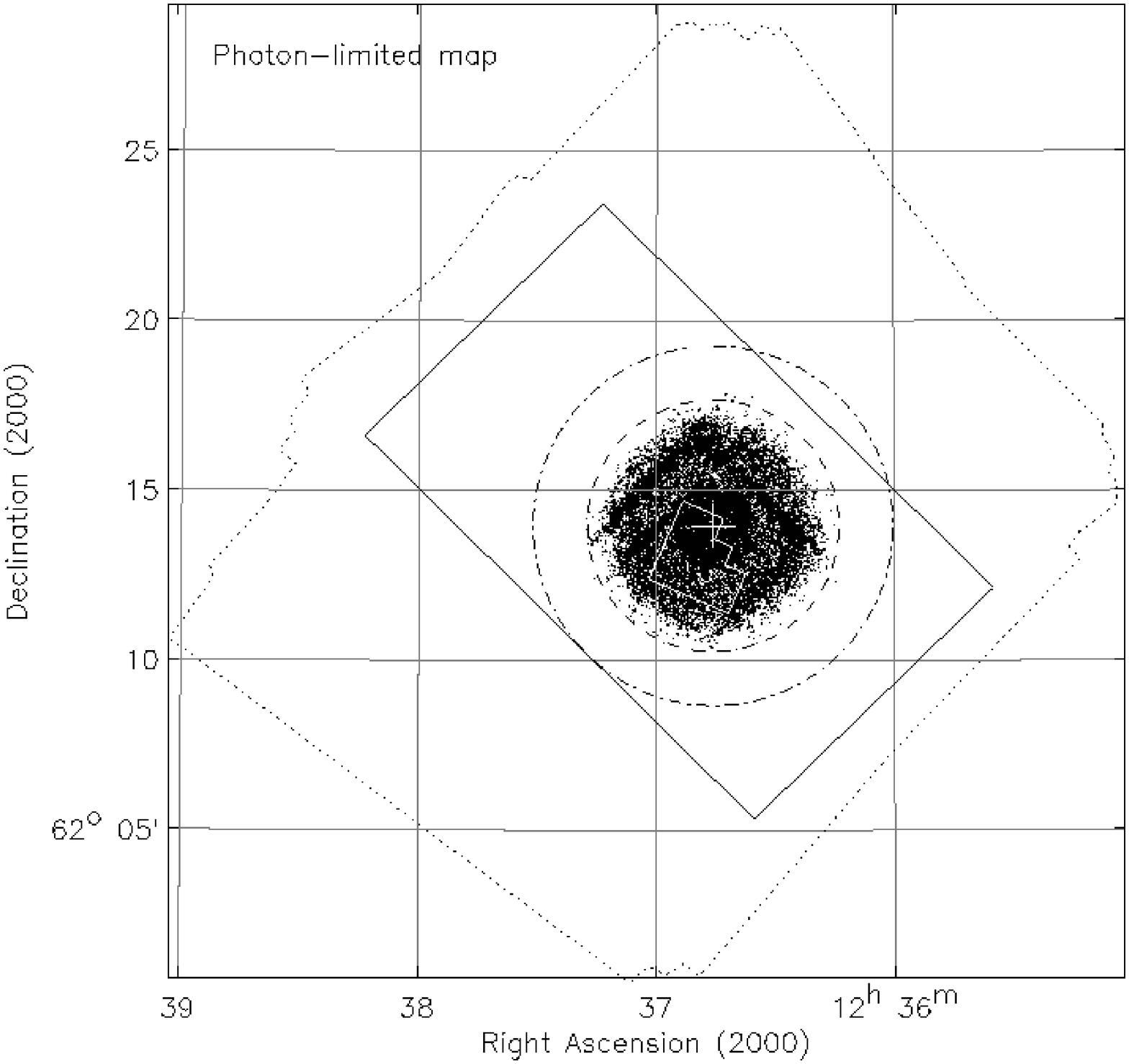}}
\vspace{-0.0truein}
\caption{Map showing the photon-limited regions in the full, soft, and hard bands.
These regions have been calculated following \S4.3 using the standard
\asca\ grade set (see Table~2). The symbols and regions have the
same meaning as those given in Figure~2. The dotted outline indicates
the extent of the 2~Ms CDF-N field, the black region corresponds to the
photon-limited area in the full band, the dot-dashed circle indicates the
approximate extent of the soft-band photon-limited region, and the
dashed circle indicates the approximate extent of the hard-band
photon-limited region (see Table~9).}

\end{figure}

\clearpage

%
%

\begin{figure}
\vspace{-0.5truein}
\epsscale{0.9}
\figurenum{21}
\centerline{\includegraphics[width=12.0cm]{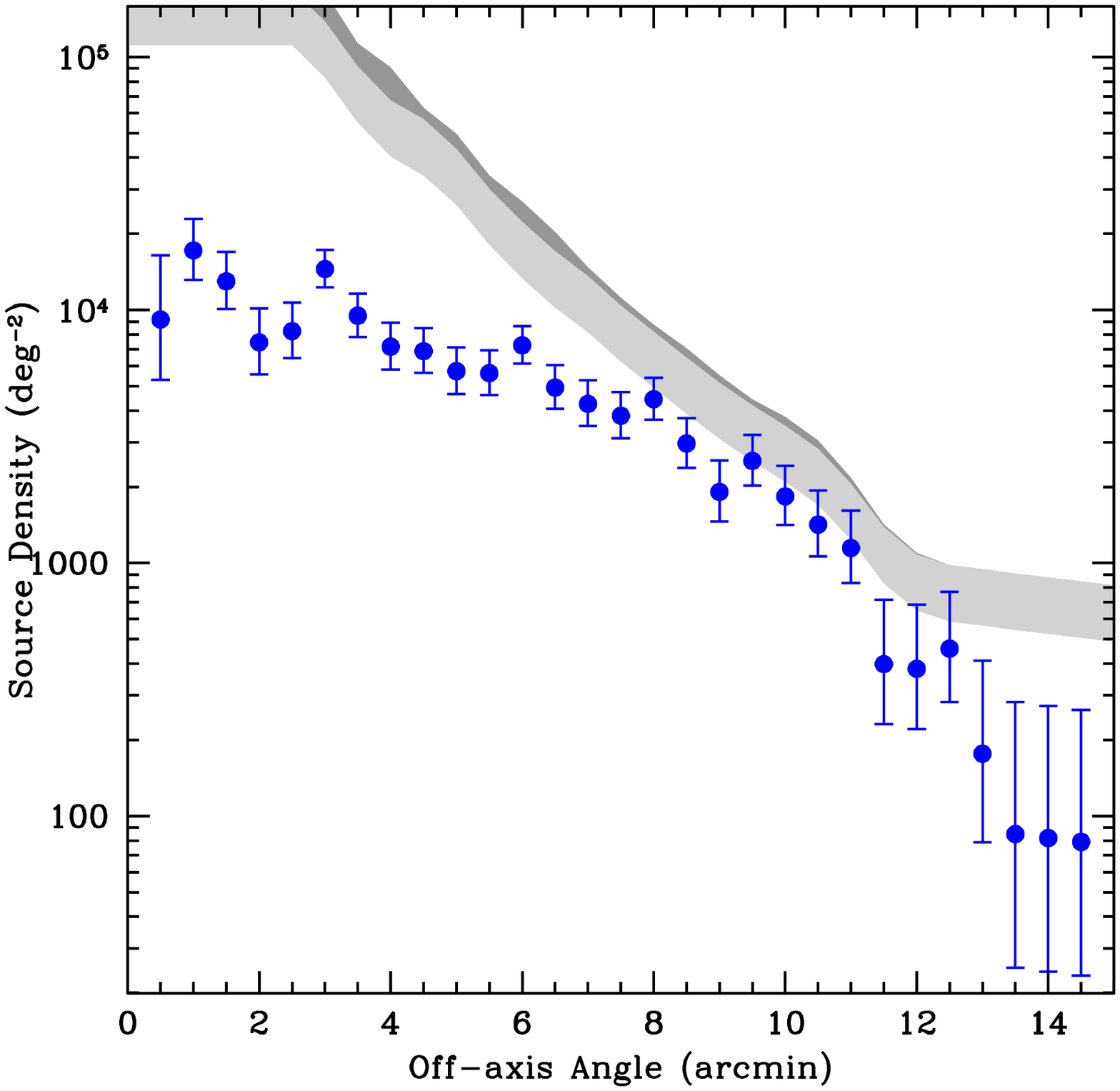}}
\vspace{-0.0truein}
\caption{Source density versus off-axis angle for the sources in the main \chandra\ catalog. The 1$\sigma$ uncertainties in the source density are shown as bars in the $y$-axis direction. The extent of the shaded regions correspond to the confusion limits assuming 30--50 beams per source; the confusion limits are calculated assuming the 70\% encircled-energy radii of the 1.5~keV (dark gray) and 4.5~keV (light gray) PSFs (see \S4.3). Only one source per bin is detected at off-axis angles $>$13$\arcmin$.}
\end{figure}

\clearpage

%
%

\begin{figure}
\epsscale{0.9}
\figurenum{A1}
\centerline{\includegraphics[width=12.0cm]{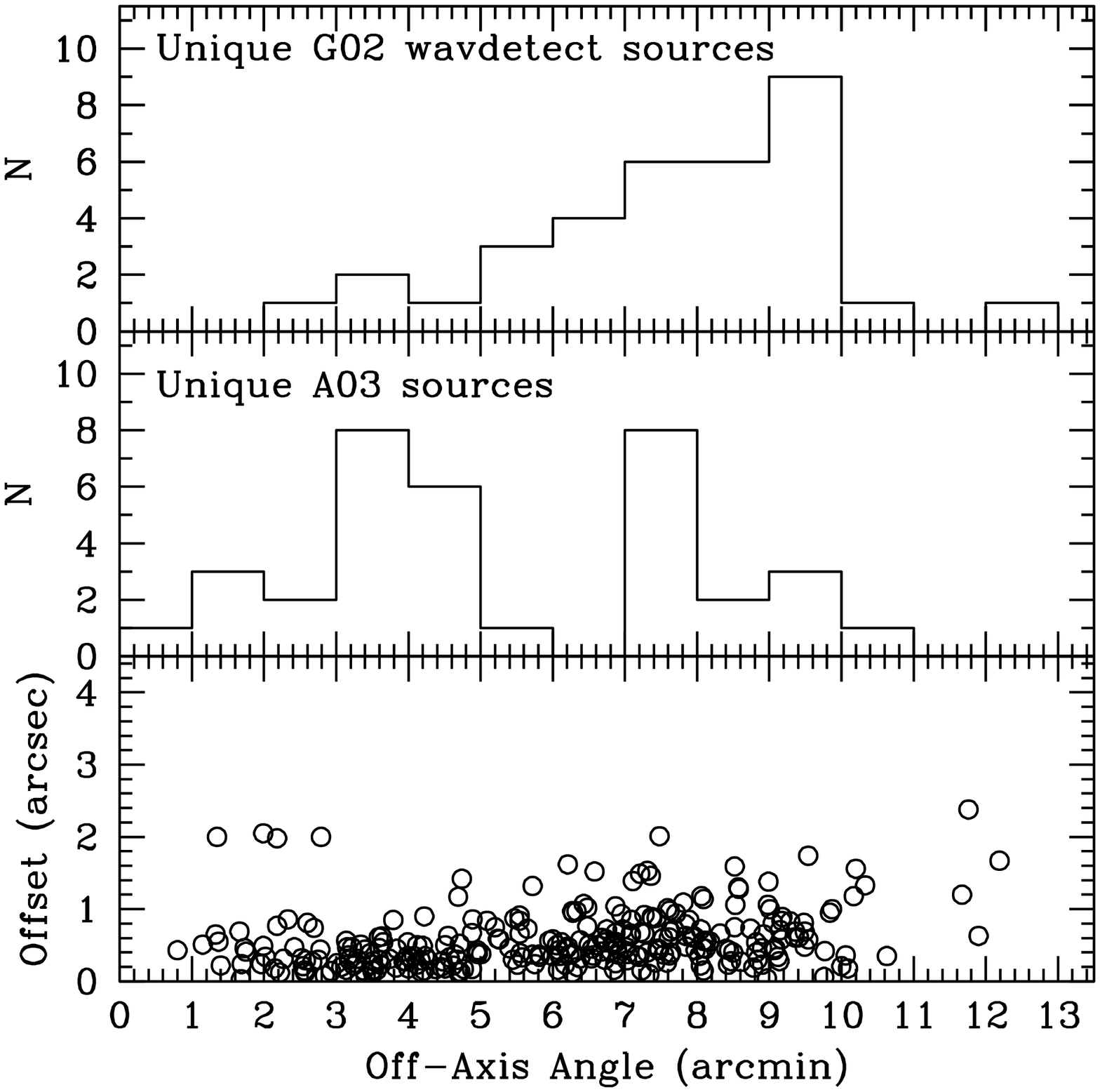}}
\caption{Comparison of the sources in the G02 {\sc wavdetect} catalog 
and the main \chandra\ catalog presented here (A03). The top and
middle panels show the distribution of off-axis angle for the sources
unique to each catalog. The absence of these sources in both catalogs
is mainly due to the adopted source detection strategies. The bottom
panel shows the offset in the source positions versus off-axis angle
for the sources common to both catalogs. When performing source
matching, we removed the source position offsets present in the G02
catalogs ($-1\farcs 2$ in RA and $0\farcs 8$ in Dec).}
\end{figure}

\clearpage

%
%

\begin{figure}
\epsscale{0.9}
\figurenum{A2}
\centerline{\includegraphics[width=12.0cm]{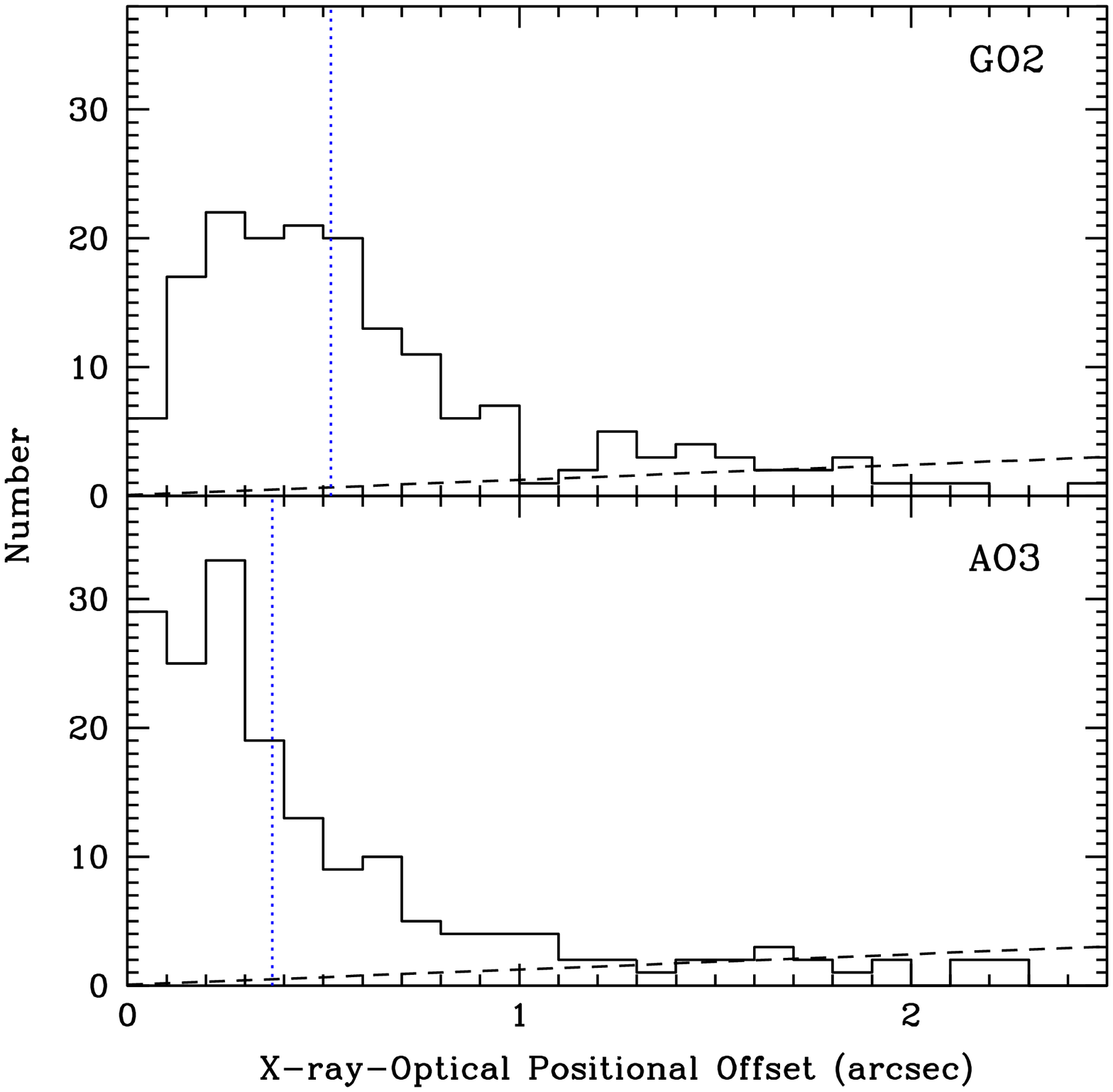}}
\caption{Histograms showing the distributions of X-ray-optical 
positional offset for the {\sc wavdetect} sources with $R<24$
counterparts common to both studies. The dotted lines indicate the
median positional offsets ($0\farcs 52$ for G02 and $0\farcs 37$ for
A03). The dashed line indicates the expected number of false matches;
these are calculated for each positional offset bin using the $R=24$
source density directly measured from the WFI observations.}
\end{figure}

\clearpage

%
%

\begin{figure}
\epsscale{0.9}
\figurenum{A3}
\centerline{\includegraphics[width=12.0cm]{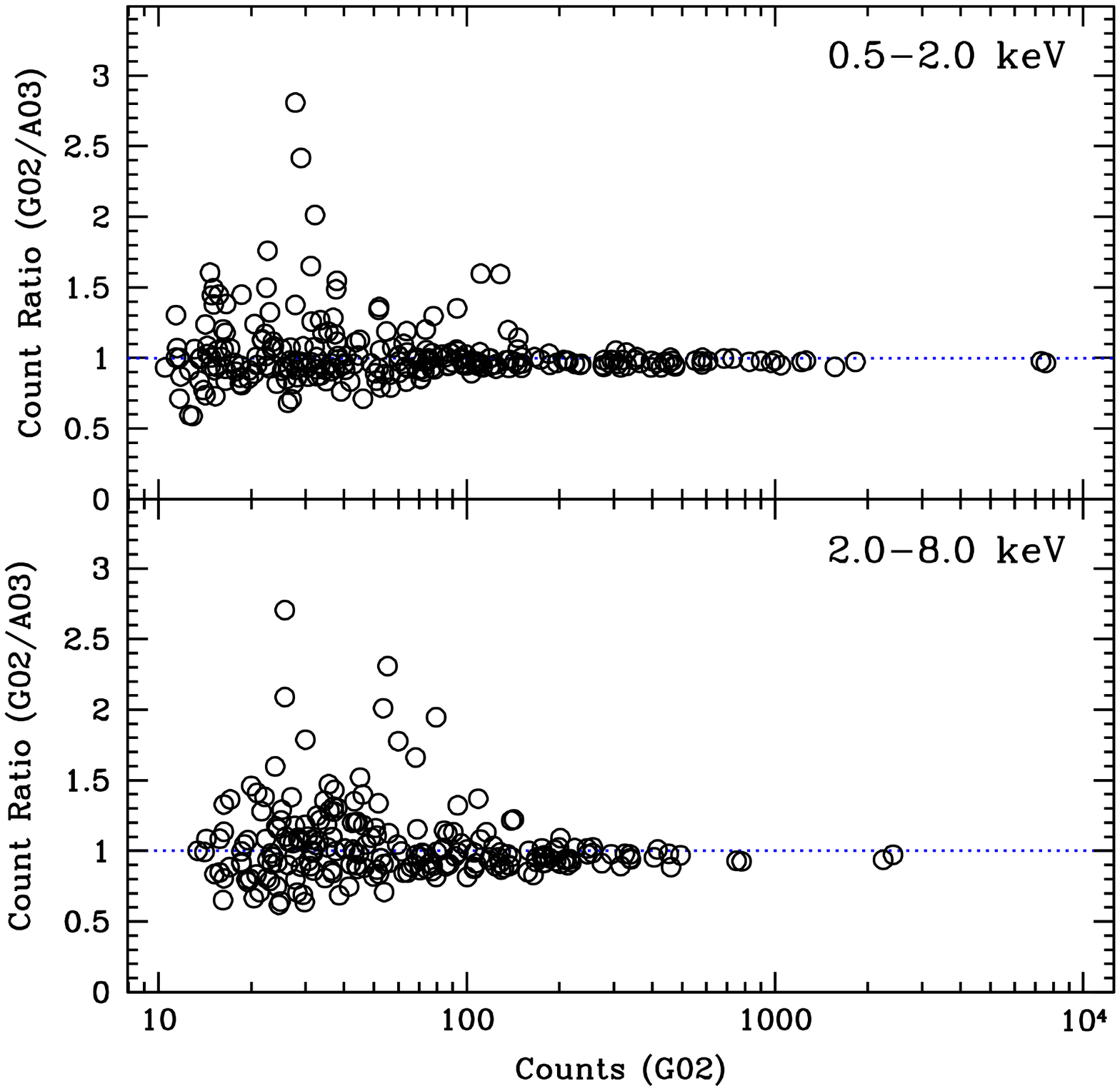}}
\caption{Source count ratio (G02/A03) versus G02 source counts
in the 0.5--2.0~keV (top) and 2--8~keV (bottom) bands for the {\sc
wavdetect} sources detected in both studies. Our source photometry
(A03) is systematically higher than that of G02 by $\approx$~2\% in
the 0.5--2.0~keV band and $\approx$~3\% in the 2--10~keV band. These
increases are in good agreement with the photometry-correction factor
estimated by G02 for their sources ($\approx$~4\%). The few outliers
(i.e.,\ those sources with count ratio disagreements of $>$50\%) are
generally sources that lie in regions where the background estimation
is problematic.}
\end{figure}

\clearpage

%
%

\begin{figure}
\epsscale{0.9}
\figurenum{A4}
\centerline{\includegraphics[width=12.0cm]{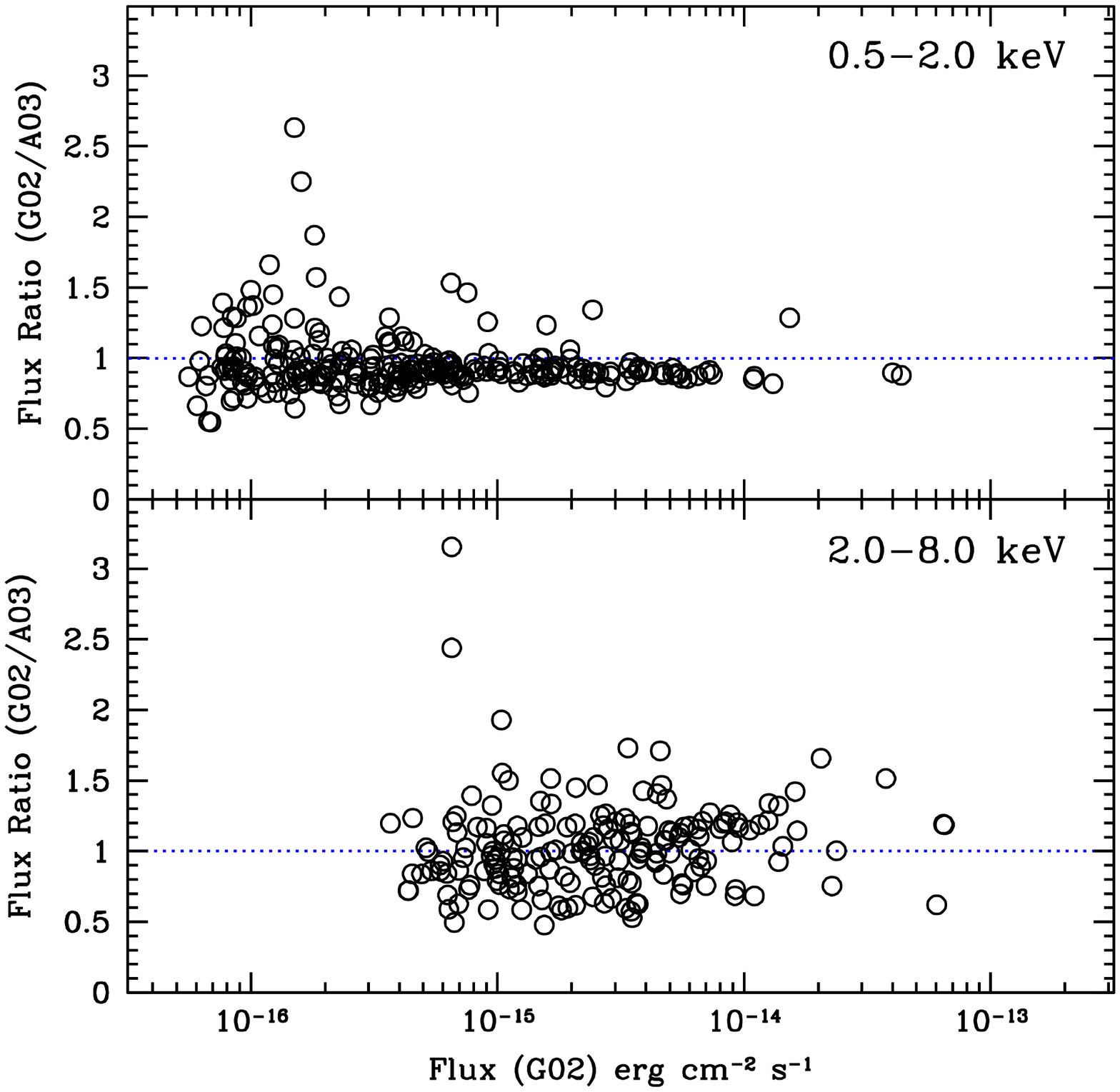}}
\caption{X-ray flux ratio (G02/A03) versus G02 X-ray flux in the 0.5--2.0~keV 
(top) and 2--8~keV (bottom) bands for the {\sc wavdetect} sources
detected in both studies. Our average X-ray fluxes (A03) are higher
than those of G02 by $\approx$~9\% in the 0.5--2.0~keV band and
$\approx$~18\% in the 2--8~keV band. There is some scatter,
particularly in the 2--8~keV band, due to differences in the adopted
spectral slopes.}
\end{figure}

\clearpage


\end{document}